\documentclass[%
 reprint,
superscriptaddress,
 amsmath,amssymb,
 aps,
prl,
]{revtex4-1}

\usepackage[T1]{fontenc}
\usepackage{graphicx}% Include figure files
\usepackage{dcolumn}% Align table columns on decimal point
\usepackage{bm}% bold math
\usepackage{hyperref}% add hypertext capabilities
\usepackage{xcolor}
\begin{document}

\preprint{APS/123-QED}

\title{Orbital Optical Raman Lattice}% Force line breaks with \\

\author{Zhi-Hao Huang}
\thanks{These authors contributed equally to this work.}
  %\altaffiliation[Also at ]{
 % }%Lines break automatically or can be forced with \\
 \affiliation{International Center for Quantum Materials and School of Physics, Peking University, Beijing 100871, China }

\affiliation{Hefei National Laboratory, Hefei 230088, China }
\author{Kou-Han Ma}
\thanks{These authors contributed equally to this work.}
\affiliation{International Center for Quantum Materials and School of Physics, Peking University, Beijing 100871, China }

\affiliation{Hefei National Laboratory, Hefei 230088, China }

\author{Bao-Zong Wang}
\affiliation{International Center for Quantum Materials and School of Physics, Peking University, Beijing 100871, China }

\affiliation{Hefei National Laboratory, Hefei 230088, China }

\author{W. Vincent Liu}%
\affiliation{Department of Physics and Astronomy and IQ Initiative, University of Pittsburgh, Pittsburgh, PA 15260, USA}

\affiliation{International Quantum Academy, Shenzhen 518048, China}

\author{Xiong-Jun Liu}%
  \email{xiongjunliu@pku.edu.cn}
\affiliation{International Center for Quantum Materials and School of Physics, Peking University, Beijing 100871, China }

\affiliation{Hefei National Laboratory, Hefei 230088, China }

\affiliation{International Quantum Academy, Shenzhen 518048, China}


%  \homepage{http://www.Second.institution.edu/~Charlie.Author}
% \affiliation{
%  Second institution and/or address\\
%  This line break forced% with \\
% }%
% \affiliation{
%  Third institution, the second for Charlie Author
% \affiliation{%
%  Authors' institution and/or address\\
%  This line break forced with \textbackslash\textbackslash
% }%

% \collaboration{CLEO Collaboration}%\noaffiliation

%              %  but any date may be explicitly specified

\begin{abstract}
    Spin and orbital are two basic degrees of freedom that play significant roles in exploring exotic quantum phases in optical lattices with synthetic spin-orbit coupling (SOC) and high orbital bands, respectively. Here, we combine these two crucial ingredients for the first time by proposing a completely new orbital optical Raman lattice scheme to explore exotic high-orbital Bose condensates with Raman-induced SOC in a square lattice. We find that both the SOC and $p$-orbital interactions influence the condensed state of bosons. Their interplay results in two novel high-orbital many-body quantum phases: the uniform angular momentum superfluid phase, which exhibits a global topological chiral orbital current characterized by a uniform Chern number, and the two-dimensional topological spin-orbital supersolid phase, which is characterized by the spin and orbital angular momentum density wave patterns and topological excitations with opposite Chern numbers, respectively protecting the chiral and antichiral edge modes in the neighboring supersolid clusters.
    Our scheme may open a new avenue for exploring exotic SOC and high-orbital physics in optical lattices, and is expected to advance the experimental realization of novel supersolids in higher dimensions.
\end{abstract}

\maketitle

%------------------------------------------------------------------------------------------------------
%
\paragraph*{Introduction.}
Ultracold atomic gases provide ideal platforms for quantum simulation due to their pristine nature and full controllability~\cite{bloch2008many,gross2017quantum}. Among them, investigations into synthetic gauge fields~\cite{Jaksch2003NJP,JuzeliPRL2004,OsterlohPRL2005,RuseckasPRL2005,LiuTPPP2006,dalibard2011colloquium,goldman2014light,ChenPRL2018,aidelsburger2022cold} and spin-orbit couplings (SOCs)~\cite{LiuPRL2009,JuzeliPRA2010,LinNature2011,CampbellPRA2011,SauPRB2011,WangPRL2012,CheukPRL2012,AndersonPRL2012,galitski2013spin,LiuPRL2014,zhai2015degenerate,Wu2016Science,BurdickPX2016,SongPRA2016,PhysRevA.93.053610,KolkowitzNatue2017,zhang2018spin,BaozongPRA2018} have attracted widespread interests. Over the past decade, experimental realizations of novel SOCs in one-dimension~\cite{LinNature2011,WangPRL2012,CheukPRL2012,BurdickPX2016,SongPRA2016,KolkowitzNatue2017}, two-dimension~\cite{Wu2016Science,huang2016experimental,meng2016experimental,sunPRL2018,sunPRL2018_1}, and three-dimension~\cite{wang2021realization} have enabled the simulation of various exotic topological models with cold atoms, such as topological semimetals~\cite{wang2021realization}, quantum anomalous Hall insulator~\cite{liang2023realization}, and non-Hermitian topological phases~\cite{zhao2023two}. Additionally, extending synthetic SOC to strongly correlated regime facilitates the realization of non-Abelian dynamical gauge fields~\cite{zhou2023non}. Besides simulating topological quantum phases~\cite{GoldmanPRL2010,CooperPRL2011,Liu2013PRL,Jotzu2014Experimental,WangPRL2014,PhysRevA.95.061601,SongScienceAdvances2018,yi2019observing}, the quest for supersolids characterized by diagonal and off-diagonal long-range order in ultracold atoms has never ceased~\cite{boninsegni2012colloquium,ritsch2013cold,recati2023supersolidity}. Currently, the supersolid phase has been observed in long-range interacting dipolar quantum gases~\cite{li2017stripe,bottcher2019transient,tanzi2019observation,chomaz2019long,norcia2021two,bland2022two,casotti2024observation} and optical cavities~\cite{leonard2017supersolid,leonard2017monitoring}.

High-orbital (e.g. $p$ and $d$) systems in optical lattices also received considerable attention due to their rich degrees of freedom, which can give rise to exotic orbital physics~\cite{ZhaoPRL2008,XiongjunPRA2010,CaiPRA2011,LewensteinNP2011,LiNC2013,TomaszPRL2013,dutta2015non,li2016physics,kock2016orbital}. With the orbital degree of freedom, one can not only %It is worth noting that the study of orbital physics in optical lattices not only
simulate the behavior of electrons in realistic materials but also, more crucially, uncover new concepts and phenomena that have no prior analog in electronic systems~\cite{li2016physics}, %. One important direction in this regard is the research on
such as the high-orbital Bose-Einstein condensates (BECs) with novel orbital ordering~\cite{Isacsson2005PRA,Kuklov2006PRL,liu2006atomic,wu2006quantum,LasonPRA2009,wirth2011evidence,li2012time,xu2016pi,di2016topological,wang2017diractopologicalphononsspinorbital,li2018rotation,PhysRevLett.121.265301,PhysRevLett.125.260402,jin2021evidence,wang2021evidence,huang2022intrinsic,wang2023evidence}. To date, significant progress have been made in the study of $p$-orbital BEC in optical lattices. For instance, the %condensed ground state of
$p$-orbital BEC in a square lattice manifests as a superfluid phase with a staggered angular momentum order that breaks time-reversal symmetry~\cite{liu2006atomic}. In a triangular lattice, it results in staggered loop current orders~\cite{wu2006quantum,wang2023evidence}, and in a $sp^2$-orbital hexagonal lattice, it exhibits Potts-nematic superfluid order~\cite{jin2021evidence} or atomic chiral superfluidity with topological excitations~\cite{wang2021evidence}. Such phases reveal the remarkable richness of higher orbital coherence whether or not breaking the time-reversal symmetry.
% Moreover, the ?¡ãdouble well?¡À lattice configuration has been widely used in various theoretical and experimental schemes to stabilize the ground state of $p$-orbital BEC due to its significant extension of the lifetime of $p$-orbital systems.

\begin{figure}[htb]
\includegraphics[scale=0.72]{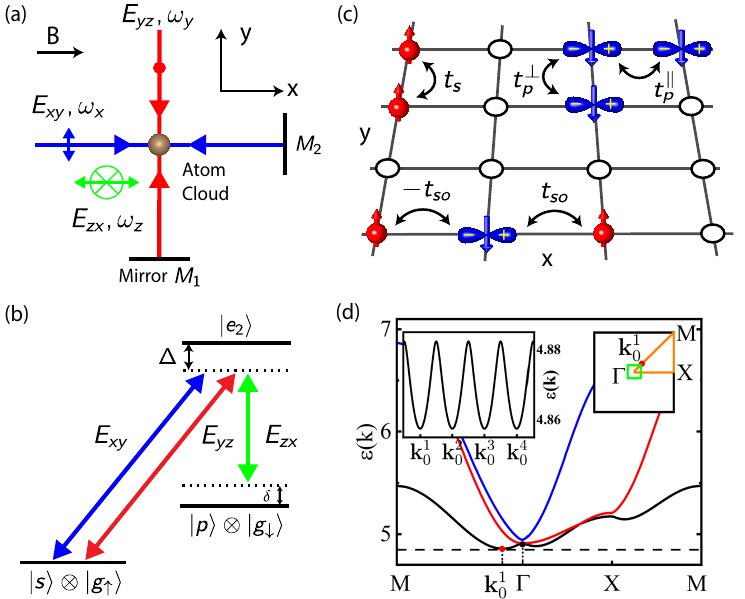}% Here is how to import EPS art
\caption{(a) Generic setup for orbital optical Raman lattice. The bias magnetic field $\boldsymbol{B}$ is parallel to the x direction. A pair of standing waves, \(\boldsymbol{E}_{xy}\) and \(\boldsymbol{E}_{yz}\), produce a square lattice. Moreover, the Raman field can be achieved by applying another plane wave $\boldsymbol{E}_{zx}$. (b) Optical-dipole transition diagram for Raman coupling in cold bosons ($^{87}\mathrm{Rb}$) coupled to two pairs of laser
beams ($\boldsymbol{E}_{xy}$, $\boldsymbol{E}_{zx}$) and ($\boldsymbol{E}_{yz}$, $\boldsymbol{E}_{zx}$). Here the hyperfine states $|1,-1\rangle$ and $|1,0\rangle$ of $^{87}\mathrm{Rb}$ are used to emulate the ground states $|g_{\uparrow,\downarrow}\rangle$. (c) The sketch for the nearest-neighbor spin-conserved hopping for $s$ and $p_{x,y}$ orbitals and the spin-flipped hopping between $s$ and $p_{x,y}$. Here, only the $p_{x}$ orbital is displayed, since the hopping terms for $p_{x}$ and $p_{y}$ are related by $C_{4}$ symmetry.
(d) Single-particle spectrum. The $C_{4}$-related points $\boldsymbol{k}_{0}^{1}$, $\boldsymbol{k}_{0}^{2}$, $\boldsymbol{k}_{0}^{3}$, $\boldsymbol{k}_{0}^{4}$ are the lowest energy points that lie on the diagonal lines in the Brillouin zone. Thereinafter, $V_0=5.0E_r$, $M_0=1.0E_r$ (tight-binding parameters: $t_{s}=0.0658E_{r}$, $t_{p}^{\parallel}=0.4228E_{r}$, $t_{p}^{\perp}=0.0658E_{r}$, $t_{so}=0.1059E_{r}$, $\mu_{s}=5.1909E_{r}$, and $\mu_{p}=5.9049E_{r}$), $m_{z}=0.0173E_{r}$. Besides, $\boldsymbol{k}_{0}^{1}=(0.154, 0.154)\pi$. $E_{r}$ is the recoil energy. }
\label{fig1}
\end{figure}

In this letter, we propose an orbital optical Raman lattice scheme that integrates the synthetic SOC  \cite{Wu2016Science,LiuPRL2014,BaozongPRA2018} and $p$-orbital condensates~\cite{Isacsson2005PRA,Kuklov2006PRL,liu2006atomic}, predicting two exotic quantum many-body phases with nontrivial interaction-induced topology in the square lattice: the uniform angular momentum superfluid (UAMSF) and the 2D topological spin-orbital supersolid (SOSS), both of which go beyond the scope of previous studies on synthetic SOC and high-orbital physics. % as characterized by different spin and orbital angular momentum orders. which both exhibit non-trivial topological excitations.
We show that these condensed states of bosons are governed by the nontrivial interplay effects between on-site $p$-orbital interactions and Raman-induced SOC, surpassing the limitations of pure orbital lattice gas models without synthetic SOC. %, whose interplay leads to two distinct quantum phases characterized by different angular momentum orders. %significantly stronger than
The UAMSF exhibits a uniform angular momentum order, supporting uniform chiral topological excitations. In contrast, the 2D topological SOSS shows staggered spin and angular momentum cluster patterns, leading to the chiral and antichiral topological edge excitations in the neighboring supersolid clusters. Our work offers a new perspective for the combined study of nontrivial synthetic SOC and high-orbital physics, which were previously considered separate fields in optical lattice research, and will promote the realization of novel supersolids without long-range interactions in ultracold atom systems.

\paragraph*{Model.}
We start with the Hamiltonian for ultracold bosons trapped in an orbital optical Raman lattice, given by $H_{0}= \frac{p_{x}^{2}}{2m}+\frac{p_{y}^{2}}{2m}+V(x,y) +M(x,y)\sigma_x+m_{z}\sigma_z$. Here, $V(x,y)=-V_{0}[\cos^{2}(\frac{\pi}{a}x)+\cos^{2}(\frac{\pi}{a}y)]$ and $M(x,y)=-M_{0}[\cos(\frac{\pi}{a}x)+\cos(\frac{\pi}{a}y)]$ represent the normal lattice potential and Raman field, respectively. The Pauli matrices $\boldsymbol{\sigma}$ act on the subspace spanned by $|s\rangle\otimes|g_{\uparrow}\rangle$ and $|p\rangle\otimes|g_{\downarrow}\rangle$, where the spin $|g_{\uparrow,\downarrow}\rangle$ is defined from hyperfine ground states, $|s\rangle$ and $|p\rangle$ represent local orbital states. The model Hamiltonian $H_{0}$ can be realized with high feasibility, through the three-laser configuration shown in Fig.~\ref{fig1}(a). %, three lasers are used to generate the orbital optical Raman lattice.
Two lasers form standing waves, \(\boldsymbol{E}_{xy}=\hat{y}E_{0}\cos(k_{0}x)\) and \(\boldsymbol{E}_{yz}=\hat{z}E_{0}\cos(k_{0}y)\), linearly polarized along \(y\) and \(z\) axes, respectively, and propagating in \(x\)-\(y\) plane. The third laser forms a plane wave, \(\boldsymbol{E}_{zx}=\hat{x}E_{z}e^{-\mathrm{i}k_{z}z}\) propagating along \(z\) direction and is \(x\)-polarized. The standing waves \(\boldsymbol{E}_{xy}\) and \(\boldsymbol{E}_{yz}\) coupling to excited states with a red detuning \(\Delta\) %induce transitions from the ground states \(|g_{\uparrow,\downarrow}\rangle\) to the relevant excited states,
contribute to the diagonal square optical lattice potentials \(V(x,y)=\hbar(|\boldsymbol{E}_{xy}|^{2}/\Delta+|\boldsymbol{E}_{yz}|^{2}/\Delta)\). Additionally, \(\boldsymbol{E}_{xy}\), \(\boldsymbol{E}_{yz}\), and \(\boldsymbol{E}_{zx}\) induce the Raman field components \(M(x)\) and \(M(y)\) via two two-photon transitions as illustrated in a \(\boldsymbol{\varLambda}\)-type configuration [Fig.\ref{fig1}(b)]. In experiment one can easily set that the Raman field only couples \(|s\rangle\otimes|g_{\uparrow}\rangle\) and \(|p\rangle\otimes|g_{\downarrow}\rangle\) by putting such two states be nearly resonant for the two-photon transitions %and the excited state $|e_2\rangle$ \cite{SM}.
except for a tunable two-photon detuning \(\delta\), which defines an effective Zeeman splitting \(m_{z}=\hbar\delta/2\), while all other orbital states are far detuned~\cite{SM}.

%To derive the tight-binding model of $H_{0}$, bosons are taken to occupy the $\phi_{s\uparrow}$ and $\phi_{p_{x,y}\downarrow}$ orbitals and only the nearest-neighbor hopping terms are retained. Thus the tight-binding Hamiltonian is given by
With the above implementation scheme, bosons can occupy the spin-orbital locking states $\phi_{s_{\uparrow}}$, $\phi_{p_{x\downarrow}}$, and $\phi_{p_{y\downarrow}}$. As shown in Fig.\ref{fig1}(c), the lattice potential $V(x,y)$ governs spin-conserved nearest-neighbour hopping ($t_s$ and $t_p$), while the Raman field $M(x,y)$ contributes to spin-flipped nearest-neighbour hopping ($t_{so}$).
% the tight-binding form of Hamiltonian $H_{0}$ is given by
%  $\hat{H}_{0}=-t_{s}\sum_{\langle\boldsymbol{i},\boldsymbol{j}\rangle}b_{\boldsymbol{i},s,\uparrow}^{\dagger}b_{\boldsymbol{j},s,\uparrow}+\sum_{\boldsymbol{i},\mu,\nu}t_{p}^{\mu\nu}(b_{\boldsymbol{i},p_{\mu},\downarrow}^{\dagger}b_{\boldsymbol{i}+\boldsymbol{e}_{\nu},p_{\mu},\downarrow}+\mathrm{H.c.})+\sum_{\boldsymbol{i},\nu}[(\mu_{s}+m_{z})n_{\boldsymbol{i},s,\uparrow}+(\mu_{p}-m_{z})n_{\boldsymbol{i},p_{\nu},\downarrow}]+\sum_{\boldsymbol{i},\nu,\boldsymbol{\delta}_{\nu}}[-t_{so}^{(\boldsymbol{i},\boldsymbol{i}+\boldsymbol{\delta}_{\nu})}b_{\boldsymbol{i},s,\uparrow}^{\dagger}b_{\boldsymbol{i}+\boldsymbol{\delta}_{\nu},p_{\nu},\downarrow}+\mathrm{H.c.}]$, where $t_{s}$ and $t_p^{\mu\nu}=t_p^{\parallel}\cdot \delta_{\mu\nu} - t_p^{\perp}\cdot ( 1 - \delta_{\mu\nu} )$ ($\mu$, $\nu$=$x$, $y$) represent the spin-conserved hopping shown in Fig.\ref{fig1}(c), $n_{\boldsymbol{i},s,\uparrow}=b_{\boldsymbol{i},s,\uparrow}^{\dagger}b_{\boldsymbol{i},s,\uparrow}$ ($n_{\boldsymbol{i},p_{\nu},\downarrow}=b_{\boldsymbol{i},p_{\nu},\downarrow}^{\dagger}b_{\boldsymbol{i},p_{\nu},\downarrow}$) is particle number operator, and $\boldsymbol{\delta}_{\nu}=\pm \boldsymbol{e}_{\nu}$. $\mu_{s}$ and $\mu_{p}$ denote the $s$ and $p$ orbitals effective on-site energies, respectively.
Since the Raman field $M(x,y)$ has twice the period of the optical lattice potential $V(x,y)$ and is symmetric with respect to each lattice site center of $V(x,y)$, the spin-flipped hopping is staggered in the $\nu$ ($\nu=x,y$) direction with $t_{so}^{(\boldsymbol{i},\boldsymbol{i}\pm\boldsymbol{e}_{v})}=\pm(-1)^{i_{v}}t_{\mathrm{SO}}$ \cite{SM}. %Absorbing this staggered property
The staggered factor $(-1)^{i_{\nu}}$ represents a nontrivial momentum transfer between spin-up and spin-down states in the Raman coupling, and can be absorbed by the transformation $b_{\boldsymbol{i},p_{\nu},\downarrow}\rightarrow -(-1)^{i_{\nu}}b_{\boldsymbol{i},p_{\nu},\downarrow}$, yielding the tight-binding Hamiltonian
\begin{align}
\hat{H}_{0}= & (\mu_{s}+m_{z})\sum_{\boldsymbol{i}}n_{\boldsymbol{i},s,\uparrow}+(\mu_{p}-m_{z})\sum_{\boldsymbol{i},\nu}n_{\boldsymbol{i},p_{\nu},\downarrow}\nonumber \\
- & t_{s}\sum_{\langle\boldsymbol{i},\boldsymbol{j}\rangle}b_{\boldsymbol{i},s,\uparrow}^{\dagger}b_{\boldsymbol{j},s,\uparrow}-\sum_{\boldsymbol{i},\mu,\nu}(\bar{t}_{p}^{\mu \nu}b_{\boldsymbol{i},p_{\mu},\downarrow}^{\dagger}b_{\boldsymbol{i}+\boldsymbol{e}_{\nu},p_{\mu},\downarrow}+\mathrm{H.c.})\nonumber \\
+ & t_{\mathrm{SO}}\sum_{\boldsymbol{i},\nu}\sum_{\delta_{\nu}=\pm\boldsymbol{e}_{v}} \Big[ \text{sgn}(\delta_{\nu})
b_{\boldsymbol{i},p_{\nu},\downarrow}^{\dagger}b_{\boldsymbol{i}+\delta_{\nu},s,\uparrow} +\mathrm{H}.\mathrm{c.}\Big],\label{2}
\end{align}
where $b_{\boldsymbol{i},l,\sigma}^{\dagger}$ ($b_{\boldsymbol{i},l,\sigma}$) ($l=s,p_{x},p_{y}$) denotes the creation (annihilation) operator of boson. $\bar{t}_{p}^{\mu\nu}=t_{p}^{\parallel}\delta_{\mu\nu}+t_{p}^{\perp}(1-\delta_{\mu\nu})$ ($\mu$, $\nu$=$x$, $y$), $n_{\boldsymbol{i},l,\sigma}=b_{\boldsymbol{i},l,\sigma}^{\dagger}b_{\boldsymbol{i},l,\sigma}$, and $\mu_{s,p}$ denote the on-site energies. %and ?¡ãsgn?¡À is a sign function.
A key feature is that the momentum transfer by Raman field forces the band minima of $\epsilon_{p_{x}\downarrow}(\boldsymbol{k})$ ($\epsilon_{p_{y}\downarrow}(\boldsymbol{k})$) to move from $\boldsymbol{Q}_{x}=(\pi, 0)$ ($\boldsymbol{Q}_{y}=(0,\pi)$) to $\boldsymbol{\Gamma}$ \cite{SM}.
% Remarkably, the longitudinal hopping of the p-orbital, with $-t_{p}^{\parallel}<0$, can be attributed to the relative momentum shift between the s and p orbitals induced by SOC.
Further, the remaining 2D SOC effect can result in a single-particle spectrum with four $C_{4}$-connected minima at $\boldsymbol{k}_{0}^{1}$, $\boldsymbol{k}_{0}^{2}$, $\boldsymbol{k}_{0}^{3}$, and $\boldsymbol{k}_{0}^{4}$ [Fig.\ref{fig1}(d)], leading to the major novel physics predicted in the present orbital optical Raman lattice. A quadratic band touch point protected by time-reversal symmetry exists at $\boldsymbol{\Gamma}$ \cite{SunNP2012}.
\begin{figure}[t]
\centering
\includegraphics[scale=0.44]{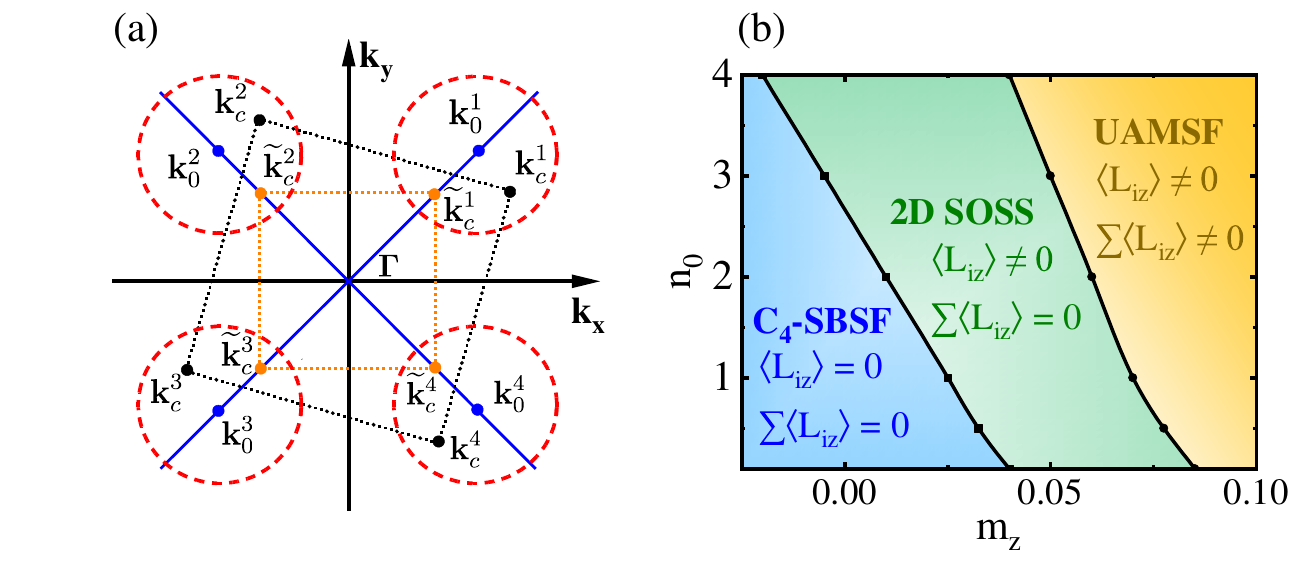}%
\caption{(a) Schematic of the variational calculation. Here, $C_4$-connected points $\boldsymbol{k}_{0}^{n}$, $\boldsymbol{k}_{c}^{n}$, and $\tilde{\boldsymbol{k}}_{c}^{n}$ ($n=1\sim4$) represent the energy minima of the single-particle spectrum, the variational momenta within the areas surrounded by red circles, and the condensation momenta along the blue diagonal lines, respectively.
(b) Ground state phase diagram. $m_{z}$ and $n_0$ denote the effective Zeeman splitting and boson condensation density, respectively. There are three types of ground states: $C_4$-symmetry-broken superfluid phase ($C_4$-SBSF), two-dimensional spin-orbital supersolid phase (2D SOSS), and uniform angular momentum superfluid phase (UAMSF).
Parameters: $V_0=5.0E_r$, $M_0=1.0E_r$, and $g=0.01E_r$ ($U_{s}=0.1044E_{r}$, $U_{sp}=0.0428E_{r}$, $U_{p}=0.0618E_{r}$, and $\widetilde{U}_{p}=0.0175E_{r}$). }
\label{fig2}
\end{figure}

The total Hamiltonian includes also the contact interaction $V(\boldsymbol{r}-\boldsymbol{r}^{\prime})=g_{\sigma\sigma^{\prime}}\delta(\boldsymbol{r}-\boldsymbol{r}^{\prime})$ ($\sigma,\sigma^{\prime}=\uparrow,\downarrow$), whose form in the lattice model reads \cite{SM}
\begin{align}
\hat{H}_{\mathrm{int}}= & \sum_{\boldsymbol{i}}\Big[\frac{U_{s}}{2}n_{\boldsymbol{i},s,\uparrow}(n_{\boldsymbol{i},s,\uparrow}-1)+U_{sp}n_{\boldsymbol{i},s,\uparrow}\sum_{\nu}n_{\boldsymbol{i},p_{\nu},\downarrow}\nonumber \\
+ & \frac{U_{p}}{2}\sum_{\nu}n_{\boldsymbol{i},p_{\nu},\downarrow}(n_{\boldsymbol{i},p_{\nu},\downarrow}-1)+2\widetilde{U}_{p}n_{\boldsymbol{i},p_{x},\downarrow}n_{\boldsymbol{i},p_{y},\downarrow}\nonumber \\
+ & \frac{\widetilde{U}_{p}}{2}(b_{\boldsymbol{i},p_{x},\downarrow}^{\dagger}b_{\boldsymbol{i},p_{x},\downarrow}^{\dagger}b_{\boldsymbol{i},p_{y},\downarrow}b_{\boldsymbol{i},p_{y},\downarrow}+\mathrm{H.c.})\Big],\label{3}
\end{align}
where $U_{s}=g_{\uparrow\uparrow}\int\mathrm{d}\boldsymbol{r}|\phi_{s,\uparrow}(\boldsymbol{r}-\boldsymbol{r}_{\boldsymbol{i}})|^{4}$, and the coefficients $U_{sp}$, $U_{p}$, and $\widetilde{U}_{p}$ can also be similarly defined with $g_{\sigma\sigma^{\prime}}\approx g$. Note that $U_{p}$ refers to the intra-orbital interaction for the $p_{\nu\downarrow}$ ($\nu=x,y$) orbitals, while $\widetilde{U}_{p}$ refers to the inter-orbital interaction between the $p_{x\downarrow}$ and $p_{y\downarrow}$ orbitals.

%where  $U_{l,\tilde{l}}=g\int\mathrm{d}\boldsymbol{r}\phi_{l\sigma}^{2}(\boldsymbol{r}-\boldsymbol{r}_{i})\phi_{\tilde{l}\tilde{\sigma}}^{2}(\boldsymbol{r}-\boldsymbol{r}_{i})$ ($l\sigma,\tilde{l}\tilde{\sigma}=s_{\uparrow},p_{x\downarrow},p_{y\downarrow}$) are interaction coefficients and $g_{\uparrow\uparrow}\approx g_{\uparrow\downarrow}\approx g_{\downarrow\downarrow}=g$.
% Note that the $p$-orbital interaction in Eq.(\ref{3}) can be recast as  $\hat{H}_{\mathrm{int}}^{p}=\sum_{\boldsymbol{i}}\frac{U_{p}}{2}[n_{\boldsymbol{i},p,\downarrow}(n_{\boldsymbol{i},p,\downarrow}-\frac{2}{3})-\frac{1}{3}L_{\boldsymbol{i},z}^{2}]$ under the harmonic approximation \cite{liu2006atomic,SM}, where the angular momentum operator is given by $L_{\boldsymbol{i},z}=-\mathrm{i}b_{\boldsymbol{i},p_{x},\downarrow}^{\dagger}b_{\boldsymbol{i},p_{y},\downarrow}+\mathrm{H.c.}$.
% It is worth mentioning that the $p$-interaction prefers the angular momentum order to be independent of the harmonic approximation, and the generation of the angular momentum order is mainly due to the pair transition interaction between $p$-orbitals [the last term in Eq.\eqref{3}].

\paragraph*{Ground state ansatz and phase diagram.}
Usually, the non-interacting BEC will take place at the single-particle band minimum points $\boldsymbol{k}_{0}^{n}$ ($n=1\sim4$). However, since these lowest energy points of Hamiltonian $\hat{H}_{\mathrm{0}}$ do not generally correspond to the energy minima of the total Hamiltonian $\hat{H}_{\mathrm{0}} + \hat{H}_{\mathrm{int}}$, the interaction changes the populations in the $s_{\uparrow}$, $p_{x\downarrow}$, and $p_{y\downarrow}$ orbitals, which shifts the condensation momenta around $\boldsymbol{k}_{0}^{n}$ ($n=1\sim4$). Under weak interaction, the condensed ground state ansatz can be given by $|g\rangle \sim e^{\sqrt{N_{0}}b^{\dagger}}|\mathrm{vac}\rangle$, where $b^{\dagger}$, expressed as $b^{\dagger}=\sum_{n=1}^{4}\sum_{l_{\sigma}}\gamma_{\boldsymbol{k}_{c}^{n}}\beta_{\boldsymbol{k}_{c}^{n},l_{\sigma}}b_{\boldsymbol{k}_{c}^{n},l_\sigma}^{\dagger}$ ($l_{\sigma}=s_{\uparrow}, p_{x\downarrow}, p_{y\downarrow}$),  represents the quasi-particle creation operator and $N_0$ is the particle number of BEC.
Here, $C_{4}$-related points $\boldsymbol{k}_{c}^{n}$ [Fig.\ref{fig2}(a)] denote the variational momenta, with the variational parameters $\gamma_{\boldsymbol{k}_{c}^{n}}$ and $\beta_{\boldsymbol{k}_{c}^{n}}$ satisfying $\sum_{n}|\gamma_{\boldsymbol{k}_{c}^{n}}|^{2}=1$ and $\sum_{l_{\sigma}}|\beta_{\boldsymbol{k}_{c}^{n},l_{\sigma}}|^{2}=1$, respectively. Note that $|\gamma_{\boldsymbol{k}_{c}^{n}}|^{2}$ represents the condensation distribution probability of bosons at $\boldsymbol{k}_{c}^{n}$, and $|\beta_{\boldsymbol{k}_{c}^{n},l_{\sigma}}|^{2}$ denotes the orbital population probability. To identify the condensation momenta $\tilde{\boldsymbol{k}}_{c}^{n}$ and the condensation parameters ($\gamma_{\tilde{\boldsymbol{k}}_{c}^{n}}$, $\beta_{\tilde{\boldsymbol{k}}_{c}^{n},l_{\sigma}}$), we search for the minimum of the energy density functional $\langle g|\hat{\mathcal{H}}_{0} + \hat{\mathcal{H}}_{\mathrm{int}}|g\rangle / N$ by simulated annealing algorithm, with $N$ the number of lattice sites~[see Supplementary for details~\cite{SM}].
As depicted in Fig.\ref{fig2}(a),
$\tilde{\boldsymbol{k}}_{c}^{n}$ ($n=1\sim4$) lie on the blue diagonal lines and move toward $\boldsymbol{\Gamma}$ (or coincide with it), and are connected by $C_{4}$ symmetry.

Next, our calculation finds that the interplay
between the $p$-orbital interaction in Eq.(\ref{3}) and the Raman-induced 2D SOC is highly nontrivial and leads to the emergence of exotic orbital orders with distinct novel topology in the condensate state $|G\rangle$. The $p$-orbital interaction in the present orbital optical Raman lattice tends to generate uniform non-zero orbital angular momentum (OAM) order $\langle G|L_{i,z}|G\rangle$ ($L_{i,z}=-\mathrm{i}b_{\boldsymbol{i},p_{x},\downarrow}^{\dagger}b_{\boldsymbol{i},p_{y},\downarrow}+\mathrm{H.c.}$) \cite{liu2006atomic,SM} to reduce the interaction energy, leading to $\beta_{\tilde{\boldsymbol{k}}_{c}^{n},p_{x\downarrow}}=\mathrm{\pm i}\beta_{\tilde{\boldsymbol{k}}_{c}^{n},p_{y\downarrow}}$. However, the Raman-induced 2D SOC in Eq.(\ref{2}) prefers $\beta_{\tilde{\boldsymbol{k}}_{c}^{n},p_{x\downarrow}}=\pm\beta_{\tilde{\boldsymbol{k}}_{c}^{n},p_{y\downarrow}}$, suppressing the uniform OAM $\langle L_{i,z}\rangle$. Thus, it is this competitive relationship between them that gives rise to three distinct phases:
(1) When the SOC dominates over the $p$-orbital interaction, the condensate will occur at one of the $\tilde{\boldsymbol{k}}_{c}^{n}$ ($n=1\sim4$) with $\beta_{\tilde{\boldsymbol{k}}_{c},p_{x\downarrow}}=\pm\beta_{\tilde{\boldsymbol{k}}_{c},p_{y\downarrow}}$, resulting in a $C_{4}$-symmetry-broken superfluid phase without OAM order; (2) In contrast, if the $p$-orbital interaction plays the leading role, the bosons condense at the $\boldsymbol{\Gamma}$ point with $\beta_{\boldsymbol{\Gamma},p_{x\downarrow}}=\mathrm{-i}\beta_{\boldsymbol{\Gamma},p_{y\downarrow}}$, and the condensate exhibits uniform OAM order; (3) Importantly, in the intermediate regime, where the $p$-orbital interaction and SOC are comparably strong, the bosons condense equally at the $\tilde{\boldsymbol{k}}_{c}^{n}$ ($n=1$ $\sim$ 4), with the orbital parameters given by
    $\beta_{\tilde{\boldsymbol{k}}_{c}^{n},p_{x\downarrow}}=-(-1)^{n}\beta_{\tilde{\boldsymbol{k}}_{c}^{n},p_{y\downarrow}}$, rendering the total system a novel OAM density-wave pattern known as the SOSS phase. Fig.\ref{fig2}(b) shows the phase diagram plotted versus condensate density $n_0$ and $m_z$, which govern the $p$-orbital interaction. %depends on the effective Zeeman splitting $m_{z}=\hbar\delta/2$ (the relative spacing between $\epsilon_{s\uparrow}$ and $\epsilon_{p\downarrow}$ bands, can be controlled by $m_{z}$), the boson condensation density $n_{0}=N_{0}/N$, and the contact interaction coefficient $g$. Therefore, we can obtain a phase diagram that includes the three exotic ground states in the $m_z-n_0$ plane [Fig.\ref{fig2}(b)].
    A most important feature of the phases in (2) and (3) [orange and green areas in Fig.\ref{fig2}(b)] is that they host different types of topological quasiparticles, whose emergence necessitates both the $p$-orbital interaction and the 2D SOC. We examine them below in detail. %their unique OAM orders, as well as the topological quasiparticle excitations.
% and the energy spacing $\epsilon_{s_{\uparrow}}(\boldsymbol{\Gamma})-\epsilon_{p_{x(y)\downarrow}}(\boldsymbol{\Gamma})$ (effective Zeeman splitting $m_{z}$)

\paragraph*{Uniform angular momentum superfluid.}
\begin{figure}[b]
\centering
\includegraphics[scale=0.73]{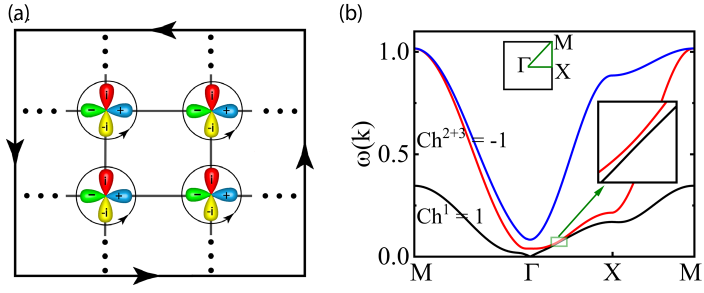}%
\caption{ (a) OAM order $\langle L_{i,z} \rangle$ and (b) topological quasiparticle excitation spectrum for UAMSF. The parameters: $m_{z}=0.08E_{r}$, $n_{0}=2.0$. Other parameters are taken as the same as those in Fig.\ref{fig2}.  %$\mu=4.9270E_{r}$
}
\label{fig3}
 \end{figure}
Adjust the Zeeman splitting $m_{z}$ such that onsite energy $\epsilon_{p\downarrow}(\boldsymbol{\mathrm{\Gamma}})$ is well below $\epsilon_{s\uparrow}(\boldsymbol{\mathrm{\Gamma}})$. With the dominant $p$-orbital population, the $p$-orbital interaction governs the ground state. Consequently, bosons condense at the $\boldsymbol{\mathrm{\Gamma}}$ with orbital parameters $\beta_{\boldsymbol{\mathrm{\Gamma}},s_{\uparrow}}=0$ and $\beta_{\boldsymbol{\mathrm{\Gamma}},p_{x\downarrow}}=\mathrm{-i}\beta_{\boldsymbol{\mathrm{\Gamma}},p_{y\downarrow}}$. As shown in Fig.\ref{fig3}(a), this ground state exhibits a uniform OAM order $\langle L_{\boldsymbol{i},z}\rangle= 2n_{0}|\beta_{\boldsymbol{\mathrm{\Gamma}},p_{y\downarrow}}|^2$ in real space, rendering the UAMSF which breaks time-reversal symmetry. While the ground state is dominated by $p$-orbital interaction, the quasiparticle excitations are actually governed by both the $p$-orbital interaction and the 2D SOC, and exhibit nontrivial topology characterized by Chern numbers.
\begin{figure}[t]
%\centering
\includegraphics[scale=0.16]{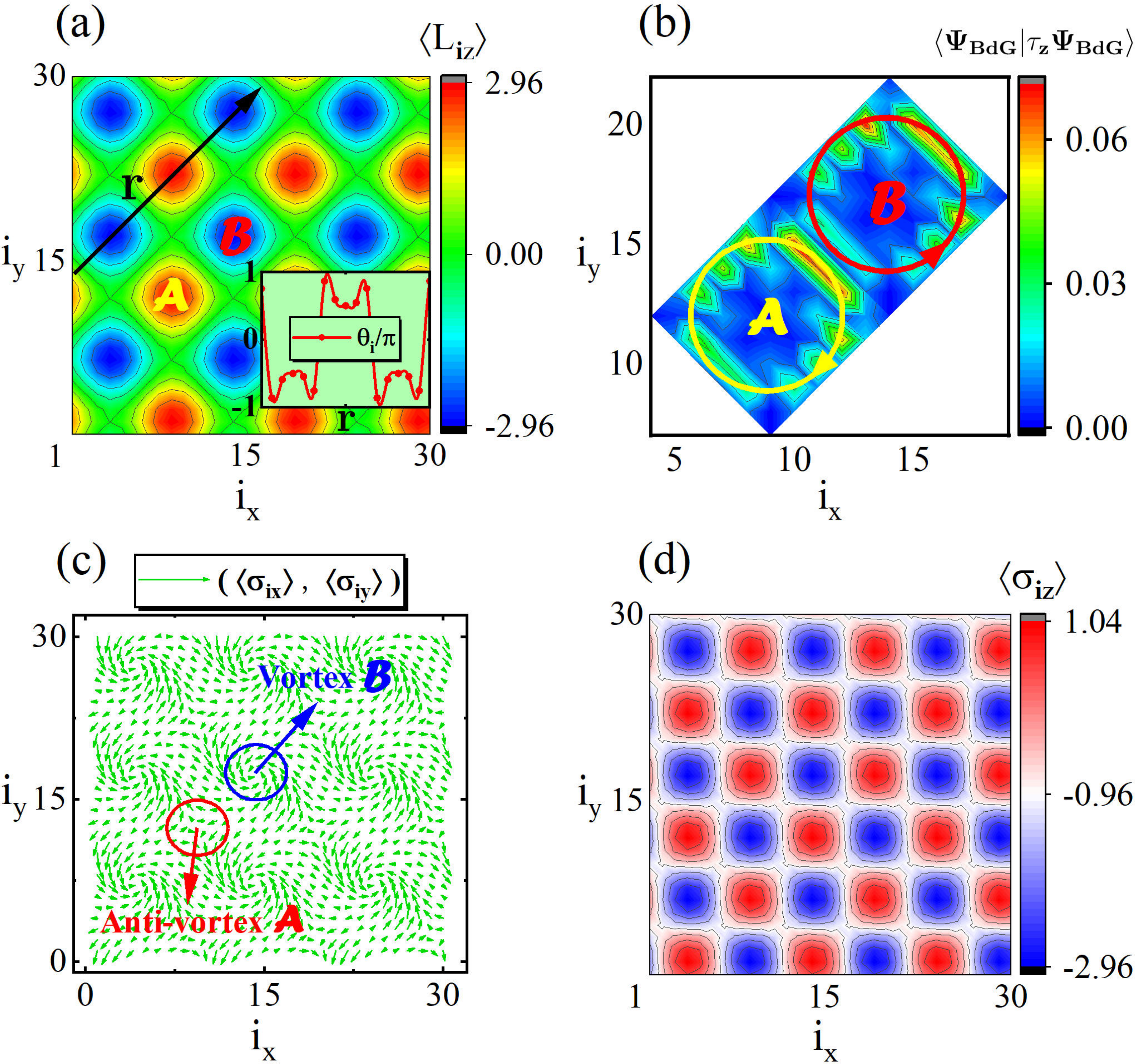}%
\caption{(a) Spatial distribution of $\langle L_{\boldsymbol{i},z}\rangle$ for 2D SOSS phase. The positive ($\langle L_{\boldsymbol{i},z}\rangle>0$) and negative ($\langle L_{\boldsymbol{i},z}\rangle<0$) OAM clusters are staggered, with $\sum_{\mathrm{Global}} \langle L_{\boldsymbol{i},z}\rangle = 0$. Inset: the relative phase difference $\theta_i=\text{Im}\Big[\mathrm{ln}\langle b_{\boldsymbol{i},p_{x},\downarrow}^{\dagger}b_{\boldsymbol{i},p_{y},\downarrow}\rangle\Big]$ between the $p_x$ and $p_y$ orbitals along path $\bf{r}$. (b) Topological edge excitations of the positive A and negative B clusters. The arrows represent the flow directions of the edge currents. (c) Spin field $(\langle \sigma_{\boldsymbol{i},x} \rangle, \langle \sigma_{\boldsymbol{i},y} \rangle)$ displays a topological pattern (skyrmion). (d) Spin component $\langle \sigma_{\boldsymbol{i},z} \rangle$ presents another staggered high-and low-density clusters pattern. Here, parameters are set as $m_{z}=0.06E_{r}$, $n_{0}=2.0$. Besides, the condensation momenta are $\widetilde{\boldsymbol{k}}_{c}^{1}=(k_c,k_c)$, $\widetilde{\boldsymbol{k}}_{c}^{2}=(-k_c,k_c)$, $\widetilde{\boldsymbol{k}}_{c}^{3}=(-k_c,-k_c)$, and $\widetilde{\boldsymbol{k}}_{c}^{4}=(k_c,-k_c)$, where  $k_c=\pi/10$. %$\mu=4.8416E_{r}$
}
\label{fig4}
\end{figure}

In order to obtain the quasiparticle excitation spectrum, here we employ the Bogoliubov theory to derive the Bogoliubov-de Gennes (BdG) Hamiltonian $\hat{\mathcal{H}}_{\mathrm{BdG}}=\frac{1}{2}\sum_{\boldsymbol{k}}\Psi_{\boldsymbol{k}}^{\dagger}\mathcal{H}_{\mathrm{BdG}}\Psi_{\boldsymbol{k}}$~\cite{SM}, where $\Psi_{\boldsymbol{k}}^{\dagger}=(b_{\boldsymbol{\Gamma}+\boldsymbol{k},l,\sigma}^{\dagger},b_{\boldsymbol{\Gamma}-\boldsymbol{k},l,\sigma})$ denotes the Nambu basis. Performing the Bogoliubov transformation, we have $T_{\boldsymbol{k}}^{\dagger}\mathcal{H}_{\mathrm{BdG}}(\boldsymbol{k})T_{\boldsymbol{k}}=E_{\boldsymbol{k}}$. Here para-unitary matrix $T_{\boldsymbol{k}}$ satisfies $T_{\boldsymbol{k}}^{\dagger}\tau_{z}T_{\boldsymbol{k}}=\tau_{z}$ (\textbf{$\tau_{z}=\sigma_{z}\otimes\mathrm{\boldsymbol{I}}_{3\times3}$}) and the diagonal terms of $E_{\boldsymbol{k}}$ represent the excitation spectrum~\cite{shindou2013topological}. As shown in Fig.\ref{fig3} (b), an interaction-induced topological gap opens at the $\boldsymbol{\Gamma}$ point \cite{SunNP2012} and, together with the 2D SOC, separates the first band from the second and third bands, leading to the Chern number of the first band $\mathrm{Ch}^{1}=1$, while the total topological number of the second and third bands $\mathrm{Ch}^{2+3}=-1$. Consequently, these topological excitations are chiral.

\paragraph*{Two-dimensional topological spin-orbital supersolid.}
In the intermediate regime with moderate $m_z$, the $p$-orbital interaction and the Raman-induced SOC are competitive, and lead to an equal condensation at momenta $\widetilde{\boldsymbol{k}}_{c}^{n}$ ($n=1\sim4$) with the condensation parameters satisfying $|\gamma_{\widetilde{\boldsymbol{k}}_{c}^{n}}|^{2}=\frac{1}{4}$ and $\beta_{\widetilde{\boldsymbol{k}}_{c}^{n},p_{x\downarrow}}=-(-1)^n\beta_{\widetilde{\boldsymbol{k}}_{c}^{n},p_{y\downarrow}}$. Here the corresponding OAM order, which is given by $\langle L_{\boldsymbol{i},z} \rangle=2n_{0}\sum_{n,n^{\prime}=1}^{4}\mathrm{Im}[e^{\mathrm{-i}(\tilde{\boldsymbol{k}}_{c}^{n}-\tilde{\boldsymbol{k}}_{c}^{n^{\prime}})\cdot\boldsymbol{i}}\gamma_{\tilde{\boldsymbol{k}}_{c}^{n}}^{\ast}\beta_{\tilde{\boldsymbol{k}}_{c}^{n},p_{x\downarrow}}^{\ast}\gamma_{\tilde{\boldsymbol{k}}_{c}^{n^{\prime}}}\beta_{\tilde{\boldsymbol{k}}_{c}^{n^{\prime}},p_{y\downarrow}}]$, exhibits a staggered positive and negative clusters distribution shown in Fig.\ref{fig4}(a), with the periods of $A(B)$-type clusters being $\pi/k_c$ in $x$ and $y$ directions. The staggered clusters pattern satisfies $\sum_{\mathrm{Cluster}}\langle L_{\boldsymbol{i},z}\rangle\neq0$ but $\sum_{\mathrm{Global}}\langle L_{\boldsymbol{i},z}\rangle=0$, which means that the time-reversal symmetry is broken within each cluster but recovered by averaging for the global system. The supersolid phase exhibits nontrivial topology at each OAM cluster, as characterized by the chiral and anti-chiral edge states of topological excitations on the boundaries of the OAM clusters A ($\langle L_{\boldsymbol{i},z}\rangle>0$) and B ($\langle L_{\boldsymbol{i},z}\rangle<0$), respectively [see Fig.\ref{fig4}(b)]~\cite{SM}. The chiral and anti-chiral edge states are protected by the opposite Chern numbers of the bulk topological excitations in the A and B clusters, and they are robust under impurities or disorders. %Moreover, as shown in Fig.\ref{fig4}(b), the edge excitations of A and B flow clockwise and anticlockwise, respectively, which suggests that the topological edge excitations of positive and negative OAM clusters have opposite chiralities, leading to a zero net orbital current globally. In conclusion, the quasiparticle excitations embody the distribution of topological in each cluster, trivial in the overall bulk.

The present supersolid phase embodies rich and exotic features beyond the OAM order. In particular, we investigate the real-space spin texture $\langle\boldsymbol{\sigma}_{i}\rangle=(\langle\sigma_{i,x}\rangle,\langle\sigma_{i,y}\rangle,\langle\sigma_{i,z}\rangle)$ of the condensate, with the three components given by $(\langle b_{i,s,\uparrow}^{\dagger}\sum_{\nu}b_{i,p_{\nu},\downarrow}+\mathrm{H.c.}\rangle,\langle-\mathrm{i}b_{i,s,\uparrow}^{\dagger}\sum_{\nu}b_{i,p_{\nu},\downarrow}+\mathrm{H.c.}\rangle,\langle n_{i,s,\uparrow}-\sum_{\nu}n_{i,p_{\nu},\downarrow}\rangle)$ with $\nu=x,y$. Here, the spin texture in $x$$-$$y$ plane $(\langle\sigma_{i,x}\rangle,\langle\sigma_{i,y}\rangle)$ given in Fig.\ref{fig4}(c) shows a topological vortex lattice structure characterized by vortex ($B$) and anti-vortex ($A$) distributions, corresponding to the negative (B) and positive (A) OAM cluster centers, respectively. Further, the spin component along $z$ direction $\langle \sigma_{iz} \rangle $ in Fig.\ref{fig4}(d) exhibits another staggered clusters pattern, with a period of $1/\sqrt{2}$ that of OAM patterns. From the three spin components the combined whole spin texture $\langle\boldsymbol{\sigma}_{i}\rangle$ forms a topological skyrmion lattice structure whose periodicity matches the OAM pattern. %rendering an unprecedented spin-orbital supersolid phase predicted in the present orbital optical Raman lattice.
With the nontrivial topology encoded in the real space and in the momentum space (for quasiparticles), the superfluidity of $|G\rangle$ renders an unprecedented 2D topological SOSS phase breaking both lattice translation symmetry and $U(1)$ symmetry, as predicted in the present orbital optical Raman lattice. 

\paragraph*{Conclusion.}
We explore exotic many-body quantum phases by introducing a completely new $p$-band SOC lattice model, which is distinct from those studied in earlier works on synthetic SOC and high-orbital physics. Through systematic analysis of the mean-field ground states and beyond-mean-field excitations, we find that the competition between the Raman-induced SOC and $p$-orbital interaction can give rise to two novel high-orbital condensed phases: the UAMSF phase, which features uniform OAM order and global chiral edge excitations, and the 2D SOSS phase, which exhibits a staggered OAM pattern and a topological skyrmion structure in the spin texture, with opposite topological boundary excitations appearing in neighboring OAM clusters.
Another key distinction of the present model compared to previous realized $p$-orbital optical lattice gases lies in its full controllability. Unlike the double-well lattice systems~\cite{wirth2011evidence,kock2016orbital,xu2016pi,di2016topological,PhysRevLett.125.260402,jin2021evidence,wang2021evidence,wang2023evidence,huang2022intrinsic}, the orbital optical Raman lattice allows independent and precise tuning of both the strength and form of the Raman-induced $s$¨C$p$ coupling. This flexibility enables tunable competition between SOC and interaction, going beyond what can be achieved in pure orbital lattice gases without synthetic SOC. Furthermore, we propose a minimal experimental scheme based on a cold atom platform, and the lifetime of the high-orbital many-body quantum phases is discussed in \cite{SM}. Our work offers new insight into the exploration of exotic high-orbital physics with nontrivial SOC effects and paves the way for the experimental realization of supersolid phases without long-range interactions.

\paragraph*{Acknowledgments.}
We thank Xin-Chi Zhou, Ting-Fung Jeffrey Poon, and Ye-Bing Zhang for their valuable discussions. This work was supported by National Key Research and Development Program of China (No. 2021YFA1400900), the National Natural Science Foundation of China (Grants No. 12104205, No. 12261160368, and No. 11921005), and the Innovation Program for Quantum Science and Technology (Grant No. 2021ZD0302000) (Z.H.H., K.H.M., B.Z.W., X.J.L.), and by U.S. AFOSR Grant No. FA9550-23-1-0598 (W.V.L.).

%-----------------------------------------------------------------------------------------------------------

\nocite{*}

\bibliography{OORL_II}% Produces the bibliography via BibTeX.

\providecommand{\noopsort}[1]{}\providecommand{\singleletter}[1]{#1}%
\begin{thebibliography}{86}%
\makeatletter
\providecommand \@ifxundefined [1]{%
 \@ifx{#1\undefined}
}%
\providecommand \@ifnum [1]{%
 \ifnum #1\expandafter \@firstoftwo
 \else \expandafter \@secondoftwo
 \fi
}%
\providecommand \@ifx [1]{%
 \ifx #1\expandafter \@firstoftwo
 \else \expandafter \@secondoftwo
 \fi
}%
\providecommand \natexlab [1]{#1}%
\providecommand \enquote  [1]{``#1''}%
\providecommand \bibnamefont  [1]{#1}%
\providecommand \bibfnamefont [1]{#1}%
\providecommand \citenamefont [1]{#1}%
\providecommand \href@noop [0]{\@secondoftwo}%
\providecommand \href [0]{\begingroup \@sanitize@url \@href}%
\providecommand \@href[1]{\@@startlink{#1}\@@href}%
\providecommand \@@href[1]{\endgroup#1\@@endlink}%
\providecommand \@sanitize@url [0]{\catcode `\\12\catcode `\$12\catcode `\&12\catcode `\#12\catcode `\^12\catcode `\_12\catcode `\%12\relax}%
\providecommand \@@startlink[1]{}%
\providecommand \@@endlink[0]{}%
\providecommand \url  [0]{\begingroup\@sanitize@url \@url }%
\providecommand \@url [1]{\endgroup\@href {#1}{\urlprefix }}%
\providecommand \urlprefix  [0]{URL }%
\providecommand \Eprint [0]{\href }%
\providecommand \doibase [0]{http://dx.doi.org/}%
\providecommand \selectlanguage [0]{\@gobble}%
\providecommand \bibinfo  [0]{\@secondoftwo}%
\providecommand \bibfield  [0]{\@secondoftwo}%
\providecommand \translation [1]{[#1]}%
\providecommand \BibitemOpen [0]{}%
\providecommand \bibitemStop [0]{}%
\providecommand \bibitemNoStop [0]{.\EOS\space}%
\providecommand \EOS [0]{\spacefactor3000\relax}%
\providecommand \BibitemShut  [1]{\csname bibitem#1\endcsname}%
\let\auto@bib@innerbib\@empty
%</preamble>
\bibitem [{\citenamefont {Bloch}\ \emph {et~al.}(2008)\citenamefont {Bloch}, \citenamefont {Dalibard},\ and\ \citenamefont {Zwerger}}]{bloch2008many}%
  \BibitemOpen
  \bibfield  {author} {\bibinfo {author} {\bibfnamefont {I.}~\bibnamefont {Bloch}}, \bibinfo {author} {\bibfnamefont {J.}~\bibnamefont {Dalibard}}, \ and\ \bibinfo {author} {\bibfnamefont {W.}~\bibnamefont {Zwerger}},\ }\href@noop {} {\bibfield  {journal} {\bibinfo  {journal} {Rev. Mod. Phys.}\ }\textbf {\bibinfo {volume} {80}},\ \bibinfo {pages} {885} (\bibinfo {year} {2008})}\BibitemShut {NoStop}%
\bibitem [{\citenamefont {Gross}\ and\ \citenamefont {Bloch}(2017)}]{gross2017quantum}%
  \BibitemOpen
  \bibfield  {author} {\bibinfo {author} {\bibfnamefont {C.}~\bibnamefont {Gross}}\ and\ \bibinfo {author} {\bibfnamefont {I.}~\bibnamefont {Bloch}},\ }\href@noop {} {\bibfield  {journal} {\bibinfo  {journal} {Science}\ }\textbf {\bibinfo {volume} {357}},\ \bibinfo {pages} {995} (\bibinfo {year} {2017})}\BibitemShut {NoStop}%
\bibitem [{\citenamefont {Jaksch}\ and\ \citenamefont {Zoller}(2003)}]{Jaksch2003NJP}%
  \BibitemOpen
  \bibfield  {author} {\bibinfo {author} {\bibfnamefont {D.}~\bibnamefont {Jaksch}}\ and\ \bibinfo {author} {\bibfnamefont {P.}~\bibnamefont {Zoller}},\ }\href@noop {} {\bibfield  {journal} {\bibinfo  {journal} {New Journal of Physics}\ }\textbf {\bibinfo {volume} {5}},\ \bibinfo {pages} {56} (\bibinfo {year} {2003})}\BibitemShut {NoStop}%
\bibitem [{\citenamefont {Juzeli{\=u}nas}\ and\ \citenamefont {{\"O}hberg}(2004)}]{JuzeliPRL2004}%
  \BibitemOpen
  \bibfield  {author} {\bibinfo {author} {\bibfnamefont {G.}~\bibnamefont {Juzeli{\=u}nas}}\ and\ \bibinfo {author} {\bibfnamefont {P.}~\bibnamefont {{\"O}hberg}},\ }\href@noop {} {\bibfield  {journal} {\bibinfo  {journal} {Phys. Rev. Lett.}\ }\textbf {\bibinfo {volume} {93}},\ \bibinfo {pages} {033602} (\bibinfo {year} {2004})}\BibitemShut {NoStop}%
\bibitem [{\citenamefont {Osterloh}\ \emph {et~al.}(2005)\citenamefont {Osterloh}, \citenamefont {Baig}, \citenamefont {Santos}, \citenamefont {Zoller},\ and\ \citenamefont {Lewenstein}}]{OsterlohPRL2005}%
  \BibitemOpen
  \bibfield  {author} {\bibinfo {author} {\bibfnamefont {K.}~\bibnamefont {Osterloh}}, \bibinfo {author} {\bibfnamefont {M.}~\bibnamefont {Baig}}, \bibinfo {author} {\bibfnamefont {L.}~\bibnamefont {Santos}}, \bibinfo {author} {\bibfnamefont {P.}~\bibnamefont {Zoller}}, \ and\ \bibinfo {author} {\bibfnamefont {M.}~\bibnamefont {Lewenstein}},\ }\href@noop {} {\bibfield  {journal} {\bibinfo  {journal} {Phys. Rev. Lett.}\ }\textbf {\bibinfo {volume} {95}},\ \bibinfo {pages} {010403} (\bibinfo {year} {2005})}\BibitemShut {NoStop}%
\bibitem [{\citenamefont {Ruseckas}\ \emph {et~al.}(2005)\citenamefont {Ruseckas}, \citenamefont {Juzeli{\=u}nas}, \citenamefont {{\"O}hberg},\ and\ \citenamefont {Fleischhauer}}]{RuseckasPRL2005}%
  \BibitemOpen
  \bibfield  {author} {\bibinfo {author} {\bibfnamefont {J.}~\bibnamefont {Ruseckas}}, \bibinfo {author} {\bibfnamefont {G.}~\bibnamefont {Juzeli{\=u}nas}}, \bibinfo {author} {\bibfnamefont {P.}~\bibnamefont {{\"O}hberg}}, \ and\ \bibinfo {author} {\bibfnamefont {M.}~\bibnamefont {Fleischhauer}},\ }\href@noop {} {\bibfield  {journal} {\bibinfo  {journal} {Phys. Rev. Lett.}\ }\textbf {\bibinfo {volume} {95}},\ \bibinfo {pages} {010404} (\bibinfo {year} {2005})}\BibitemShut {NoStop}%
\bibitem [{\citenamefont {Liu}\ \emph {et~al.}(2006)\citenamefont {Liu}, \citenamefont {Jing}, \citenamefont {Liu},\ and\ \citenamefont {Ge}}]{LiuTPPP2006}%
  \BibitemOpen
  \bibfield  {author} {\bibinfo {author} {\bibfnamefont {X.-J.}\ \bibnamefont {Liu}}, \bibinfo {author} {\bibfnamefont {H.}~\bibnamefont {Jing}}, \bibinfo {author} {\bibfnamefont {X.}~\bibnamefont {Liu}}, \ and\ \bibinfo {author} {\bibfnamefont {M.-L.}\ \bibnamefont {Ge}},\ }\href@noop {} {\bibfield  {journal} {\bibinfo  {journal} {The European Physical Journal D-Atomic, Molecular, Optical and Plasma Physics}\ }\textbf {\bibinfo {volume} {37}},\ \bibinfo {pages} {261} (\bibinfo {year} {2006})}\BibitemShut {NoStop}%
\bibitem [{\citenamefont {Dalibard}\ \emph {et~al.}(2011)\citenamefont {Dalibard}, \citenamefont {Gerbier}, \citenamefont {Juzeli{\=u}nas},\ and\ \citenamefont {{\"O}hberg}}]{dalibard2011colloquium}%
  \BibitemOpen
  \bibfield  {author} {\bibinfo {author} {\bibfnamefont {J.}~\bibnamefont {Dalibard}}, \bibinfo {author} {\bibfnamefont {F.}~\bibnamefont {Gerbier}}, \bibinfo {author} {\bibfnamefont {G.}~\bibnamefont {Juzeli{\=u}nas}}, \ and\ \bibinfo {author} {\bibfnamefont {P.}~\bibnamefont {{\"O}hberg}},\ }\href@noop {} {\bibfield  {journal} {\bibinfo  {journal} {Rev. Mod. Phys.}\ }\textbf {\bibinfo {volume} {83}},\ \bibinfo {pages} {1523} (\bibinfo {year} {2011})}\BibitemShut {NoStop}%
\bibitem [{\citenamefont {Goldman}\ \emph {et~al.}(2014)\citenamefont {Goldman}, \citenamefont {Juzeli{\=u}nas}, \citenamefont {{\"O}hberg},\ and\ \citenamefont {Spielman}}]{goldman2014light}%
  \BibitemOpen
  \bibfield  {author} {\bibinfo {author} {\bibfnamefont {N.}~\bibnamefont {Goldman}}, \bibinfo {author} {\bibfnamefont {G.}~\bibnamefont {Juzeli{\=u}nas}}, \bibinfo {author} {\bibfnamefont {P.}~\bibnamefont {{\"O}hberg}}, \ and\ \bibinfo {author} {\bibfnamefont {I.~B.}\ \bibnamefont {Spielman}},\ }\href@noop {} {\bibfield  {journal} {\bibinfo  {journal} {Reports on Progress in Physics}\ }\textbf {\bibinfo {volume} {77}},\ \bibinfo {pages} {126401} (\bibinfo {year} {2014})}\BibitemShut {NoStop}%
\bibitem [{\citenamefont {Chen}\ \emph {et~al.}(2018)\citenamefont {Chen}, \citenamefont {Lin}, \citenamefont {Chen}, \citenamefont {Chiu}, \citenamefont {Wang}, \citenamefont {Chen}, \citenamefont {Huang}, \citenamefont {Yip}, \citenamefont {Kawaguchi},\ and\ \citenamefont {Lin}}]{ChenPRL2018}%
  \BibitemOpen
  \bibfield  {author} {\bibinfo {author} {\bibfnamefont {H.-R.}\ \bibnamefont {Chen}}, \bibinfo {author} {\bibfnamefont {K.-Y.}\ \bibnamefont {Lin}}, \bibinfo {author} {\bibfnamefont {P.-K.}\ \bibnamefont {Chen}}, \bibinfo {author} {\bibfnamefont {N.-C.}\ \bibnamefont {Chiu}}, \bibinfo {author} {\bibfnamefont {J.-B.}\ \bibnamefont {Wang}}, \bibinfo {author} {\bibfnamefont {C.-A.}\ \bibnamefont {Chen}}, \bibinfo {author} {\bibfnamefont {P.}~\bibnamefont {Huang}}, \bibinfo {author} {\bibfnamefont {S.-K.}\ \bibnamefont {Yip}}, \bibinfo {author} {\bibfnamefont {Y.}~\bibnamefont {Kawaguchi}}, \ and\ \bibinfo {author} {\bibfnamefont {Y.-J.}\ \bibnamefont {Lin}},\ }\href@noop {} {\bibfield  {journal} {\bibinfo  {journal} {Phys. Rev. Lett.}\ }\textbf {\bibinfo {volume} {121}},\ \bibinfo {pages} {113204} (\bibinfo {year} {2018})}\BibitemShut {NoStop}%
\bibitem [{\citenamefont {Aidelsburger}\ \emph {et~al.}(2022)\citenamefont {Aidelsburger}, \citenamefont {Barbiero}, \citenamefont {Bermudez}, \citenamefont {Chanda}, \citenamefont {Dauphin}, \citenamefont {Gonz{\'a}lez-Cuadra}, \citenamefont {Grzybowski}, \citenamefont {Hands}, \citenamefont {Jendrzejewski}, \citenamefont {J{\"u}nemann} \emph {et~al.}}]{aidelsburger2022cold}%
  \BibitemOpen
  \bibfield  {author} {\bibinfo {author} {\bibfnamefont {M.}~\bibnamefont {Aidelsburger}}, \bibinfo {author} {\bibfnamefont {L.}~\bibnamefont {Barbiero}}, \bibinfo {author} {\bibfnamefont {A.}~\bibnamefont {Bermudez}}, \bibinfo {author} {\bibfnamefont {T.}~\bibnamefont {Chanda}}, \bibinfo {author} {\bibfnamefont {A.}~\bibnamefont {Dauphin}}, \bibinfo {author} {\bibfnamefont {D.}~\bibnamefont {Gonz{\'a}lez-Cuadra}}, \bibinfo {author} {\bibfnamefont {P.~R.}\ \bibnamefont {Grzybowski}}, \bibinfo {author} {\bibfnamefont {S.}~\bibnamefont {Hands}}, \bibinfo {author} {\bibfnamefont {F.}~\bibnamefont {Jendrzejewski}}, \bibinfo {author} {\bibfnamefont {J.}~\bibnamefont {J{\"u}nemann}},  \emph {et~al.},\ }\href@noop {} {\bibfield  {journal} {\bibinfo  {journal} {Philosophical Transactions of the Royal Society A}\ }\textbf {\bibinfo {volume} {380}},\ \bibinfo {pages} {20210064} (\bibinfo {year} {2022})}\BibitemShut {NoStop}%
\bibitem [{\citenamefont {Liu}\ \emph {et~al.}(2009)\citenamefont {Liu}, \citenamefont {Borunda}, \citenamefont {Liu},\ and\ \citenamefont {Sinova}}]{LiuPRL2009}%
  \BibitemOpen
  \bibfield  {author} {\bibinfo {author} {\bibfnamefont {X.-J.}\ \bibnamefont {Liu}}, \bibinfo {author} {\bibfnamefont {M.~F.}\ \bibnamefont {Borunda}}, \bibinfo {author} {\bibfnamefont {X.}~\bibnamefont {Liu}}, \ and\ \bibinfo {author} {\bibfnamefont {J.}~\bibnamefont {Sinova}},\ }\href@noop {} {\bibfield  {journal} {\bibinfo  {journal} {Phys. Rev. Lett.}\ }\textbf {\bibinfo {volume} {102}},\ \bibinfo {pages} {046402} (\bibinfo {year} {2009})}\BibitemShut {NoStop}%
\bibitem [{\citenamefont {Juzeli{\=u}nas}\ \emph {et~al.}(2010)\citenamefont {Juzeli{\=u}nas}, \citenamefont {Ruseckas},\ and\ \citenamefont {Dalibard}}]{JuzeliPRA2010}%
  \BibitemOpen
  \bibfield  {author} {\bibinfo {author} {\bibfnamefont {G.}~\bibnamefont {Juzeli{\=u}nas}}, \bibinfo {author} {\bibfnamefont {J.}~\bibnamefont {Ruseckas}}, \ and\ \bibinfo {author} {\bibfnamefont {J.}~\bibnamefont {Dalibard}},\ }\href@noop {} {\bibfield  {journal} {\bibinfo  {journal} {Phys. Rev. A}\ }\textbf {\bibinfo {volume} {81}},\ \bibinfo {pages} {053403} (\bibinfo {year} {2010})}\BibitemShut {NoStop}%
\bibitem [{\citenamefont {Lin}\ \emph {et~al.}(2011)\citenamefont {Lin}, \citenamefont {Jim{\'e}nez-Garc{\'\i}a},\ and\ \citenamefont {Spielman}}]{LinNature2011}%
  \BibitemOpen
  \bibfield  {author} {\bibinfo {author} {\bibfnamefont {Y.-J.}\ \bibnamefont {Lin}}, \bibinfo {author} {\bibfnamefont {K.}~\bibnamefont {Jim{\'e}nez-Garc{\'\i}a}}, \ and\ \bibinfo {author} {\bibfnamefont {I.~B.}\ \bibnamefont {Spielman}},\ }\href@noop {} {\bibfield  {journal} {\bibinfo  {journal} {Nature}\ }\textbf {\bibinfo {volume} {471}},\ \bibinfo {pages} {83} (\bibinfo {year} {2011})}\BibitemShut {NoStop}%
\bibitem [{\citenamefont {Campbell}\ \emph {et~al.}(2011)\citenamefont {Campbell}, \citenamefont {Juzeli{\=u}nas},\ and\ \citenamefont {Spielman}}]{CampbellPRA2011}%
  \BibitemOpen
  \bibfield  {author} {\bibinfo {author} {\bibfnamefont {D.~L.}\ \bibnamefont {Campbell}}, \bibinfo {author} {\bibfnamefont {G.}~\bibnamefont {Juzeli{\=u}nas}}, \ and\ \bibinfo {author} {\bibfnamefont {I.~B.}\ \bibnamefont {Spielman}},\ }\href@noop {} {\bibfield  {journal} {\bibinfo  {journal} {Phys. Rev. A}\ }\textbf {\bibinfo {volume} {84}},\ \bibinfo {pages} {025602} (\bibinfo {year} {2011})}\BibitemShut {NoStop}%
\bibitem [{\citenamefont {Sau}\ \emph {et~al.}(2011)\citenamefont {Sau}, \citenamefont {Sensarma}, \citenamefont {Powell}, \citenamefont {Spielman},\ and\ \citenamefont {Sarma}}]{SauPRB2011}%
  \BibitemOpen
  \bibfield  {author} {\bibinfo {author} {\bibfnamefont {J.~D.}\ \bibnamefont {Sau}}, \bibinfo {author} {\bibfnamefont {R.}~\bibnamefont {Sensarma}}, \bibinfo {author} {\bibfnamefont {S.}~\bibnamefont {Powell}}, \bibinfo {author} {\bibfnamefont {I.}~\bibnamefont {Spielman}}, \ and\ \bibinfo {author} {\bibfnamefont {S.~D.}\ \bibnamefont {Sarma}},\ }\href@noop {} {\bibfield  {journal} {\bibinfo  {journal} {Phys. Rev. B}\ }\textbf {\bibinfo {volume} {83}},\ \bibinfo {pages} {140510} (\bibinfo {year} {2011})}\BibitemShut {NoStop}%
\bibitem [{\citenamefont {Wang}\ \emph {et~al.}(2012)\citenamefont {Wang}, \citenamefont {Yu}, \citenamefont {Fu}, \citenamefont {Miao}, \citenamefont {Huang}, \citenamefont {Chai}, \citenamefont {Zhai},\ and\ \citenamefont {Zhang}}]{WangPRL2012}%
  \BibitemOpen
  \bibfield  {author} {\bibinfo {author} {\bibfnamefont {P.}~\bibnamefont {Wang}}, \bibinfo {author} {\bibfnamefont {Z.-Q.}\ \bibnamefont {Yu}}, \bibinfo {author} {\bibfnamefont {Z.}~\bibnamefont {Fu}}, \bibinfo {author} {\bibfnamefont {J.}~\bibnamefont {Miao}}, \bibinfo {author} {\bibfnamefont {L.}~\bibnamefont {Huang}}, \bibinfo {author} {\bibfnamefont {S.}~\bibnamefont {Chai}}, \bibinfo {author} {\bibfnamefont {H.}~\bibnamefont {Zhai}}, \ and\ \bibinfo {author} {\bibfnamefont {J.}~\bibnamefont {Zhang}},\ }\href@noop {} {\bibfield  {journal} {\bibinfo  {journal} {Phys. Rev. Lett.}\ }\textbf {\bibinfo {volume} {109}},\ \bibinfo {pages} {095301} (\bibinfo {year} {2012})}\BibitemShut {NoStop}%
\bibitem [{\citenamefont {Cheuk}\ \emph {et~al.}(2012)\citenamefont {Cheuk}, \citenamefont {Sommer}, \citenamefont {Hadzibabic}, \citenamefont {Yefsah}, \citenamefont {Bakr},\ and\ \citenamefont {Zwierlein}}]{CheukPRL2012}%
  \BibitemOpen
  \bibfield  {author} {\bibinfo {author} {\bibfnamefont {L.~W.}\ \bibnamefont {Cheuk}}, \bibinfo {author} {\bibfnamefont {A.~T.}\ \bibnamefont {Sommer}}, \bibinfo {author} {\bibfnamefont {Z.}~\bibnamefont {Hadzibabic}}, \bibinfo {author} {\bibfnamefont {T.}~\bibnamefont {Yefsah}}, \bibinfo {author} {\bibfnamefont {W.~S.}\ \bibnamefont {Bakr}}, \ and\ \bibinfo {author} {\bibfnamefont {M.~W.}\ \bibnamefont {Zwierlein}},\ }\href@noop {} {\bibfield  {journal} {\bibinfo  {journal} {Phys. Rev. Lett.}\ }\textbf {\bibinfo {volume} {109}},\ \bibinfo {pages} {095302} (\bibinfo {year} {2012})}\BibitemShut {NoStop}%
\bibitem [{\citenamefont {Anderson}\ \emph {et~al.}(2012)\citenamefont {Anderson}, \citenamefont {Juzeli{\=u}nas}, \citenamefont {Galitski},\ and\ \citenamefont {Spielman}}]{AndersonPRL2012}%
  \BibitemOpen
  \bibfield  {author} {\bibinfo {author} {\bibfnamefont {B.~M.}\ \bibnamefont {Anderson}}, \bibinfo {author} {\bibfnamefont {G.}~\bibnamefont {Juzeli{\=u}nas}}, \bibinfo {author} {\bibfnamefont {V.~M.}\ \bibnamefont {Galitski}}, \ and\ \bibinfo {author} {\bibfnamefont {I.~B.}\ \bibnamefont {Spielman}},\ }\href@noop {} {\bibfield  {journal} {\bibinfo  {journal} {Phys. Rev. Lett.}\ }\textbf {\bibinfo {volume} {108}},\ \bibinfo {pages} {235301} (\bibinfo {year} {2012})}\BibitemShut {NoStop}%
\bibitem [{\citenamefont {Galitski}\ and\ \citenamefont {Spielman}(2013)}]{galitski2013spin}%
  \BibitemOpen
  \bibfield  {author} {\bibinfo {author} {\bibfnamefont {V.}~\bibnamefont {Galitski}}\ and\ \bibinfo {author} {\bibfnamefont {I.~B.}\ \bibnamefont {Spielman}},\ }\href@noop {} {\bibfield  {journal} {\bibinfo  {journal} {Nature}\ }\textbf {\bibinfo {volume} {494}},\ \bibinfo {pages} {49} (\bibinfo {year} {2013})}\BibitemShut {NoStop}%
\bibitem [{\citenamefont {Liu}\ \emph {et~al.}(2014)\citenamefont {Liu}, \citenamefont {Law},\ and\ \citenamefont {Ng}}]{LiuPRL2014}%
  \BibitemOpen
  \bibfield  {author} {\bibinfo {author} {\bibfnamefont {X.-J.}\ \bibnamefont {Liu}}, \bibinfo {author} {\bibfnamefont {K.~T.}\ \bibnamefont {Law}}, \ and\ \bibinfo {author} {\bibfnamefont {T.~K.}\ \bibnamefont {Ng}},\ }\href@noop {} {\bibfield  {journal} {\bibinfo  {journal} {Phys. Rev. Lett.}\ }\textbf {\bibinfo {volume} {112}},\ \bibinfo {pages} {086401} (\bibinfo {year} {2014})}\BibitemShut {NoStop}%
\bibitem [{\citenamefont {Zhai}(2015)}]{zhai2015degenerate}%
  \BibitemOpen
  \bibfield  {author} {\bibinfo {author} {\bibfnamefont {H.}~\bibnamefont {Zhai}},\ }\href@noop {} {\bibfield  {journal} {\bibinfo  {journal} {Reports on Progress in Physics}\ }\textbf {\bibinfo {volume} {78}},\ \bibinfo {pages} {026001} (\bibinfo {year} {2015})}\BibitemShut {NoStop}%
\bibitem [{\citenamefont {Wu}\ \emph {et~al.}(2016)\citenamefont {Wu}, \citenamefont {Zhang}, \citenamefont {Sun}, \citenamefont {Xu}, \citenamefont {Wang}, \citenamefont {Ji}, \citenamefont {Deng}, \citenamefont {Chen}, \citenamefont {Liu},\ and\ \citenamefont {Pan}}]{Wu2016Science}%
  \BibitemOpen
  \bibfield  {author} {\bibinfo {author} {\bibfnamefont {Z.}~\bibnamefont {Wu}}, \bibinfo {author} {\bibfnamefont {L.}~\bibnamefont {Zhang}}, \bibinfo {author} {\bibfnamefont {W.}~\bibnamefont {Sun}}, \bibinfo {author} {\bibfnamefont {X.-T.}\ \bibnamefont {Xu}}, \bibinfo {author} {\bibfnamefont {B.-Z.}\ \bibnamefont {Wang}}, \bibinfo {author} {\bibfnamefont {S.-C.}\ \bibnamefont {Ji}}, \bibinfo {author} {\bibfnamefont {Y.}~\bibnamefont {Deng}}, \bibinfo {author} {\bibfnamefont {S.}~\bibnamefont {Chen}}, \bibinfo {author} {\bibfnamefont {X.-J.}\ \bibnamefont {Liu}}, \ and\ \bibinfo {author} {\bibfnamefont {J.-W.}\ \bibnamefont {Pan}},\ }\href@noop {} {\bibfield  {journal} {\bibinfo  {journal} {Science}\ }\textbf {\bibinfo {volume} {354}},\ \bibinfo {pages} {83} (\bibinfo {year} {2016})}\BibitemShut {NoStop}%
\bibitem [{\citenamefont {Burdick}\ \emph {et~al.}(2016)\citenamefont {Burdick}, \citenamefont {Tang},\ and\ \citenamefont {Lev}}]{BurdickPX2016}%
  \BibitemOpen
  \bibfield  {author} {\bibinfo {author} {\bibfnamefont {N.~Q.}\ \bibnamefont {Burdick}}, \bibinfo {author} {\bibfnamefont {Y.}~\bibnamefont {Tang}}, \ and\ \bibinfo {author} {\bibfnamefont {B.~L.}\ \bibnamefont {Lev}},\ }\href@noop {} {\bibfield  {journal} {\bibinfo  {journal} {Phys. Rev. X}\ }\textbf {\bibinfo {volume} {6}},\ \bibinfo {pages} {031022} (\bibinfo {year} {2016})}\BibitemShut {NoStop}%
\bibitem [{\citenamefont {Song}\ \emph {et~al.}(2016)\citenamefont {Song}, \citenamefont {He}, \citenamefont {Zhang}, \citenamefont {Hajiyev}, \citenamefont {Huang}, \citenamefont {Liu},\ and\ \citenamefont {Jo}}]{SongPRA2016}%
  \BibitemOpen
  \bibfield  {author} {\bibinfo {author} {\bibfnamefont {B.}~\bibnamefont {Song}}, \bibinfo {author} {\bibfnamefont {C.}~\bibnamefont {He}}, \bibinfo {author} {\bibfnamefont {S.}~\bibnamefont {Zhang}}, \bibinfo {author} {\bibfnamefont {E.}~\bibnamefont {Hajiyev}}, \bibinfo {author} {\bibfnamefont {W.}~\bibnamefont {Huang}}, \bibinfo {author} {\bibfnamefont {X.-J.}\ \bibnamefont {Liu}}, \ and\ \bibinfo {author} {\bibfnamefont {G.-B.}\ \bibnamefont {Jo}},\ }\href@noop {} {\bibfield  {journal} {\bibinfo  {journal} {Phys. Rev. A}\ }\textbf {\bibinfo {volume} {94}},\ \bibinfo {pages} {061604} (\bibinfo {year} {2016})}\BibitemShut {NoStop}%
\bibitem [{\citenamefont {Chen}\ \emph {et~al.}(2016)\citenamefont {Chen}, \citenamefont {Liu},\ and\ \citenamefont {Xie}}]{PhysRevA.93.053610}%
  \BibitemOpen
  \bibfield  {author} {\bibinfo {author} {\bibfnamefont {H.}~\bibnamefont {Chen}}, \bibinfo {author} {\bibfnamefont {X.-J.}\ \bibnamefont {Liu}}, \ and\ \bibinfo {author} {\bibfnamefont {X.~C.}\ \bibnamefont {Xie}},\ }\href@noop {} {\bibfield  {journal} {\bibinfo  {journal} {Phys. Rev. A}\ }\textbf {\bibinfo {volume} {93}},\ \bibinfo {pages} {053610} (\bibinfo {year} {2016})}\BibitemShut {NoStop}%
\bibitem [{\citenamefont {Kolkowitz}\ \emph {et~al.}(2017)\citenamefont {Kolkowitz}, \citenamefont {Bromley}, \citenamefont {Bothwell}, \citenamefont {Wall}, \citenamefont {Marti}, \citenamefont {Koller}, \citenamefont {Zhang}, \citenamefont {Rey},\ and\ \citenamefont {Ye}}]{KolkowitzNatue2017}%
  \BibitemOpen
  \bibfield  {author} {\bibinfo {author} {\bibfnamefont {S.}~\bibnamefont {Kolkowitz}}, \bibinfo {author} {\bibfnamefont {S.}~\bibnamefont {Bromley}}, \bibinfo {author} {\bibfnamefont {T.}~\bibnamefont {Bothwell}}, \bibinfo {author} {\bibfnamefont {M.}~\bibnamefont {Wall}}, \bibinfo {author} {\bibfnamefont {G.}~\bibnamefont {Marti}}, \bibinfo {author} {\bibfnamefont {A.}~\bibnamefont {Koller}}, \bibinfo {author} {\bibfnamefont {X.}~\bibnamefont {Zhang}}, \bibinfo {author} {\bibfnamefont {A.}~\bibnamefont {Rey}}, \ and\ \bibinfo {author} {\bibfnamefont {J.}~\bibnamefont {Ye}},\ }\href@noop {} {\bibfield  {journal} {\bibinfo  {journal} {Nature}\ }\textbf {\bibinfo {volume} {542}},\ \bibinfo {pages} {66} (\bibinfo {year} {2017})}\BibitemShut {NoStop}%
\bibitem [{\citenamefont {Zhang}\ and\ \citenamefont {Liu}(2018)}]{zhang2018spin}%
  \BibitemOpen
  \bibfield  {author} {\bibinfo {author} {\bibfnamefont {L.}~\bibnamefont {Zhang}}\ and\ \bibinfo {author} {\bibfnamefont {X.-J.}\ \bibnamefont {Liu}},\ }in\ \href@noop {} {\emph {\bibinfo {booktitle} {Synthetic Spin-Orbit Coupling in Cold Atoms}}}\ (\bibinfo  {publisher} {World Scientific},\ \bibinfo {year} {2018})\ pp.\ \bibinfo {pages} {1--87}\BibitemShut {NoStop}%
\bibitem [{\citenamefont {Wang}\ \emph {et~al.}(2018)\citenamefont {Wang}, \citenamefont {Lu}, \citenamefont {Sun}, \citenamefont {Chen}, \citenamefont {Deng},\ and\ \citenamefont {Liu}}]{BaozongPRA2018}%
  \BibitemOpen
  \bibfield  {author} {\bibinfo {author} {\bibfnamefont {B.-Z.}\ \bibnamefont {Wang}}, \bibinfo {author} {\bibfnamefont {Y.-H.}\ \bibnamefont {Lu}}, \bibinfo {author} {\bibfnamefont {W.}~\bibnamefont {Sun}}, \bibinfo {author} {\bibfnamefont {S.}~\bibnamefont {Chen}}, \bibinfo {author} {\bibfnamefont {Y.}~\bibnamefont {Deng}}, \ and\ \bibinfo {author} {\bibfnamefont {X.-J.}\ \bibnamefont {Liu}},\ }\href@noop {} {\bibfield  {journal} {\bibinfo  {journal} {Phys. Rev. A}\ }\textbf {\bibinfo {volume} {97}},\ \bibinfo {pages} {011605} (\bibinfo {year} {2018})}\BibitemShut {NoStop}%
\bibitem [{\citenamefont {Huang}\ \emph {et~al.}(2016)\citenamefont {Huang}, \citenamefont {Meng}, \citenamefont {Wang}, \citenamefont {Peng}, \citenamefont {Zhang}, \citenamefont {Chen}, \citenamefont {Li}, \citenamefont {Zhou},\ and\ \citenamefont {Zhang}}]{huang2016experimental}%
  \BibitemOpen
  \bibfield  {author} {\bibinfo {author} {\bibfnamefont {L.}~\bibnamefont {Huang}}, \bibinfo {author} {\bibfnamefont {Z.}~\bibnamefont {Meng}}, \bibinfo {author} {\bibfnamefont {P.}~\bibnamefont {Wang}}, \bibinfo {author} {\bibfnamefont {P.}~\bibnamefont {Peng}}, \bibinfo {author} {\bibfnamefont {S.-L.}\ \bibnamefont {Zhang}}, \bibinfo {author} {\bibfnamefont {L.}~\bibnamefont {Chen}}, \bibinfo {author} {\bibfnamefont {D.}~\bibnamefont {Li}}, \bibinfo {author} {\bibfnamefont {Q.}~\bibnamefont {Zhou}}, \ and\ \bibinfo {author} {\bibfnamefont {J.}~\bibnamefont {Zhang}},\ }\href@noop {} {\bibfield  {journal} {\bibinfo  {journal} {Nat. Phys.}\ }\textbf {\bibinfo {volume} {12}},\ \bibinfo {pages} {540} (\bibinfo {year} {2016})}\BibitemShut {NoStop}%
\bibitem [{\citenamefont {Meng}\ \emph {et~al.}(2016)\citenamefont {Meng}, \citenamefont {Huang}, \citenamefont {Peng}, \citenamefont {Li}, \citenamefont {Chen}, \citenamefont {Xu}, \citenamefont {Zhang}, \citenamefont {Wang},\ and\ \citenamefont {Zhang}}]{meng2016experimental}%
  \BibitemOpen
  \bibfield  {author} {\bibinfo {author} {\bibfnamefont {Z.}~\bibnamefont {Meng}}, \bibinfo {author} {\bibfnamefont {L.}~\bibnamefont {Huang}}, \bibinfo {author} {\bibfnamefont {P.}~\bibnamefont {Peng}}, \bibinfo {author} {\bibfnamefont {D.}~\bibnamefont {Li}}, \bibinfo {author} {\bibfnamefont {L.}~\bibnamefont {Chen}}, \bibinfo {author} {\bibfnamefont {Y.}~\bibnamefont {Xu}}, \bibinfo {author} {\bibfnamefont {C.}~\bibnamefont {Zhang}}, \bibinfo {author} {\bibfnamefont {P.}~\bibnamefont {Wang}}, \ and\ \bibinfo {author} {\bibfnamefont {J.}~\bibnamefont {Zhang}},\ }\href@noop {} {\bibfield  {journal} {\bibinfo  {journal} {Phys. Rev. Lett.}\ }\textbf {\bibinfo {volume} {117}},\ \bibinfo {pages} {235304} (\bibinfo {year} {2016})}\BibitemShut {NoStop}%
\bibitem [{\citenamefont {Sun}\ \emph {et~al.}(2018{\natexlab{a}})\citenamefont {Sun}, \citenamefont {Wang}, \citenamefont {Xu}, \citenamefont {Yi}, \citenamefont {Zhang}, \citenamefont {Wu}, \citenamefont {Deng}, \citenamefont {Liu}, \citenamefont {Chen},\ and\ \citenamefont {Pan}}]{sunPRL2018}%
  \BibitemOpen
  \bibfield  {author} {\bibinfo {author} {\bibfnamefont {W.}~\bibnamefont {Sun}}, \bibinfo {author} {\bibfnamefont {B.-Z.}\ \bibnamefont {Wang}}, \bibinfo {author} {\bibfnamefont {X.-T.}\ \bibnamefont {Xu}}, \bibinfo {author} {\bibfnamefont {C.-R.}\ \bibnamefont {Yi}}, \bibinfo {author} {\bibfnamefont {L.}~\bibnamefont {Zhang}}, \bibinfo {author} {\bibfnamefont {Z.}~\bibnamefont {Wu}}, \bibinfo {author} {\bibfnamefont {Y.}~\bibnamefont {Deng}}, \bibinfo {author} {\bibfnamefont {X.-J.}\ \bibnamefont {Liu}}, \bibinfo {author} {\bibfnamefont {S.}~\bibnamefont {Chen}}, \ and\ \bibinfo {author} {\bibfnamefont {J.-W.}\ \bibnamefont {Pan}},\ }\href@noop {} {\bibfield  {journal} {\bibinfo  {journal} {Phys. Rev. Lett.}\ }\textbf {\bibinfo {volume} {121}},\ \bibinfo {pages} {150401} (\bibinfo {year} {2018}{\natexlab{a}})}\BibitemShut {NoStop}%
\bibitem [{\citenamefont {Sun}\ \emph {et~al.}(2018{\natexlab{b}})\citenamefont {Sun}, \citenamefont {Yi}, \citenamefont {Wang}, \citenamefont {Zhang}, \citenamefont {Sanders}, \citenamefont {Xu}, \citenamefont {Wang}, \citenamefont {Schmiedmayer}, \citenamefont {Deng}, \citenamefont {Liu}, \citenamefont {Chen},\ and\ \citenamefont {Pan}}]{sunPRL2018_1}%
  \BibitemOpen
  \bibfield  {author} {\bibinfo {author} {\bibfnamefont {W.}~\bibnamefont {Sun}}, \bibinfo {author} {\bibfnamefont {C.-R.}\ \bibnamefont {Yi}}, \bibinfo {author} {\bibfnamefont {B.-Z.}\ \bibnamefont {Wang}}, \bibinfo {author} {\bibfnamefont {W.-W.}\ \bibnamefont {Zhang}}, \bibinfo {author} {\bibfnamefont {B.~C.}\ \bibnamefont {Sanders}}, \bibinfo {author} {\bibfnamefont {X.-T.}\ \bibnamefont {Xu}}, \bibinfo {author} {\bibfnamefont {Z.-Y.}\ \bibnamefont {Wang}}, \bibinfo {author} {\bibfnamefont {J.}~\bibnamefont {Schmiedmayer}}, \bibinfo {author} {\bibfnamefont {Y.}~\bibnamefont {Deng}}, \bibinfo {author} {\bibfnamefont {X.-J.}\ \bibnamefont {Liu}}, \bibinfo {author} {\bibfnamefont {S.}~\bibnamefont {Chen}}, \ and\ \bibinfo {author} {\bibfnamefont {J.-W.}\ \bibnamefont {Pan}},\ }\href@noop {} {\bibfield  {journal} {\bibinfo  {journal} {Phys. Rev. Lett.}\ }\textbf {\bibinfo {volume} {121}},\ \bibinfo {pages} {250403} (\bibinfo {year} {2018}{\natexlab{b}})}\BibitemShut {NoStop}%
\bibitem [{\citenamefont {Wang}\ \emph {et~al.}(2021{\natexlab{a}})\citenamefont {Wang}, \citenamefont {Cheng}, \citenamefont {Wang}, \citenamefont {Zhang}, \citenamefont {Lu}, \citenamefont {Yi}, \citenamefont {Niu}, \citenamefont {Deng}, \citenamefont {Liu}, \citenamefont {Chen} \emph {et~al.}}]{wang2021realization}%
  \BibitemOpen
  \bibfield  {author} {\bibinfo {author} {\bibfnamefont {Z.-Y.}\ \bibnamefont {Wang}}, \bibinfo {author} {\bibfnamefont {X.-C.}\ \bibnamefont {Cheng}}, \bibinfo {author} {\bibfnamefont {B.-Z.}\ \bibnamefont {Wang}}, \bibinfo {author} {\bibfnamefont {J.-Y.}\ \bibnamefont {Zhang}}, \bibinfo {author} {\bibfnamefont {Y.-H.}\ \bibnamefont {Lu}}, \bibinfo {author} {\bibfnamefont {C.-R.}\ \bibnamefont {Yi}}, \bibinfo {author} {\bibfnamefont {S.}~\bibnamefont {Niu}}, \bibinfo {author} {\bibfnamefont {Y.}~\bibnamefont {Deng}}, \bibinfo {author} {\bibfnamefont {X.-J.}\ \bibnamefont {Liu}}, \bibinfo {author} {\bibfnamefont {S.}~\bibnamefont {Chen}},  \emph {et~al.},\ }\href@noop {} {\bibfield  {journal} {\bibinfo  {journal} {Science}\ }\textbf {\bibinfo {volume} {372}},\ \bibinfo {pages} {271} (\bibinfo {year} {2021}{\natexlab{a}})}\BibitemShut {NoStop}%
\bibitem [{\citenamefont {Liang}\ \emph {et~al.}(2023)\citenamefont {Liang}, \citenamefont {Wei}, \citenamefont {Zhang}, \citenamefont {Wang}, \citenamefont {Zhang}, \citenamefont {Wang}, \citenamefont {Qi}, \citenamefont {Liu},\ and\ \citenamefont {Zhang}}]{liang2023realization}%
  \BibitemOpen
  \bibfield  {author} {\bibinfo {author} {\bibfnamefont {M.-C.}\ \bibnamefont {Liang}}, \bibinfo {author} {\bibfnamefont {Y.-D.}\ \bibnamefont {Wei}}, \bibinfo {author} {\bibfnamefont {L.}~\bibnamefont {Zhang}}, \bibinfo {author} {\bibfnamefont {X.-J.}\ \bibnamefont {Wang}}, \bibinfo {author} {\bibfnamefont {H.}~\bibnamefont {Zhang}}, \bibinfo {author} {\bibfnamefont {W.-W.}\ \bibnamefont {Wang}}, \bibinfo {author} {\bibfnamefont {W.}~\bibnamefont {Qi}}, \bibinfo {author} {\bibfnamefont {X.-J.}\ \bibnamefont {Liu}}, \ and\ \bibinfo {author} {\bibfnamefont {X.}~\bibnamefont {Zhang}},\ }\href@noop {} {\bibfield  {journal} {\bibinfo  {journal} {Phys. Rev. Research}\ }\textbf {\bibinfo {volume} {5}},\ \bibinfo {pages} {L012006} (\bibinfo {year} {2023})}\BibitemShut {NoStop}%
\bibitem [{\citenamefont {Zhao}\ \emph {et~al.}(2023)\citenamefont {Zhao}, \citenamefont {Wang}, \citenamefont {He}, \citenamefont {Poon}, \citenamefont {Pak}, \citenamefont {Liu}, \citenamefont {Ren}, \citenamefont {Liu},\ and\ \citenamefont {Jo}}]{zhao2023two}%
  \BibitemOpen
  \bibfield  {author} {\bibinfo {author} {\bibfnamefont {E.}~\bibnamefont {Zhao}}, \bibinfo {author} {\bibfnamefont {Z.}~\bibnamefont {Wang}}, \bibinfo {author} {\bibfnamefont {C.}~\bibnamefont {He}}, \bibinfo {author} {\bibfnamefont {T.~F.~J.}\ \bibnamefont {Poon}}, \bibinfo {author} {\bibfnamefont {K.~K.}\ \bibnamefont {Pak}}, \bibinfo {author} {\bibfnamefont {Y.-J.}\ \bibnamefont {Liu}}, \bibinfo {author} {\bibfnamefont {P.}~\bibnamefont {Ren}}, \bibinfo {author} {\bibfnamefont {X.-J.}\ \bibnamefont {Liu}}, \ and\ \bibinfo {author} {\bibfnamefont {G.-B.}\ \bibnamefont {Jo}},\ }\href@noop {} {\bibfield  {journal} {\bibinfo  {journal} {arXiv preprint arXiv:2311.07931}\ } (\bibinfo {year} {2023})}\BibitemShut {NoStop}%
\bibitem [{\citenamefont {Zhou}\ \emph {et~al.}(2023)\citenamefont {Zhou}, \citenamefont {Yang}, \citenamefont {Wang},\ and\ \citenamefont {Liu}}]{zhou2023non}%
  \BibitemOpen
  \bibfield  {author} {\bibinfo {author} {\bibfnamefont {X.-C.}\ \bibnamefont {Zhou}}, \bibinfo {author} {\bibfnamefont {T.-H.}\ \bibnamefont {Yang}}, \bibinfo {author} {\bibfnamefont {Z.-Y.}\ \bibnamefont {Wang}}, \ and\ \bibinfo {author} {\bibfnamefont {X.-J.}\ \bibnamefont {Liu}},\ }\href@noop {} {\bibfield  {journal} {\bibinfo  {journal} {arXiv preprint arXiv:2309.12923}\ } (\bibinfo {year} {2023})}\BibitemShut {NoStop}%
\bibitem [{\citenamefont {Goldman}\ \emph {et~al.}(2010)\citenamefont {Goldman}, \citenamefont {Satija}, \citenamefont {Nikolic}, \citenamefont {Bermudez}, \citenamefont {Martin-Delgado}, \citenamefont {Lewenstein},\ and\ \citenamefont {Spielman}}]{GoldmanPRL2010}%
  \BibitemOpen
  \bibfield  {author} {\bibinfo {author} {\bibfnamefont {N.}~\bibnamefont {Goldman}}, \bibinfo {author} {\bibfnamefont {I.}~\bibnamefont {Satija}}, \bibinfo {author} {\bibfnamefont {P.}~\bibnamefont {Nikolic}}, \bibinfo {author} {\bibfnamefont {A.}~\bibnamefont {Bermudez}}, \bibinfo {author} {\bibfnamefont {M.~A.}\ \bibnamefont {Martin-Delgado}}, \bibinfo {author} {\bibfnamefont {M.}~\bibnamefont {Lewenstein}}, \ and\ \bibinfo {author} {\bibfnamefont {I.}~\bibnamefont {Spielman}},\ }\href@noop {} {\bibfield  {journal} {\bibinfo  {journal} {Phys. Rev. Lett.}\ }\textbf {\bibinfo {volume} {105}},\ \bibinfo {pages} {255302} (\bibinfo {year} {2010})}\BibitemShut {NoStop}%
\bibitem [{\citenamefont {B{\'e}ri}\ and\ \citenamefont {Cooper}(2011)}]{CooperPRL2011}%
  \BibitemOpen
  \bibfield  {author} {\bibinfo {author} {\bibfnamefont {B.}~\bibnamefont {B{\'e}ri}}\ and\ \bibinfo {author} {\bibfnamefont {N.}~\bibnamefont {Cooper}},\ }\href@noop {} {\bibfield  {journal} {\bibinfo  {journal} {Phys. Rev. Lett.}\ }\textbf {\bibinfo {volume} {107}},\ \bibinfo {pages} {145301} (\bibinfo {year} {2011})}\BibitemShut {NoStop}%
\bibitem [{\citenamefont {Liu}\ \emph {et~al.}(2013)\citenamefont {Liu}, \citenamefont {Liu},\ and\ \citenamefont {Cheng}}]{Liu2013PRL}%
  \BibitemOpen
  \bibfield  {author} {\bibinfo {author} {\bibfnamefont {X.-J.}\ \bibnamefont {Liu}}, \bibinfo {author} {\bibfnamefont {Z.-X.}\ \bibnamefont {Liu}}, \ and\ \bibinfo {author} {\bibfnamefont {M.}~\bibnamefont {Cheng}},\ }\href@noop {} {\bibfield  {journal} {\bibinfo  {journal} {Phys. Rev. Lett.}\ }\textbf {\bibinfo {volume} {110}},\ \bibinfo {pages} {076401} (\bibinfo {year} {2013})}\BibitemShut {NoStop}%
\bibitem [{\citenamefont {Jotzu}\ \emph {et~al.}(2014)\citenamefont {Jotzu}, \citenamefont {Messer}, \citenamefont {Desbuquois}, \citenamefont {Lebrat}, \citenamefont {Uehlinger}, \citenamefont {Greif},\ and\ \citenamefont {Esslinger}}]{Jotzu2014Experimental}%
  \BibitemOpen
  \bibfield  {author} {\bibinfo {author} {\bibfnamefont {G.}~\bibnamefont {Jotzu}}, \bibinfo {author} {\bibfnamefont {M.}~\bibnamefont {Messer}}, \bibinfo {author} {\bibfnamefont {R.}~\bibnamefont {Desbuquois}}, \bibinfo {author} {\bibfnamefont {M.}~\bibnamefont {Lebrat}}, \bibinfo {author} {\bibfnamefont {T.}~\bibnamefont {Uehlinger}}, \bibinfo {author} {\bibfnamefont {D.}~\bibnamefont {Greif}}, \ and\ \bibinfo {author} {\bibfnamefont {T.}~\bibnamefont {Esslinger}},\ }\href@noop {} {\bibfield  {journal} {\bibinfo  {journal} {Nature}\ }\textbf {\bibinfo {volume} {515}},\ \bibinfo {pages} {237} (\bibinfo {year} {2014})}\BibitemShut {NoStop}%
\bibitem [{\citenamefont {Wang}\ \emph {et~al.}(2014)\citenamefont {Wang}, \citenamefont {Deng},\ and\ \citenamefont {Duan}}]{WangPRL2014}%
  \BibitemOpen
  \bibfield  {author} {\bibinfo {author} {\bibfnamefont {S.-T.}\ \bibnamefont {Wang}}, \bibinfo {author} {\bibfnamefont {D.-L.}\ \bibnamefont {Deng}}, \ and\ \bibinfo {author} {\bibfnamefont {L.-M.}\ \bibnamefont {Duan}},\ }\href@noop {} {\bibfield  {journal} {\bibinfo  {journal} {Phys. Rev. Lett.}\ }\textbf {\bibinfo {volume} {113}},\ \bibinfo {pages} {033002} (\bibinfo {year} {2014})}\BibitemShut {NoStop}%
\bibitem [{\citenamefont {Zhang}\ and\ \citenamefont {Zhou}(2017)}]{PhysRevA.95.061601}%
  \BibitemOpen
  \bibfield  {author} {\bibinfo {author} {\bibfnamefont {S.-L.}\ \bibnamefont {Zhang}}\ and\ \bibinfo {author} {\bibfnamefont {Q.}~\bibnamefont {Zhou}},\ }\href@noop {} {\bibfield  {journal} {\bibinfo  {journal} {Phys. Rev. A}\ }\textbf {\bibinfo {volume} {95}},\ \bibinfo {pages} {061601} (\bibinfo {year} {2017})}\BibitemShut {NoStop}%
\bibitem [{\citenamefont {Song}\ \emph {et~al.}(2018)\citenamefont {Song}, \citenamefont {Zhang}, \citenamefont {He}, \citenamefont {Poon}, \citenamefont {Hajiyev}, \citenamefont {Zhang}, \citenamefont {Liu},\ and\ \citenamefont {Jo}}]{SongScienceAdvances2018}%
  \BibitemOpen
  \bibfield  {author} {\bibinfo {author} {\bibfnamefont {B.}~\bibnamefont {Song}}, \bibinfo {author} {\bibfnamefont {L.}~\bibnamefont {Zhang}}, \bibinfo {author} {\bibfnamefont {C.}~\bibnamefont {He}}, \bibinfo {author} {\bibfnamefont {T.~F.~J.}\ \bibnamefont {Poon}}, \bibinfo {author} {\bibfnamefont {E.}~\bibnamefont {Hajiyev}}, \bibinfo {author} {\bibfnamefont {S.}~\bibnamefont {Zhang}}, \bibinfo {author} {\bibfnamefont {X.-J.}\ \bibnamefont {Liu}}, \ and\ \bibinfo {author} {\bibfnamefont {G.-B.}\ \bibnamefont {Jo}},\ }\href@noop {} {\bibfield  {journal} {\bibinfo  {journal} {Sci. Adv}\ }\textbf {\bibinfo {volume} {4}} (\bibinfo {year} {2018})}\BibitemShut {NoStop}%
\bibitem [{\citenamefont {Yi}\ \emph {et~al.}(2019)\citenamefont {Yi}, \citenamefont {Zhang}, \citenamefont {Zhang}, \citenamefont {Jiao}, \citenamefont {Cheng}, \citenamefont {Wang}, \citenamefont {Xu}, \citenamefont {Sun}, \citenamefont {Liu}, \citenamefont {Chen} \emph {et~al.}}]{yi2019observing}%
  \BibitemOpen
  \bibfield  {author} {\bibinfo {author} {\bibfnamefont {C.-R.}\ \bibnamefont {Yi}}, \bibinfo {author} {\bibfnamefont {L.}~\bibnamefont {Zhang}}, \bibinfo {author} {\bibfnamefont {L.}~\bibnamefont {Zhang}}, \bibinfo {author} {\bibfnamefont {R.-H.}\ \bibnamefont {Jiao}}, \bibinfo {author} {\bibfnamefont {X.-C.}\ \bibnamefont {Cheng}}, \bibinfo {author} {\bibfnamefont {Z.-Y.}\ \bibnamefont {Wang}}, \bibinfo {author} {\bibfnamefont {X.-T.}\ \bibnamefont {Xu}}, \bibinfo {author} {\bibfnamefont {W.}~\bibnamefont {Sun}}, \bibinfo {author} {\bibfnamefont {X.-J.}\ \bibnamefont {Liu}}, \bibinfo {author} {\bibfnamefont {S.}~\bibnamefont {Chen}},  \emph {et~al.},\ }\href@noop {} {\bibfield  {journal} {\bibinfo  {journal} {Phys. Rev. Lett.}\ }\textbf {\bibinfo {volume} {123}},\ \bibinfo {pages} {190603} (\bibinfo {year} {2019})}\BibitemShut {NoStop}%
\bibitem [{\citenamefont {Boninsegni}\ and\ \citenamefont {Prokof’ev}(2012)}]{boninsegni2012colloquium}%
  \BibitemOpen
  \bibfield  {author} {\bibinfo {author} {\bibfnamefont {M.}~\bibnamefont {Boninsegni}}\ and\ \bibinfo {author} {\bibfnamefont {N.~V.}\ \bibnamefont {Prokof’ev}},\ }\href@noop {} {\bibfield  {journal} {\bibinfo  {journal} {Rev. Mod. Phys.}\ }\textbf {\bibinfo {volume} {84}},\ \bibinfo {pages} {759} (\bibinfo {year} {2012})}\BibitemShut {NoStop}%
\bibitem [{\citenamefont {Ritsch}\ \emph {et~al.}(2013)\citenamefont {Ritsch}, \citenamefont {Domokos}, \citenamefont {Brennecke},\ and\ \citenamefont {Esslinger}}]{ritsch2013cold}%
  \BibitemOpen
  \bibfield  {author} {\bibinfo {author} {\bibfnamefont {H.}~\bibnamefont {Ritsch}}, \bibinfo {author} {\bibfnamefont {P.}~\bibnamefont {Domokos}}, \bibinfo {author} {\bibfnamefont {F.}~\bibnamefont {Brennecke}}, \ and\ \bibinfo {author} {\bibfnamefont {T.}~\bibnamefont {Esslinger}},\ }\href@noop {} {\bibfield  {journal} {\bibinfo  {journal} {Rev. Mod. Phys.}\ }\textbf {\bibinfo {volume} {85}},\ \bibinfo {pages} {553} (\bibinfo {year} {2013})}\BibitemShut {NoStop}%
\bibitem [{\citenamefont {Recati}\ and\ \citenamefont {Stringari}(2023)}]{recati2023supersolidity}%
  \BibitemOpen
  \bibfield  {author} {\bibinfo {author} {\bibfnamefont {A.}~\bibnamefont {Recati}}\ and\ \bibinfo {author} {\bibfnamefont {S.}~\bibnamefont {Stringari}},\ }\href@noop {} {\bibfield  {journal} {\bibinfo  {journal} {Nature Reviews Physics}\ }\textbf {\bibinfo {volume} {5}},\ \bibinfo {pages} {735} (\bibinfo {year} {2023})}\BibitemShut {NoStop}%
\bibitem [{\citenamefont {Li}\ \emph {et~al.}(2017)\citenamefont {Li}, \citenamefont {Lee}, \citenamefont {Huang}, \citenamefont {Burchesky}, \citenamefont {Shteynas}, \citenamefont {Top}, \citenamefont {Jamison},\ and\ \citenamefont {Ketterle}}]{li2017stripe}%
  \BibitemOpen
  \bibfield  {author} {\bibinfo {author} {\bibfnamefont {J.-R.}\ \bibnamefont {Li}}, \bibinfo {author} {\bibfnamefont {J.}~\bibnamefont {Lee}}, \bibinfo {author} {\bibfnamefont {W.}~\bibnamefont {Huang}}, \bibinfo {author} {\bibfnamefont {S.}~\bibnamefont {Burchesky}}, \bibinfo {author} {\bibfnamefont {B.}~\bibnamefont {Shteynas}}, \bibinfo {author} {\bibfnamefont {F.~{\c{C}}.}\ \bibnamefont {Top}}, \bibinfo {author} {\bibfnamefont {A.~O.}\ \bibnamefont {Jamison}}, \ and\ \bibinfo {author} {\bibfnamefont {W.}~\bibnamefont {Ketterle}},\ }\href@noop {} {\bibfield  {journal} {\bibinfo  {journal} {Nature}\ }\textbf {\bibinfo {volume} {543}},\ \bibinfo {pages} {91} (\bibinfo {year} {2017})}\BibitemShut {NoStop}%
\bibitem [{\citenamefont {B{\"o}ttcher}\ \emph {et~al.}(2019)\citenamefont {B{\"o}ttcher}, \citenamefont {Schmidt}, \citenamefont {Wenzel}, \citenamefont {Hertkorn}, \citenamefont {Guo}, \citenamefont {Langen},\ and\ \citenamefont {Pfau}}]{bottcher2019transient}%
  \BibitemOpen
  \bibfield  {author} {\bibinfo {author} {\bibfnamefont {F.}~\bibnamefont {B{\"o}ttcher}}, \bibinfo {author} {\bibfnamefont {J.-N.}\ \bibnamefont {Schmidt}}, \bibinfo {author} {\bibfnamefont {M.}~\bibnamefont {Wenzel}}, \bibinfo {author} {\bibfnamefont {J.}~\bibnamefont {Hertkorn}}, \bibinfo {author} {\bibfnamefont {M.}~\bibnamefont {Guo}}, \bibinfo {author} {\bibfnamefont {T.}~\bibnamefont {Langen}}, \ and\ \bibinfo {author} {\bibfnamefont {T.}~\bibnamefont {Pfau}},\ }\href@noop {} {\bibfield  {journal} {\bibinfo  {journal} {Phys. Rev. X}\ }\textbf {\bibinfo {volume} {9}},\ \bibinfo {pages} {011051} (\bibinfo {year} {2019})}\BibitemShut {NoStop}%
\bibitem [{\citenamefont {Tanzi}\ \emph {et~al.}(2019)\citenamefont {Tanzi}, \citenamefont {Lucioni}, \citenamefont {Fam{\`a}}, \citenamefont {Catani}, \citenamefont {Fioretti}, \citenamefont {Gabbanini}, \citenamefont {Bisset}, \citenamefont {Santos},\ and\ \citenamefont {Modugno}}]{tanzi2019observation}%
  \BibitemOpen
  \bibfield  {author} {\bibinfo {author} {\bibfnamefont {L.}~\bibnamefont {Tanzi}}, \bibinfo {author} {\bibfnamefont {E.}~\bibnamefont {Lucioni}}, \bibinfo {author} {\bibfnamefont {F.}~\bibnamefont {Fam{\`a}}}, \bibinfo {author} {\bibfnamefont {J.}~\bibnamefont {Catani}}, \bibinfo {author} {\bibfnamefont {A.}~\bibnamefont {Fioretti}}, \bibinfo {author} {\bibfnamefont {C.}~\bibnamefont {Gabbanini}}, \bibinfo {author} {\bibfnamefont {R.~N.}\ \bibnamefont {Bisset}}, \bibinfo {author} {\bibfnamefont {L.}~\bibnamefont {Santos}}, \ and\ \bibinfo {author} {\bibfnamefont {G.}~\bibnamefont {Modugno}},\ }\href@noop {} {\bibfield  {journal} {\bibinfo  {journal} {Phys. Rev. Lett.}\ }\textbf {\bibinfo {volume} {122}},\ \bibinfo {pages} {130405} (\bibinfo {year} {2019})}\BibitemShut {NoStop}%
\bibitem [{\citenamefont {Chomaz}\ \emph {et~al.}(2019)\citenamefont {Chomaz}, \citenamefont {Petter}, \citenamefont {Ilzh{\"o}fer}, \citenamefont {Natale}, \citenamefont {Trautmann}, \citenamefont {Politi}, \citenamefont {Durastante}, \citenamefont {Van~Bijnen}, \citenamefont {Patscheider}, \citenamefont {Sohmen} \emph {et~al.}}]{chomaz2019long}%
  \BibitemOpen
  \bibfield  {author} {\bibinfo {author} {\bibfnamefont {L.}~\bibnamefont {Chomaz}}, \bibinfo {author} {\bibfnamefont {D.}~\bibnamefont {Petter}}, \bibinfo {author} {\bibfnamefont {P.}~\bibnamefont {Ilzh{\"o}fer}}, \bibinfo {author} {\bibfnamefont {G.}~\bibnamefont {Natale}}, \bibinfo {author} {\bibfnamefont {A.}~\bibnamefont {Trautmann}}, \bibinfo {author} {\bibfnamefont {C.}~\bibnamefont {Politi}}, \bibinfo {author} {\bibfnamefont {G.}~\bibnamefont {Durastante}}, \bibinfo {author} {\bibfnamefont {R.}~\bibnamefont {Van~Bijnen}}, \bibinfo {author} {\bibfnamefont {A.}~\bibnamefont {Patscheider}}, \bibinfo {author} {\bibfnamefont {M.}~\bibnamefont {Sohmen}},  \emph {et~al.},\ }\href@noop {} {\bibfield  {journal} {\bibinfo  {journal} {Phys. Rev. X}\ }\textbf {\bibinfo {volume} {9}},\ \bibinfo {pages} {021012} (\bibinfo {year} {2019})}\BibitemShut {NoStop}%
\bibitem [{\citenamefont {Norcia}\ \emph {et~al.}(2021)\citenamefont {Norcia}, \citenamefont {Politi}, \citenamefont {Klaus}, \citenamefont {Poli}, \citenamefont {Sohmen}, \citenamefont {Mark}, \citenamefont {Bisset}, \citenamefont {Santos},\ and\ \citenamefont {Ferlaino}}]{norcia2021two}%
  \BibitemOpen
  \bibfield  {author} {\bibinfo {author} {\bibfnamefont {M.~A.}\ \bibnamefont {Norcia}}, \bibinfo {author} {\bibfnamefont {C.}~\bibnamefont {Politi}}, \bibinfo {author} {\bibfnamefont {L.}~\bibnamefont {Klaus}}, \bibinfo {author} {\bibfnamefont {E.}~\bibnamefont {Poli}}, \bibinfo {author} {\bibfnamefont {M.}~\bibnamefont {Sohmen}}, \bibinfo {author} {\bibfnamefont {M.~J.}\ \bibnamefont {Mark}}, \bibinfo {author} {\bibfnamefont {R.~N.}\ \bibnamefont {Bisset}}, \bibinfo {author} {\bibfnamefont {L.}~\bibnamefont {Santos}}, \ and\ \bibinfo {author} {\bibfnamefont {F.}~\bibnamefont {Ferlaino}},\ }\href@noop {} {\bibfield  {journal} {\bibinfo  {journal} {Nature}\ }\textbf {\bibinfo {volume} {596}},\ \bibinfo {pages} {357} (\bibinfo {year} {2021})}\BibitemShut {NoStop}%
\bibitem [{\citenamefont {Bland}\ \emph {et~al.}(2022)\citenamefont {Bland}, \citenamefont {Poli}, \citenamefont {Politi}, \citenamefont {Klaus}, \citenamefont {Norcia}, \citenamefont {Ferlaino}, \citenamefont {Santos},\ and\ \citenamefont {Bisset}}]{bland2022two}%
  \BibitemOpen
  \bibfield  {author} {\bibinfo {author} {\bibfnamefont {T.}~\bibnamefont {Bland}}, \bibinfo {author} {\bibfnamefont {E.}~\bibnamefont {Poli}}, \bibinfo {author} {\bibfnamefont {C.}~\bibnamefont {Politi}}, \bibinfo {author} {\bibfnamefont {L.}~\bibnamefont {Klaus}}, \bibinfo {author} {\bibfnamefont {M.~A.}\ \bibnamefont {Norcia}}, \bibinfo {author} {\bibfnamefont {F.}~\bibnamefont {Ferlaino}}, \bibinfo {author} {\bibfnamefont {L.}~\bibnamefont {Santos}}, \ and\ \bibinfo {author} {\bibfnamefont {R.~N.}\ \bibnamefont {Bisset}},\ }\href@noop {} {\bibfield  {journal} {\bibinfo  {journal} {Phys. Rev. Lett.}\ }\textbf {\bibinfo {volume} {128}},\ \bibinfo {pages} {195302} (\bibinfo {year} {2022})}\BibitemShut {NoStop}%
\bibitem [{\citenamefont {Casotti}\ \emph {et~al.}(2024)\citenamefont {Casotti}, \citenamefont {Poli}, \citenamefont {Klaus}, \citenamefont {Litvinov}, \citenamefont {Ulm}, \citenamefont {Politi}, \citenamefont {Mark}, \citenamefont {Bland},\ and\ \citenamefont {Ferlaino}}]{casotti2024observation}%
  \BibitemOpen
  \bibfield  {author} {\bibinfo {author} {\bibfnamefont {E.}~\bibnamefont {Casotti}}, \bibinfo {author} {\bibfnamefont {E.}~\bibnamefont {Poli}}, \bibinfo {author} {\bibfnamefont {L.}~\bibnamefont {Klaus}}, \bibinfo {author} {\bibfnamefont {A.}~\bibnamefont {Litvinov}}, \bibinfo {author} {\bibfnamefont {C.}~\bibnamefont {Ulm}}, \bibinfo {author} {\bibfnamefont {C.}~\bibnamefont {Politi}}, \bibinfo {author} {\bibfnamefont {M.~J.}\ \bibnamefont {Mark}}, \bibinfo {author} {\bibfnamefont {T.}~\bibnamefont {Bland}}, \ and\ \bibinfo {author} {\bibfnamefont {F.}~\bibnamefont {Ferlaino}},\ }\href@noop {} {\bibfield  {journal} {\bibinfo  {journal} {Nature}\ }\textbf {\bibinfo {volume} {635}},\ \bibinfo {pages} {327} (\bibinfo {year} {2024})}\BibitemShut {NoStop}%
\bibitem [{\citenamefont {L{\'e}onard}\ \emph {et~al.}(2017{\natexlab{a}})\citenamefont {L{\'e}onard}, \citenamefont {Morales}, \citenamefont {Zupancic}, \citenamefont {Esslinger},\ and\ \citenamefont {Donner}}]{leonard2017supersolid}%
  \BibitemOpen
  \bibfield  {author} {\bibinfo {author} {\bibfnamefont {J.}~\bibnamefont {L{\'e}onard}}, \bibinfo {author} {\bibfnamefont {A.}~\bibnamefont {Morales}}, \bibinfo {author} {\bibfnamefont {P.}~\bibnamefont {Zupancic}}, \bibinfo {author} {\bibfnamefont {T.}~\bibnamefont {Esslinger}}, \ and\ \bibinfo {author} {\bibfnamefont {T.}~\bibnamefont {Donner}},\ }\href@noop {} {\bibfield  {journal} {\bibinfo  {journal} {Nature}\ }\textbf {\bibinfo {volume} {543}},\ \bibinfo {pages} {87} (\bibinfo {year} {2017}{\natexlab{a}})}\BibitemShut {NoStop}%
\bibitem [{\citenamefont {L{\'e}onard}\ \emph {et~al.}(2017{\natexlab{b}})\citenamefont {L{\'e}onard}, \citenamefont {Morales}, \citenamefont {Zupancic}, \citenamefont {Donner},\ and\ \citenamefont {Esslinger}}]{leonard2017monitoring}%
  \BibitemOpen
  \bibfield  {author} {\bibinfo {author} {\bibfnamefont {J.}~\bibnamefont {L{\'e}onard}}, \bibinfo {author} {\bibfnamefont {A.}~\bibnamefont {Morales}}, \bibinfo {author} {\bibfnamefont {P.}~\bibnamefont {Zupancic}}, \bibinfo {author} {\bibfnamefont {T.}~\bibnamefont {Donner}}, \ and\ \bibinfo {author} {\bibfnamefont {T.}~\bibnamefont {Esslinger}},\ }\href@noop {} {\bibfield  {journal} {\bibinfo  {journal} {Science}\ }\textbf {\bibinfo {volume} {358}},\ \bibinfo {pages} {1415} (\bibinfo {year} {2017}{\natexlab{b}})}\BibitemShut {NoStop}%
\bibitem [{\citenamefont {Zhao}\ and\ \citenamefont {Liu}(2008)}]{ZhaoPRL2008}%
  \BibitemOpen
  \bibfield  {author} {\bibinfo {author} {\bibfnamefont {E.}~\bibnamefont {Zhao}}\ and\ \bibinfo {author} {\bibfnamefont {W.~V.}\ \bibnamefont {Liu}},\ }\href@noop {} {\bibfield  {journal} {\bibinfo  {journal} {Phys. Rev. Lett.}\ }\textbf {\bibinfo {volume} {100}},\ \bibinfo {pages} {160403} (\bibinfo {year} {2008})}\BibitemShut {NoStop}%
\bibitem [{\citenamefont {Liu}\ \emph {et~al.}(2010)\citenamefont {Liu}, \citenamefont {Liu}, \citenamefont {Wu},\ and\ \citenamefont {Sinova}}]{XiongjunPRA2010}%
  \BibitemOpen
  \bibfield  {author} {\bibinfo {author} {\bibfnamefont {X.-J.}\ \bibnamefont {Liu}}, \bibinfo {author} {\bibfnamefont {X.}~\bibnamefont {Liu}}, \bibinfo {author} {\bibfnamefont {C.}~\bibnamefont {Wu}}, \ and\ \bibinfo {author} {\bibfnamefont {J.}~\bibnamefont {Sinova}},\ }\href@noop {} {\bibfield  {journal} {\bibinfo  {journal} {Phys. Rev. A}\ }\textbf {\bibinfo {volume} {81}},\ \bibinfo {pages} {033622} (\bibinfo {year} {2010})}\BibitemShut {NoStop}%
\bibitem [{\citenamefont {Cai}\ \emph {et~al.}(2011)\citenamefont {Cai}, \citenamefont {Wang},\ and\ \citenamefont {Wu}}]{CaiPRA2011}%
  \BibitemOpen
  \bibfield  {author} {\bibinfo {author} {\bibfnamefont {Z.}~\bibnamefont {Cai}}, \bibinfo {author} {\bibfnamefont {Y.}~\bibnamefont {Wang}}, \ and\ \bibinfo {author} {\bibfnamefont {C.}~\bibnamefont {Wu}},\ }\href@noop {} {\bibfield  {journal} {\bibinfo  {journal} {Phys. Rev. A}\ }\textbf {\bibinfo {volume} {83}},\ \bibinfo {pages} {063621} (\bibinfo {year} {2011})}\BibitemShut {NoStop}%
\bibitem [{\citenamefont {Lewenstein}\ and\ \citenamefont {Liu}(2011)}]{LewensteinNP2011}%
  \BibitemOpen
  \bibfield  {author} {\bibinfo {author} {\bibfnamefont {M.}~\bibnamefont {Lewenstein}}\ and\ \bibinfo {author} {\bibfnamefont {W.~V.}\ \bibnamefont {Liu}},\ }\href@noop {} {\bibfield  {journal} {\bibinfo  {journal} {Nat. Phys.}\ }\textbf {\bibinfo {volume} {7}},\ \bibinfo {pages} {101} (\bibinfo {year} {2011})}\BibitemShut {NoStop}%
\bibitem [{\citenamefont {Li}\ \emph {et~al.}(2013)\citenamefont {Li}, \citenamefont {Zhao},\ and\ \citenamefont {Vincent~Liu}}]{LiNC2013}%
  \BibitemOpen
  \bibfield  {author} {\bibinfo {author} {\bibfnamefont {X.}~\bibnamefont {Li}}, \bibinfo {author} {\bibfnamefont {E.}~\bibnamefont {Zhao}}, \ and\ \bibinfo {author} {\bibfnamefont {W.}~\bibnamefont {Vincent~Liu}},\ }\href@noop {} {\bibfield  {journal} {\bibinfo  {journal} {Nat Commun}\ }\textbf {\bibinfo {volume} {4}},\ \bibinfo {pages} {1523} (\bibinfo {year} {2013})}\BibitemShut {NoStop}%
\bibitem [{\citenamefont {Sowi{\'n}ski}\ \emph {et~al.}(2013)\citenamefont {Sowi{\'n}ski}, \citenamefont {{\L}{\k{a}}cki}, \citenamefont {Dutta}, \citenamefont {Pietraszewicz}, \citenamefont {Sierant}, \citenamefont {Gajda}, \citenamefont {Zakrzewski},\ and\ \citenamefont {Lewenstein}}]{TomaszPRL2013}%
  \BibitemOpen
  \bibfield  {author} {\bibinfo {author} {\bibfnamefont {T.}~\bibnamefont {Sowi{\'n}ski}}, \bibinfo {author} {\bibfnamefont {M.}~\bibnamefont {{\L}{\k{a}}cki}}, \bibinfo {author} {\bibfnamefont {O.}~\bibnamefont {Dutta}}, \bibinfo {author} {\bibfnamefont {J.}~\bibnamefont {Pietraszewicz}}, \bibinfo {author} {\bibfnamefont {P.}~\bibnamefont {Sierant}}, \bibinfo {author} {\bibfnamefont {M.}~\bibnamefont {Gajda}}, \bibinfo {author} {\bibfnamefont {J.}~\bibnamefont {Zakrzewski}}, \ and\ \bibinfo {author} {\bibfnamefont {M.}~\bibnamefont {Lewenstein}},\ }\href@noop {} {\bibfield  {journal} {\bibinfo  {journal} {Phys. Rev. Lett.}\ }\textbf {\bibinfo {volume} {111}},\ \bibinfo {pages} {215302} (\bibinfo {year} {2013})}\BibitemShut {NoStop}%
\bibitem [{\citenamefont {Dutta}\ \emph {et~al.}(2015)\citenamefont {Dutta}, \citenamefont {Gajda}, \citenamefont {Hauke}, \citenamefont {Lewenstein}, \citenamefont {L{\"u}hmann}, \citenamefont {Malomed}, \citenamefont {Sowi{\'n}ski},\ and\ \citenamefont {Zakrzewski}}]{dutta2015non}%
  \BibitemOpen
  \bibfield  {author} {\bibinfo {author} {\bibfnamefont {O.}~\bibnamefont {Dutta}}, \bibinfo {author} {\bibfnamefont {M.}~\bibnamefont {Gajda}}, \bibinfo {author} {\bibfnamefont {P.}~\bibnamefont {Hauke}}, \bibinfo {author} {\bibfnamefont {M.}~\bibnamefont {Lewenstein}}, \bibinfo {author} {\bibfnamefont {D.-S.}\ \bibnamefont {L{\"u}hmann}}, \bibinfo {author} {\bibfnamefont {B.~A.}\ \bibnamefont {Malomed}}, \bibinfo {author} {\bibfnamefont {T.}~\bibnamefont {Sowi{\'n}ski}}, \ and\ \bibinfo {author} {\bibfnamefont {J.}~\bibnamefont {Zakrzewski}},\ }\href@noop {} {\bibfield  {journal} {\bibinfo  {journal} {Reports on Progress in Physics}\ }\textbf {\bibinfo {volume} {78}},\ \bibinfo {pages} {066001} (\bibinfo {year} {2015})}\BibitemShut {NoStop}%
\bibitem [{\citenamefont {Li}\ and\ \citenamefont {Liu}(2016)}]{li2016physics}%
  \BibitemOpen
  \bibfield  {author} {\bibinfo {author} {\bibfnamefont {X.}~\bibnamefont {Li}}\ and\ \bibinfo {author} {\bibfnamefont {W.~V.}\ \bibnamefont {Liu}},\ }\href@noop {} {\bibfield  {journal} {\bibinfo  {journal} {Reports on Progress in Physics}\ }\textbf {\bibinfo {volume} {79}},\ \bibinfo {pages} {116401} (\bibinfo {year} {2016})}\BibitemShut {NoStop}%
\bibitem [{\citenamefont {Kock}\ \emph {et~al.}(2016)\citenamefont {Kock}, \citenamefont {Hippler}, \citenamefont {Ewerbeck},\ and\ \citenamefont {Hemmerich}}]{kock2016orbital}%
  \BibitemOpen
  \bibfield  {author} {\bibinfo {author} {\bibfnamefont {T.}~\bibnamefont {Kock}}, \bibinfo {author} {\bibfnamefont {C.}~\bibnamefont {Hippler}}, \bibinfo {author} {\bibfnamefont {A.}~\bibnamefont {Ewerbeck}}, \ and\ \bibinfo {author} {\bibfnamefont {A.}~\bibnamefont {Hemmerich}},\ }\href@noop {} {\bibfield  {journal} {\bibinfo  {journal} {Journal of Physics B: Atomic, Molecular and Optical Physics}\ }\textbf {\bibinfo {volume} {49}},\ \bibinfo {pages} {042001} (\bibinfo {year} {2016})}\BibitemShut {NoStop}%
\bibitem [{\citenamefont {Isacsson}\ and\ \citenamefont {Girvin}(2005)}]{Isacsson2005PRA}%
  \BibitemOpen
  \bibfield  {author} {\bibinfo {author} {\bibfnamefont {A.}~\bibnamefont {Isacsson}}\ and\ \bibinfo {author} {\bibfnamefont {S.~M.}\ \bibnamefont {Girvin}},\ }\href@noop {} {\bibfield  {journal} {\bibinfo  {journal} {Phys. Rev. A}\ }\textbf {\bibinfo {volume} {72}},\ \bibinfo {pages} {053604} (\bibinfo {year} {2005})}\BibitemShut {NoStop}%
\bibitem [{\citenamefont {Kuklov}(2006)}]{Kuklov2006PRL}%
  \BibitemOpen
  \bibfield  {author} {\bibinfo {author} {\bibfnamefont {A.~B.}\ \bibnamefont {Kuklov}},\ }\href@noop {} {\bibfield  {journal} {\bibinfo  {journal} {Phys. Rev. Lett.}\ }\textbf {\bibinfo {volume} {97}},\ \bibinfo {pages} {110405} (\bibinfo {year} {2006})}\BibitemShut {NoStop}%
\bibitem [{\citenamefont {Liu}\ and\ \citenamefont {Wu}(2006)}]{liu2006atomic}%
  \BibitemOpen
  \bibfield  {author} {\bibinfo {author} {\bibfnamefont {W.~V.}\ \bibnamefont {Liu}}\ and\ \bibinfo {author} {\bibfnamefont {C.}~\bibnamefont {Wu}},\ }\href@noop {} {\bibfield  {journal} {\bibinfo  {journal} {Phys. Rev. A}\ }\textbf {\bibinfo {volume} {74}},\ \bibinfo {pages} {013607} (\bibinfo {year} {2006})}\BibitemShut {NoStop}%
\bibitem [{\citenamefont {Wu}\ \emph {et~al.}(2006)\citenamefont {Wu}, \citenamefont {Liu}, \citenamefont {Moore},\ and\ \citenamefont {Sarma}}]{wu2006quantum}%
  \BibitemOpen
  \bibfield  {author} {\bibinfo {author} {\bibfnamefont {C.}~\bibnamefont {Wu}}, \bibinfo {author} {\bibfnamefont {W.~V.}\ \bibnamefont {Liu}}, \bibinfo {author} {\bibfnamefont {J.}~\bibnamefont {Moore}}, \ and\ \bibinfo {author} {\bibfnamefont {S.~D.}\ \bibnamefont {Sarma}},\ }\href@noop {} {\bibfield  {journal} {\bibinfo  {journal} {Phys. Rev. Lett.}\ }\textbf {\bibinfo {volume} {97}},\ \bibinfo {pages} {190406} (\bibinfo {year} {2006})}\BibitemShut {NoStop}%
\bibitem [{\citenamefont {Larson}\ \emph {et~al.}(2009)\citenamefont {Larson}, \citenamefont {Collin},\ and\ \citenamefont {Martikainen}}]{LasonPRA2009}%
  \BibitemOpen
  \bibfield  {author} {\bibinfo {author} {\bibfnamefont {J.}~\bibnamefont {Larson}}, \bibinfo {author} {\bibfnamefont {A.}~\bibnamefont {Collin}}, \ and\ \bibinfo {author} {\bibfnamefont {J.-P.}\ \bibnamefont {Martikainen}},\ }\href@noop {} {\bibfield  {journal} {\bibinfo  {journal} {Phys. Rev. A}\ }\textbf {\bibinfo {volume} {79}},\ \bibinfo {pages} {033603} (\bibinfo {year} {2009})}\BibitemShut {NoStop}%
\bibitem [{\citenamefont {Wirth}\ \emph {et~al.}(2011)\citenamefont {Wirth}, \citenamefont {{\"O}lschl{\"a}ger},\ and\ \citenamefont {Hemmerich}}]{wirth2011evidence}%
  \BibitemOpen
  \bibfield  {author} {\bibinfo {author} {\bibfnamefont {G.}~\bibnamefont {Wirth}}, \bibinfo {author} {\bibfnamefont {M.}~\bibnamefont {{\"O}lschl{\"a}ger}}, \ and\ \bibinfo {author} {\bibfnamefont {A.}~\bibnamefont {Hemmerich}},\ }\href@noop {} {\bibfield  {journal} {\bibinfo  {journal} {Nat. Phys.}\ }\textbf {\bibinfo {volume} {7}},\ \bibinfo {pages} {147} (\bibinfo {year} {2011})}\BibitemShut {NoStop}%
\bibitem [{\citenamefont {Li}\ \emph {et~al.}(2012)\citenamefont {Li}, \citenamefont {Zhang},\ and\ \citenamefont {Liu}}]{li2012time}%
  \BibitemOpen
  \bibfield  {author} {\bibinfo {author} {\bibfnamefont {X.}~\bibnamefont {Li}}, \bibinfo {author} {\bibfnamefont {Z.}~\bibnamefont {Zhang}}, \ and\ \bibinfo {author} {\bibfnamefont {W.~V.}\ \bibnamefont {Liu}},\ }\href@noop {} {\bibfield  {journal} {\bibinfo  {journal} {Phys. Rev. Lett.}\ }\textbf {\bibinfo {volume} {108}},\ \bibinfo {pages} {175302} (\bibinfo {year} {2012})}\BibitemShut {NoStop}%
\bibitem [{\citenamefont {Xu}\ \emph {et~al.}(2016)\citenamefont {Xu}, \citenamefont {You}, \citenamefont {Hemmerich},\ and\ \citenamefont {Liu}}]{xu2016pi}%
  \BibitemOpen
  \bibfield  {author} {\bibinfo {author} {\bibfnamefont {Z.-F.}\ \bibnamefont {Xu}}, \bibinfo {author} {\bibfnamefont {L.}~\bibnamefont {You}}, \bibinfo {author} {\bibfnamefont {A.}~\bibnamefont {Hemmerich}}, \ and\ \bibinfo {author} {\bibfnamefont {W.~V.}\ \bibnamefont {Liu}},\ }\href@noop {} {\bibfield  {journal} {\bibinfo  {journal} {Phys. Rev. Lett.}\ }\textbf {\bibinfo {volume} {117}},\ \bibinfo {pages} {085301} (\bibinfo {year} {2016})}\BibitemShut {NoStop}%
\bibitem [{\citenamefont {Di~Liberto}\ \emph {et~al.}(2016)\citenamefont {Di~Liberto}, \citenamefont {Hemmerich},\ and\ \citenamefont {Morais~Smith}}]{di2016topological}%
  \BibitemOpen
  \bibfield  {author} {\bibinfo {author} {\bibfnamefont {M.}~\bibnamefont {Di~Liberto}}, \bibinfo {author} {\bibfnamefont {A.}~\bibnamefont {Hemmerich}}, \ and\ \bibinfo {author} {\bibfnamefont {C.}~\bibnamefont {Morais~Smith}},\ }\href@noop {} {\bibfield  {journal} {\bibinfo  {journal} {Phys. Rev. Lett.}\ }\textbf {\bibinfo {volume} {117}},\ \bibinfo {pages} {163001} (\bibinfo {year} {2016})}\BibitemShut {NoStop}%
\bibitem [{\citenamefont {Wang}\ and\ \citenamefont {Liu}(2017)}]{wang2017diractopologicalphononsspinorbital}%
  \BibitemOpen
  \bibfield  {author} {\bibinfo {author} {\bibfnamefont {Y.-Q.}\ \bibnamefont {Wang}}\ and\ \bibinfo {author} {\bibfnamefont {X.-J.}\ \bibnamefont {Liu}},\ }\href@noop {} {\enquote {\bibinfo {title} {Dirac and topological phonons with spin-orbital entangled orders},}\ } (\bibinfo {year} {2017}),\ \Eprint {http://arxiv.org/abs/1710.02070} {arXiv:1710.02070 [cond-mat.quant-gas]} \BibitemShut {NoStop}%
\bibitem [{\citenamefont {Li}\ \emph {et~al.}(2018)\citenamefont {Li}, \citenamefont {Yuan}, \citenamefont {Hemmerich},\ and\ \citenamefont {Li}}]{li2018rotation}%
  \BibitemOpen
  \bibfield  {author} {\bibinfo {author} {\bibfnamefont {Y.}~\bibnamefont {Li}}, \bibinfo {author} {\bibfnamefont {J.}~\bibnamefont {Yuan}}, \bibinfo {author} {\bibfnamefont {A.}~\bibnamefont {Hemmerich}}, \ and\ \bibinfo {author} {\bibfnamefont {X.}~\bibnamefont {Li}},\ }\href@noop {} {\bibfield  {journal} {\bibinfo  {journal} {Phys. Rev. Lett.}\ }\textbf {\bibinfo {volume} {121}},\ \bibinfo {pages} {093401} (\bibinfo {year} {2018})}\BibitemShut {NoStop}%
\bibitem [{\citenamefont {Niu}\ \emph {et~al.}(2018)\citenamefont {Niu}, \citenamefont {Jin}, \citenamefont {Chen}, \citenamefont {Li},\ and\ \citenamefont {Zhou}}]{PhysRevLett.121.265301}%
  \BibitemOpen
  \bibfield  {author} {\bibinfo {author} {\bibfnamefont {L.}~\bibnamefont {Niu}}, \bibinfo {author} {\bibfnamefont {S.}~\bibnamefont {Jin}}, \bibinfo {author} {\bibfnamefont {X.}~\bibnamefont {Chen}}, \bibinfo {author} {\bibfnamefont {X.}~\bibnamefont {Li}}, \ and\ \bibinfo {author} {\bibfnamefont {X.}~\bibnamefont {Zhou}},\ }\href@noop {} {\bibfield  {journal} {\bibinfo  {journal} {Phys. Rev. Lett.}\ }\textbf {\bibinfo {volume} {121}},\ \bibinfo {pages} {265301} (\bibinfo {year} {2018})}\BibitemShut {NoStop}%
\bibitem [{\citenamefont {Pan}\ \emph {et~al.}(2020)\citenamefont {Pan}, \citenamefont {Liu},\ and\ \citenamefont {Liu}}]{PhysRevLett.125.260402}%
  \BibitemOpen
  \bibfield  {author} {\bibinfo {author} {\bibfnamefont {J.-S.}\ \bibnamefont {Pan}}, \bibinfo {author} {\bibfnamefont {W.~V.}\ \bibnamefont {Liu}}, \ and\ \bibinfo {author} {\bibfnamefont {X.-J.}\ \bibnamefont {Liu}},\ }\href@noop {} {\bibfield  {journal} {\bibinfo  {journal} {Phys. Rev. Lett.}\ }\textbf {\bibinfo {volume} {125}},\ \bibinfo {pages} {260402} (\bibinfo {year} {2020})}\BibitemShut {NoStop}%
\bibitem [{\citenamefont {Jin}\ \emph {et~al.}(2021)\citenamefont {Jin}, \citenamefont {Zhang}, \citenamefont {Guo}, \citenamefont {Chen}, \citenamefont {Zhou},\ and\ \citenamefont {Li}}]{jin2021evidence}%
  \BibitemOpen
  \bibfield  {author} {\bibinfo {author} {\bibfnamefont {S.}~\bibnamefont {Jin}}, \bibinfo {author} {\bibfnamefont {W.}~\bibnamefont {Zhang}}, \bibinfo {author} {\bibfnamefont {X.}~\bibnamefont {Guo}}, \bibinfo {author} {\bibfnamefont {X.}~\bibnamefont {Chen}}, \bibinfo {author} {\bibfnamefont {X.}~\bibnamefont {Zhou}}, \ and\ \bibinfo {author} {\bibfnamefont {X.}~\bibnamefont {Li}},\ }\href@noop {} {\bibfield  {journal} {\bibinfo  {journal} {Phys. Rev. Lett.}\ }\textbf {\bibinfo {volume} {126}},\ \bibinfo {pages} {035301} (\bibinfo {year} {2021})}\BibitemShut {NoStop}%
\bibitem [{\citenamefont {Wang}\ \emph {et~al.}(2021{\natexlab{b}})\citenamefont {Wang}, \citenamefont {Luo}, \citenamefont {Liu}, \citenamefont {Liu}, \citenamefont {Hemmerich},\ and\ \citenamefont {Xu}}]{wang2021evidence}%
  \BibitemOpen
  \bibfield  {author} {\bibinfo {author} {\bibfnamefont {X.-Q.}\ \bibnamefont {Wang}}, \bibinfo {author} {\bibfnamefont {G.-Q.}\ \bibnamefont {Luo}}, \bibinfo {author} {\bibfnamefont {J.-Y.}\ \bibnamefont {Liu}}, \bibinfo {author} {\bibfnamefont {W.~V.}\ \bibnamefont {Liu}}, \bibinfo {author} {\bibfnamefont {A.}~\bibnamefont {Hemmerich}}, \ and\ \bibinfo {author} {\bibfnamefont {Z.-F.}\ \bibnamefont {Xu}},\ }\href@noop {} {\bibfield  {journal} {\bibinfo  {journal} {Nature}\ }\textbf {\bibinfo {volume} {596}},\ \bibinfo {pages} {227} (\bibinfo {year} {2021}{\natexlab{b}})}\BibitemShut {NoStop}%
\bibitem [{\citenamefont {Huang}\ \emph {et~al.}(2022)\citenamefont {Huang}, \citenamefont {Xu},\ and\ \citenamefont {Wu}}]{huang2022intrinsic}%
  \BibitemOpen
  \bibfield  {author} {\bibinfo {author} {\bibfnamefont {G.-H.}\ \bibnamefont {Huang}}, \bibinfo {author} {\bibfnamefont {Z.-F.}\ \bibnamefont {Xu}}, \ and\ \bibinfo {author} {\bibfnamefont {Z.}~\bibnamefont {Wu}},\ }\href@noop {} {\bibfield  {journal} {\bibinfo  {journal} {Phys. Rev. Lett.}\ }\textbf {\bibinfo {volume} {129}},\ \bibinfo {pages} {185301} (\bibinfo {year} {2022})}\BibitemShut {NoStop}%
\bibitem [{\citenamefont {Wang}\ \emph {et~al.}(2023)\citenamefont {Wang}, \citenamefont {Luo}, \citenamefont {Liu}, \citenamefont {Huang}, \citenamefont {Li}, \citenamefont {Wu}, \citenamefont {Hemmerich},\ and\ \citenamefont {Xu}}]{wang2023evidence}%
  \BibitemOpen
  \bibfield  {author} {\bibinfo {author} {\bibfnamefont {X.-Q.}\ \bibnamefont {Wang}}, \bibinfo {author} {\bibfnamefont {G.-Q.}\ \bibnamefont {Luo}}, \bibinfo {author} {\bibfnamefont {J.-Y.}\ \bibnamefont {Liu}}, \bibinfo {author} {\bibfnamefont {G.-H.}\ \bibnamefont {Huang}}, \bibinfo {author} {\bibfnamefont {Z.-X.}\ \bibnamefont {Li}}, \bibinfo {author} {\bibfnamefont {C.}~\bibnamefont {Wu}}, \bibinfo {author} {\bibfnamefont {A.}~\bibnamefont {Hemmerich}}, \ and\ \bibinfo {author} {\bibfnamefont {Z.-F.}\ \bibnamefont {Xu}},\ }\href@noop {} {\bibfield  {journal} {\bibinfo  {journal} {Phys. Rev. Lett.}\ }\textbf {\bibinfo {volume} {131}},\ \bibinfo {pages} {226001} (\bibinfo {year} {2023})}\BibitemShut {NoStop}%
\bibitem [{SM()}]{SM}%
  \BibitemOpen
  \href@noop {} {}\bibinfo {howpublished} {See supplementary material for the details on the expermental realization scheme of the orbital optical Raman lattice, the derivation of lattice model Hamiltonian, the variational calculation of high-orbital many-body ground states, the Bogoliubov theory of quasiparticle excitations, and the discussion of lifetime.}\BibitemShut {Stop}%
\bibitem [{\citenamefont {Sun}\ \emph {et~al.}(2012)\citenamefont {Sun}, \citenamefont {Liu}, \citenamefont {Hemmerich},\ and\ \citenamefont {Das~Sarma}}]{SunNP2012}%
  \BibitemOpen
  \bibfield  {author} {\bibinfo {author} {\bibfnamefont {K.}~\bibnamefont {Sun}}, \bibinfo {author} {\bibfnamefont {W.~V.}\ \bibnamefont {Liu}}, \bibinfo {author} {\bibfnamefont {A.}~\bibnamefont {Hemmerich}}, \ and\ \bibinfo {author} {\bibfnamefont {S.}~\bibnamefont {Das~Sarma}},\ }\href@noop {} {\bibfield  {journal} {\bibinfo  {journal} {Nat. Phys.}\ }\textbf {\bibinfo {volume} {8}},\ \bibinfo {pages} {67} (\bibinfo {year} {2012})}\BibitemShut {NoStop}%
\bibitem [{\citenamefont {Shindou}\ \emph {et~al.}(2013)\citenamefont {Shindou}, \citenamefont {Matsumoto}, \citenamefont {Murakami},\ and\ \citenamefont {Ohe}}]{shindou2013topological}%
  \BibitemOpen
  \bibfield  {author} {\bibinfo {author} {\bibfnamefont {R.}~\bibnamefont {Shindou}}, \bibinfo {author} {\bibfnamefont {R.}~\bibnamefont {Matsumoto}}, \bibinfo {author} {\bibfnamefont {S.}~\bibnamefont {Murakami}}, \ and\ \bibinfo {author} {\bibfnamefont {J.-i.}\ \bibnamefont {Ohe}},\ }\href@noop {} {\bibfield  {journal} {\bibinfo  {journal} {Phys. Rev. B}\ }\textbf {\bibinfo {volume} {87}},\ \bibinfo {pages} {174427} (\bibinfo {year} {2013})}\BibitemShut {NoStop}%
\end{thebibliography}%


\begin{thebibliography}{10}

\bibitem{vanLaarhoven1987}
Peter J.~M. van Laarhoven and Emile H.~L. Aarts.
\newblock {\em Simulated annealing}, pages 7--15.
\newblock Springer Netherlands, Dordrecht, 1987.

\bibitem{zhai2021ultracold}
Hui Zhai.
\newblock {\em Ultracold atomic physics}, chapter~3, pages 88--89.
\newblock Cambridge University Press, 2021.

\bibitem{kock2016orbital}
T~Kock, C~Hippler, A~Ewerbeck, and A~Hemmerich.
\newblock Orbital optical lattices with bosons.
\newblock {\em Journal of Physics B: Atomic, Molecular and Optical Physics}, 49(4):042001, 2016.

\bibitem{wirth2011evidence}
Georg Wirth, Matthias {\"O}lschl{\"a}ger, and Andreas Hemmerich.
\newblock Evidence for orbital superfluidity in the p-band of a bipartite optical square lattice.
\newblock {\em Nat. Phys.}, 7(2):147--153, 2011.

\bibitem{PhysRevLett.121.265301}
Linxiao Niu, Shengjie Jin, Xuzong Chen, Xiaopeng Li, and Xiaoji Zhou.
\newblock Observation of a dynamical sliding phase superfluid with $p$-band bosons.
\newblock {\em Phys. Rev. Lett.}, 121:265301, Dec 2018.

\bibitem{jin2021evidence}
Shengjie Jin, Wenjun Zhang, Xinxin Guo, Xuzong Chen, Xiaoji Zhou, and Xiaopeng Li.
\newblock Evidence of potts-nematic superfluidity in a hexagonal $sp^2$ optical lattice.
\newblock {\em Phys. Rev. Lett.}, 126(3):035301, 2021.

\bibitem{wang2021evidence}
Xiao-Qiong Wang, Guang-Quan Luo, Jin-Yu Liu, W~Vincent Liu, Andreas Hemmerich, and Zhi-Fang Xu.
\newblock Evidence for an atomic chiral superfluid with topological excitations.
\newblock {\em Nature}, 596(7871):227--231, 2021.

\bibitem{Wu2016Science}
Zhan Wu, Long Zhang, Wei Sun, Xiao-Tian Xu, Bao-Zong Wang, Si-Cong Ji, Youjin Deng, Shuai Chen, Xiong-Jun Liu, and Jian-Wei Pan.
\newblock Realization of two-dimensional spin-orbit coupling for bose-einstein condensates.
\newblock {\em Science}, 354(6308):83--88, 2016.

\bibitem{sunPRL2018}
Wei Sun, Bao-Zong Wang, Xiao-Tian Xu, Chang-Rui Yi, Long Zhang, Zhan Wu, Youjin Deng, Xiong-Jun Liu, Shuai Chen, and Jian-Wei Pan.
\newblock Highly controllable and robust 2d spin-orbit coupling for quantum gases.
\newblock {\em Phys. Rev. Lett.}, 121:150401, 2018.

\bibitem{sunPRL2018_1}
Wei Sun, Chang-Rui Yi, Bao-Zong Wang, Wei-Wei Zhang, Barry~C. Sanders, Xiao-Tian Xu, Zong-Yao Wang, Joerg Schmiedmayer, Youjin Deng, Xiong-Jun Liu, Shuai Chen, and Jian-Wei Pan.
\newblock Uncover topology by quantum quench dynamics.
\newblock {\em Phys. Rev. Lett.}, 121:250403, 2018.

\bibitem{wang2021realization}
Zong-Yao Wang, Xiang-Can Cheng, Bao-Zong Wang, Jin-Yi Zhang, Yue-Hui Lu, Chang-Rui Yi, Sen Niu, Youjin Deng, Xiong-Jun Liu, Shuai Chen, et~al.
\newblock Realization of an ideal weyl semimetal band in a quantum gas with 3d spin-orbit coupling.
\newblock {\em Science}, 372(6539):271--276, 2021.

\end{thebibliography}
%\printbibliography

\end{document}

\maketitle
    In this supplementary material, we first demonstrate the experimental realization of the continuous Hamiltonian described in the main text using the orbital optical Raman lattice and derive the corresponding lattice-model Hamiltonian using appropriate parameters. Then, we discuss the variational minimization of high-orbital many-body ground state energies by applying the simulated annealing algorithm. Next, we employ the Bogoliubov theory to obtain the quasiparticle excitation spectra for the uniform angular momentum superfluid phase and the two-dimensional spin-orbital supersolid phase. Finally, we qualitatively discuss the lifetimes of the exotic many-body quantum phases.
\tableofcontents

\vspace{.25in}
\section{Orbital optical Raman lattice scheme}
The optical Raman lattice plays a significant role in investigating the use of cold atoms to simulate exotic topological quantum matters. In this section, we will introduce the orbital optical Raman lattice scheme for realizing the continuous Hamiltonian shown in the main text. 

\begin{figure}[t]
\centering
\includegraphics[width=17.5cm,height=10.5cm]{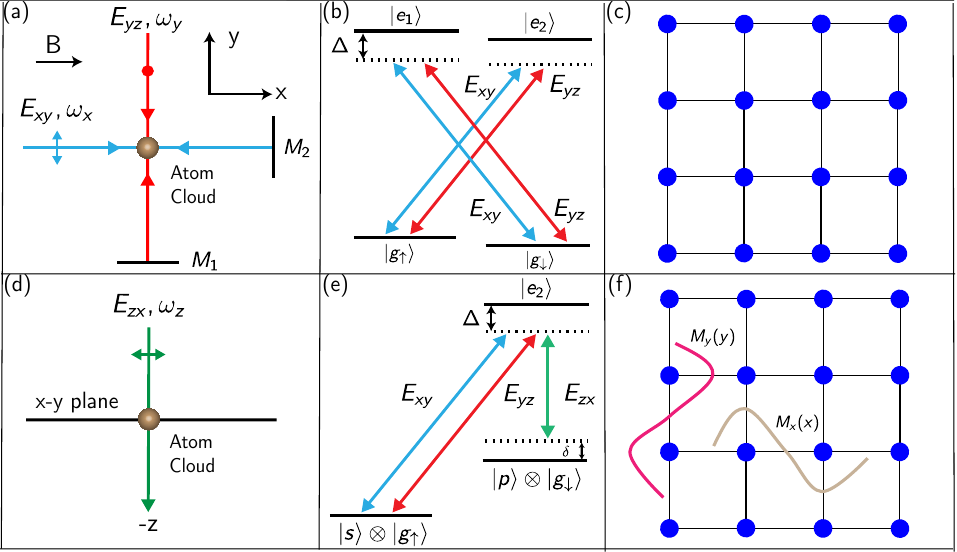}
\caption{(a) Applying two laser beams $\boldsymbol{E}_{xy}=\hat{y}E_{0}\cos(k_{0}x)$ and $\boldsymbol{E}_{yz}=\hat{z}E_{0}\cos(k_{0}y)$ can realize a square lattice. Here the bias magnetic field is set along the $x$ direction. (b) Both $\boldsymbol{E}_{xy}$ and $\boldsymbol{E}_{yz}$ can induce the transition from the ground states with $m_{F}$ to excited states with $m_{F}+1$ and $m_{F}-1$, where the hyperfine states $|1,-1\rangle$ and $|1,0\rangle$ of $^{87}\mathrm{Rb}$ are used to mimic the ground states $|g_{\uparrow,\downarrow}\rangle$. (c) Square optical lattice $V(\boldsymbol{r})$. (d) Applying another plane wave laser $\boldsymbol{E}_{zx}$ with polarization along $x$ direction based on the setup in (a). (e) There is a single $\boldsymbol{\varLambda}$ configuration where the Raman potentials $M_{x}(x)$ and $M_{y}(y)$ can be generated by the laser pairs $(\boldsymbol{E}_{xy}$, $\boldsymbol{E}_{zx})$ and $(\boldsymbol{E}_{yz}$, $\boldsymbol{E}_{zx})$, respectively. (f) Raman field $M(\boldsymbol{r})$. Notice that the Raman field has twice the periodicity of the lattice potential. }
\label{fig1}
\end{figure}

\subsection{Optical lattice}
We first generate the square lattice on $x-y$ plane using the standing waves $\boldsymbol{E}_{xy}=\hat{y}E_{0}\cos(k_{0}x)$ and $\boldsymbol{E}_{yz}=\hat{z}E_{0}\cos(k_{0}y)$ [Fig.\ref{fig1} (a)], where $k_{0}=\frac{\pi}{a}$. Here, the standing waves $\boldsymbol{E}_{xy}$ and $\boldsymbol{E}_{yz}$ are linearly polarized along $y$ and $z$ directions, respectively, eliminating the interference term between them. Note that the bias magnetic field is set along the $x$ direction. As shown in Fig.\ref{fig1} (b), both $\boldsymbol{E}_{xy}$ (blue line) and $\boldsymbol{E}_{yz}$ (red line) can induce the transition from the ground state with $m_{F}$ to the excited states with $m_{F}+1$ and $m_{F}-1$. Here, we select two hyperfine states $|1,-1\rangle$ and $|1,0\rangle$ of $^{87}\mathrm{Rb}$ to mimic the ground states $|g_{\uparrow,\downarrow}\rangle$ with $|1,-1\rangle=|g_{\uparrow}\rangle$, $|1,0\rangle=|g_{\downarrow}\rangle$. When the detuning is much larger than the Rabi frequency of the standing waves ($|\Delta|\gg|\Omega_{(\boldsymbol{E}_{xy},\boldsymbol{E}_{yz})}|$), the atom-light couplings contribute to diagonal potentials for the ground states $|g_{\uparrow,\downarrow}\rangle$, given by $V_{\uparrow}=\hbar(|\Omega_{\boldsymbol{E}_{xy}}|^{2}/\Delta+|\Omega_{\boldsymbol{E}_{yz}}|^{2}/\Delta)$ and $V_{\downarrow}=\hbar(|\Omega_{\boldsymbol{E}_{xy}}|^{2}/\Delta+|\Omega_{\boldsymbol{E}_{yz}}|^{2}/\Delta)$, where $\Delta$ denotes the red detuning for $\boldsymbol{E}_{xy}$ and $\boldsymbol{E}_{yz}$. Thus, the optical lattice potential reads 
\begin{equation}
V(\mathrm{\boldsymbol{r}})=-V_{0}[\cos^{2}(k_{0}x)+\cos^{2}(k_{0}y)],\label{39}
\end{equation}
where $V_{0}=\frac{\hbar E_{0}^{2}}{|\Delta|}$ ($\Delta<0$). The configuration of optical lattice in Eq.(\ref{39}) is shown in Fig.\ref{fig1} (c).

\subsection{Raman field}
As shown in Fig.\ref{fig1} (d), the Raman potentials can be generated by adding another plane wave laser $\boldsymbol{E}_{zx}=\hat{x}E_{z}e^{-\mathrm{i}k_{z}z}$, which propagates along the $z$ axis and is polarized along the $x$ axis. The applied plane wave laser $\boldsymbol{E}_{zx}$ can induce the $\pi$ transitions from ground states $|g_{\uparrow}\rangle$ and $|g_{\downarrow}\rangle$ to the excited states $|e_{1}\rangle$ and $|e_{2}\rangle$, respectively. As shown in Fig.\ref{fig1} (e), we tune the frequency of the plane wave $\boldsymbol{E}_{zx}$ so that the atom-light couplings can occur between $|s\rangle\otimes|g_{\uparrow}\rangle$, $|p\rangle\otimes|g_{\downarrow}\rangle$, and the excited state $|e_{2}\rangle$, forming a single \textbf{$\boldsymbol{\varLambda}$}-type configuration [Fig.\ref{fig1} (e)]. Here, $|s\rangle$ and $|p\rangle$ denote the local orbital states. Through this configuration, the Raman potentials $M_{x}(x)$ and $M_{y}(y)$ can be generated by the laser pairs ($\boldsymbol{E}_{xy}$, $\boldsymbol{E}_{zx}$) and ($\boldsymbol{E}_{yz}$, $\boldsymbol{E}_{zx}$), respectively.  When $|\Delta|\gg|\Omega_{\boldsymbol{E}_{xy}}|,|\Omega_{\boldsymbol{E}_{zx}}| 
(|\Omega_{\boldsymbol{E}_{yz}}|,|\Omega_{\boldsymbol{E}_{zx}}|)$ and the two-photon off resonance $\delta$ is smaller than the coupling strength $|\Omega_{\boldsymbol{E}_{xy}}\Omega_{\boldsymbol{E}_{zx}}/\Delta|$ ($|\Omega_{\boldsymbol{E}_{yz}}\Omega_{\boldsymbol{E}_{zx}}/\Delta|$), the two-photon processes in the \textbf{$\boldsymbol{\varLambda}$}-type configuration are dominant and induce the Raman couplings between $|s\rangle\otimes|g_{\uparrow}\rangle$  and $|p\rangle\otimes|g_{\downarrow}\rangle$, where $\Omega$ denotes the Rabi frequency of the lasers. Then the Raman field is given by [Fig.\ref{fig1} (f)] 
\begin{equation}
M(\mathrm{\boldsymbol{r}})=-M_{0}[\cos(k_{0}x)+\cos(k_{0}y)](|s\rangle\otimes|g_{\uparrow}\rangle\langle g_{\downarrow}|\otimes\langle p|+\mathrm{H.c.}),
\end{equation}
where $M_{0}=\frac{\hbar E_{0}E_{z}}{|\Delta|}$. The two-photon detuning also contributes to the Zeeman splitting $m_{z}=\hbar\delta/2$.

\section{Model Hamiltonian}
In this section, we derive the model Hamiltonian $\hat{H}=\hat{H}_{0}+\hat{H}_{\mathrm{int}}$ for the interacting ultracold Bose gases trapped in a square orbital optical Raman lattice.
 Here, $\hat{H}_{0}$ and $\hat{H}_{\mathrm{int}}$ represent the non-interacting and interacting parts, respectively. We also provide the appropriate model parameters to facilitate the experimental realization.

\subsection{Non-interacting part}
We first start from the non-interacting continuous Hamiltonian
\begin{equation} 
H_{0}=\frac{p_{\mathrm{\boldsymbol{r}}}^{2}}{2M}+V(\mathrm{\boldsymbol{r}})+M(\mathrm{\boldsymbol{r}})\sigma_{x}+m_{z}\sigma_{z},\label{1}
\end{equation}
where Pauli matrices $\boldsymbol{\sigma}$ act on the two-dimensional space spanned by $|s\rangle\otimes|g_{\uparrow}\rangle$  and $|p\rangle\otimes|g_{\downarrow}\rangle$. Moreover, $V(\mathrm{\boldsymbol{r}})$ and $M(\mathrm{\boldsymbol{r}})$ represent respectively the optical lattice potential and Raman field with 
\begin{align} 
V(\mathrm{\boldsymbol{r}})=-V_{0}[\cos^{2}(\frac{\pi}{a}x)+\cos^{2}(\frac{\pi}{a}y)], \quad & M(\mathrm{\boldsymbol{r}})=-M_{0}[\cos(\frac{\pi}{a}x)+\cos(\frac{\pi}{a}y)].\label{2}
\end{align} 
Here $V_{0}$, $M_{0}$, and $a$ denote the lattice depth, Raman strength, and lattice constant, respectively. For convenience, we take $a=1$. $m_z$ is the effective Zeeman field. According to the second quantization, Eq. (\ref{1}) can be rewritten as 
\begin{equation} 
\hat{H}_{0}=\int\mathrm{d}\mathrm{\boldsymbol{r}}\Psi^{\dagger}(\boldsymbol{\mathrm{r}})[\frac{p_{\mathrm{\boldsymbol{r}}}^{2}}{2M}+V(\mathrm{\boldsymbol{r}})+M(\mathrm{\boldsymbol{r}})\sigma_{x}+m_{z}\sigma_{z}]\Psi(\boldsymbol{\mathrm{r}}).
\end{equation} 
Here $\Psi^{\dagger}(\boldsymbol{\mathrm{r}})=\sum_{il\sigma}b_{i,l,\sigma}^{\dagger}\phi_{l\sigma}^{\ast}(\mathrm{\boldsymbol{r}}-\mathrm{\boldsymbol{r}_{i}})$ ($l=s,p_{x},p_{y};\sigma=\uparrow,\downarrow$) is the field operator and $b_{i,l,\sigma}^{\dagger}$ represents creating a Boson with orbital $l$ and spin $\sigma$ at lattice site $\mathrm{\boldsymbol{r}_{i}}$. Moreover, $\phi_{l,\sigma}(\boldsymbol{r}-\boldsymbol{r}_{\boldsymbol{i}})=\psi_{l}(\boldsymbol{r}-\boldsymbol{r}_{\boldsymbol{i}})\otimes\chi_{\sigma}$ denotes is the Wannier function at $\mathrm{\boldsymbol{r}_{i}}$, where $\psi_{l}(\boldsymbol{r}-\boldsymbol{r}_{\boldsymbol{i}})$ and $\chi_{\sigma}$ denote the corresponding spatial and spin components, respectively. 

As discussed in the previous section, Raman coupling occurs between the $|s\rangle\otimes|g_{\uparrow}\rangle$ and $|p\rangle\otimes|g_{\downarrow}\rangle$, which can effectively bring the $s_{\uparrow}$ and $p_{x\downarrow}$ ($p_{y\downarrow}$) orbitals closer together under the rotating coordinate system, separating them from other orbitals. As a consequence, within the subspace spanned by ($s_{\uparrow}$, $p_{x\downarrow}$, $p_{y\downarrow}$), the field operator can be approximated as $\Psi^{\dagger}(\boldsymbol{\mathrm{r}})\approx\sum_{i} \Big[ b_{i,s,\uparrow}^{\dagger}\phi_{s\uparrow}^{\ast}(\mathrm{\boldsymbol{r}}-\mathrm{\boldsymbol{r}_{i}}) + 
b_{i,p_x,\downarrow}^{\dagger}\phi_{p_x\downarrow}^{\ast}(\mathrm{\boldsymbol{r}}-\mathrm{\boldsymbol{r}_{i}}) + b_{i,p_y,\downarrow}^{\dagger}\phi_{p_y\downarrow}^{\ast}(\mathrm{\boldsymbol{r}}-\mathrm{\boldsymbol{r}_{i}}) \Big]$. Now we can obtain the tight-binding Hamiltonian $\hat{H}_{0}=\hat{H}_{\mathrm{sc}}+\hat{H}_{\mathrm{sf}}$ straightforwardly, where $\hat{H}_{\mathrm{sc}}$ and $\hat{H}_{\mathrm{sf}}$ represent the spin-conserved and spin-flipped parts induced by the optical lattice potential $V(\boldsymbol{r})$ and the Raman field $M(\boldsymbol{r})$, respectively. Specifically, 
\begin{align} 
\hat{H}_{\mathrm{sc}}= & -t_{s}\sum_{\langle\mathrm{\boldsymbol{i}},\mathrm{\boldsymbol{j}}\rangle}b_{\mathrm{\boldsymbol{i}},s,\uparrow}^{\dagger}b_{\mathrm{\boldsymbol{j}},s,\uparrow}
+  \sum_{\mathrm{\boldsymbol{i}\mu\nu}}t_p^{\mu\nu}\Big( b_{\mathrm{\boldsymbol{i}},p_{\mu},\downarrow}^{\dagger}b_{\mathrm{\boldsymbol{i}}+\mathrm{\boldsymbol{e}_{\nu}},p_{\mu},\downarrow} +\mathrm{H}.\mathrm{c.}\Big) \nonumber \\
& + (\mu_{s}+m_{z})\sum_{\mathrm{\boldsymbol{i}}}n_{\mathrm{\boldsymbol{i},s,\uparrow}}
+ (\mu_{p}-m_{z})\sum_{\mathrm{\boldsymbol{i}\nu}}n_{\mathrm{\boldsymbol{i}},p_{\nu},\downarrow},\\
\hat{H}_{\mathrm{sf}}= & \sum_{\mathrm{\boldsymbol{i},\nu, \delta_{\nu}}} - t_{so}^{(\boldsymbol{i},\boldsymbol{i}+\delta_{\nu})} 
b_{\mathrm{\boldsymbol{i}},p_{\nu},\downarrow}^{\dagger}b_{\mathrm{\boldsymbol{i}}+\delta_{\nu},s,\uparrow} +\mathrm{H}.\mathrm{c.},\label{5}
\end{align}
where $n_{\mathrm{\boldsymbol{i},s,\uparrow}}=b_{\mathrm{\boldsymbol{i}},s,\uparrow}^{\dagger}b_{\mathrm{\boldsymbol{i}},s,\uparrow}$ , $n_{\mathrm{\boldsymbol{i}},p_{\nu},\downarrow}=b_{\mathrm{\boldsymbol{i}},p_{\nu},\downarrow}^{\dagger}b_{\mathrm{\boldsymbol{i}},p_{\nu},\downarrow}$, and $t_p^{\mu\nu}=t_p^{\parallel}\cdot \delta_{\mu\nu} - t_p^{\perp}\cdot ( 1 - \delta_{\mu\nu} )$
($\mu$, $\nu$ $=x$, $y$). “sgn” is a sign function and $\delta_{\nu}=\pm e_{\nu}$.
$\mu_{s}$ and $\mu_{p}$ represent the effective $s$ and $p$ orbitals on-site energies, respectively. Furthermore, the corresponding spin-conserved hopping coefficients are given by 
\begin{align} 
t_{s}&=-\int\mathrm{d}\mathrm{\boldsymbol{r}}\phi_{s,\uparrow}^{\ast}(\mathrm{\boldsymbol{r}}-\mathrm{\boldsymbol{r}_{i}})\Big[\frac{p_{\mathrm{\boldsymbol{r}}}^{2}}{2M}+V(\mathrm{\boldsymbol{r}})\Big]\phi_{s,\uparrow}(\mathrm{\boldsymbol{r}}-\mathrm{\boldsymbol{r}_{i+e_{x(y)}}}),\nonumber \\
t_{p}^{\parallel} & =\int\mathrm{d}\mathrm{\boldsymbol{r}}\phi_{p_{x(y)},\downarrow}^{\ast}(\mathrm{\boldsymbol{r}}-\mathrm{\boldsymbol{r}_{i}})\Big[\frac{p_{\mathrm{\boldsymbol{r}}}^{2}}{2M}+V(\mathrm{\boldsymbol{r}})\Big]\phi_{p_{x(y)},\downarrow}(\mathrm{\boldsymbol{r}}-\mathrm{\boldsymbol{r}_{i+e_{x(y)}}}),\nonumber \\
t_{p}^{\perp} & =-\int\mathrm{d}\mathrm{\boldsymbol{r}}\phi_{p_{x(y)},\downarrow}^{\ast}(\mathrm{\boldsymbol{r}}-\mathrm{\boldsymbol{r}_{i}})\Big[\frac{p_{\mathrm{\boldsymbol{r}}}^{2}}{2M}+V(\mathrm{\boldsymbol{r}})\Big]\phi_{p_{x(y)},\downarrow}(\mathrm{\boldsymbol{r}}-\mathrm{\boldsymbol{r}_{i+e_{y(x)}}}),
\end{align}
As given in Eq.(\ref{2}), the Raman field $M(\boldsymbol{r})$ has twice the period of the optical lattice potential $V(\boldsymbol{r})$. Therefore, the spin-flipped hopping coefficient in Eq.(\ref{5}) is staggered in $\nu$ ($\nu=x,y$) direction: $t_{so}^{(\boldsymbol{i},\boldsymbol{i}\pm\boldsymbol{e}_{v})}=\pm(-1)^{i_{v}}t_{\mathrm{SO}}$, where 
\begin{align}
 t_{\mathrm{SO}}=M_{0}\int\mathrm{d}\mathrm{\boldsymbol{r}}\psi_{s}^{\ast}(\mathrm{\boldsymbol{r}}-\mathrm{\boldsymbol{r}_{0}})\cos(\pi\nu)\psi_{p_{\nu}}(\mathrm{\boldsymbol{r}}-\mathrm{\boldsymbol{r}_{\mathrm{e_{\nu}}}}).
\end{align}
Note that this staggered spin-flipped hopping has two important consequences. The staggered property implies that the coupling between $s_{\uparrow}$ and $p_{x(y)\downarrow}$ states transfers $\pi/a$ momentum, which effective shifts the Brillouin zone by half along $k_{x}$ ($k_{y}$) direction for the $p_{x(y)\downarrow}$
states (relative to the $s_{\uparrow}$ states). Here we can redefine the spin-down operator by a unitary transformation $b_{\mathrm{\boldsymbol{i}},p_{\nu},\downarrow}\rightarrow -(-1)^{\mathrm{i_{\nu}}}b_{\mathrm{\boldsymbol{i}},p_{\nu},\downarrow}$ to absorb this effect. Thus the tight-binding Hamiltonian $\hat{H}_{0}$ can be rewritten as 
\begin{align}
\hat{H}_{\mathrm{0}}= & -t_{s}\sum_{\langle\mathrm{\boldsymbol{i}},\mathrm{\boldsymbol{j}}\rangle}b_{\mathrm{\boldsymbol{i}},s,\uparrow}^{\dagger}b_{\mathrm{\boldsymbol{j}},s,\uparrow} 
-  \sum_{\mathrm{\boldsymbol{i}}}\sum_{\nu,\widetilde{\nu}\neq \nu}\Big(t_{p}^{\parallel}b_{\mathrm{\boldsymbol{i}},p_{\nu},\downarrow}^{\dagger}b_{\mathrm{\boldsymbol{i}}+\mathrm{\boldsymbol{e}_{\nu}},p_{\nu},\downarrow} + t_{p}^{\perp}b_{\mathrm{\boldsymbol{i}},p_{\nu},\downarrow}^{\dagger}b_{\mathrm{\boldsymbol{i}}+\mathrm{\boldsymbol{e}_{\widetilde{\nu}}},p_{\nu},\downarrow}+\mathrm{H}.\mathrm{c.} \Big)\nonumber \\
& + t_{\mathrm{SO}}\sum_{\mathrm{\boldsymbol{i},\nu, \delta_{\nu}}} \Big[ \text{sgn}(\delta_{\nu})
b_{\mathrm{\boldsymbol{i}},p_{\nu},\downarrow}^{\dagger}b_{\mathrm{\boldsymbol{i}}+\delta_{\nu},s,\uparrow} +\mathrm{H}.\mathrm{c.}\Big]
 + (\mu_{s}+m_{z})\sum_{\mathrm{\boldsymbol{i}}}n_{\mathrm{\boldsymbol{i},s,\uparrow}}
+ (\mu_{p}-m_{z})\sum_{\mathrm{\boldsymbol{i}\nu}}n_{\mathrm{\boldsymbol{i}},p_{\nu},\downarrow}.\label{8}
\end{align}

\begin{figure}[htb]
\centering
\includegraphics[scale=0.8]{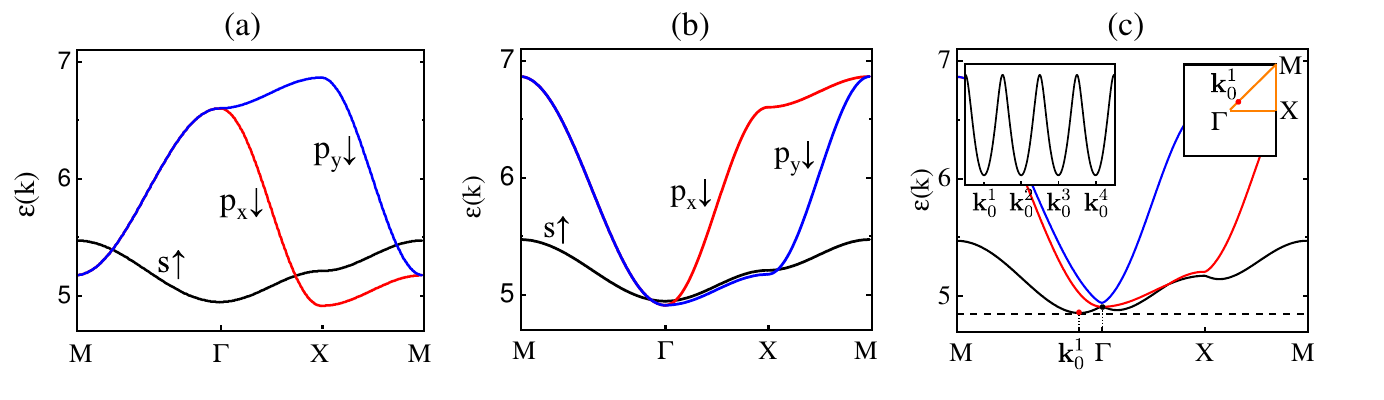}
\caption{ Single-particle spectrum. Here, $V_0=5.0E_r$, $M_0=1.0E_r$ (tight-binding parameters: $t_{s}=0.0658E_{r}$, $t_{p}^{\parallel}=0.4228E_{r}$, $t_{p}^{\perp}=0.0658E_{r}$, $t_{so}=0.1059E_{r}$, $\mu_{s}=5.1909E_{r}$, and $\mu_{p}=5.9049E_{r}$), $m_{z}=0.0173E_{r}$. $E_{r}$ is the recoil energy. (a) and (b) demonstrate that the staggered properties of Raman-induced spin-flipped hoppings can result in a $\pi/a$ momentum transfer for the $p_x$ band and $p_y$ band along the $k_x$ and $k_y$ directions, respectively. (c) shows the energy spectrum of the non-interacting Hamiltonian $\hat{\mathcal{H}}_{0}$, which features the four lowest energy points related by $C_{4}$ symmetry: $\boldsymbol{k}_{0}^{1}$, $\boldsymbol{k}_{0}^{2}$, $\boldsymbol{k}_{0}^{3}$, $\boldsymbol{k}_{0}^{4}$. Note that $\boldsymbol{k}_{0}^{1}=(0.154, 0.154)\pi$.
 }
\label{fig2}
\end{figure}

Performing a Fourier transformation, we can obtain the tight-binding Hamiltonian in momentum space
\begin{equation}
\hat{H}_{0}=\sum_{\boldsymbol{k}}\mathcal{B}_{\boldsymbol{k}}^{\dagger}\mathcal{H}_{0}\mathcal{B}_{\boldsymbol{k}}=\sum_{\boldsymbol{k}}
\mathcal{B}_{\boldsymbol{k}}^{\dagger}\left(\begin{array}{ccc}
\epsilon_{s_{\uparrow}}(\boldsymbol{k}) & \mathrm{-i}2t_{\mathrm{SO}}\sin k_{x} & \mathrm{-i}2t_{\mathrm{SO}}\sin k_{y}\\
\mathrm{i}2t_{\mathrm{SO}}\sin k_{x} & \epsilon_{p_{x\downarrow}}(\boldsymbol{k})\\
\mathrm{i}2t_{\mathrm{SO}}\sin k_{y} &  & \epsilon_{p_{y\downarrow}}(\boldsymbol{k})
\end{array}\right)\mathcal{B}_{\boldsymbol{k}},\label{9}
\end{equation}
where $\mathcal{B}_{\mathrm{\boldsymbol{k}}}^{\dagger}=\left(\begin{array}{ccc}
b_{\mathrm{\boldsymbol{k}},s_{\uparrow}}^{\dagger} & b_{\mathrm{\boldsymbol{k}},p_{x\downarrow}}^{\dagger} & b_{\mathrm{\boldsymbol{k}},p_{y\downarrow}}^{\dagger}\end{array}\right)$. The three diagonal elements $\epsilon_{s\uparrow}(\boldsymbol{k})=\mu_{s}+m_{z}-2t_{s}(\cos k_{x}+\cos k_{y})$, $\epsilon_{p_{x}\downarrow}(\boldsymbol{k})=\mu_{p}-m_{z}-2t_{p}^{\parallel}\cos k_{x}-2t_{p}^{\perp}\cos k_{y}$, and $\epsilon_{p_{y}\downarrow}(\boldsymbol{k})=\mu_{p}-m_{z}-2t_{p}^{\perp}\cos k_{x}-2t_{p}^{\parallel}\cos k_{y}$ represent the energy bands for $s_{\uparrow}$, $p_{x\downarrow}$, and $p_{y\downarrow}$ states, respectively. It's worth noting that the $\pi/a$-momentum transfer between $s_{\uparrow}$ and $p_{x(y)\downarrow}$ states effectively reverses the hopping coefficient $t_{p}^{\parallel}\rightarrow-t_{p}^{\parallel}$ for the $p_{x(y)\downarrow}$ states, shifting the lowest energy points of $\epsilon_{p_{x}\downarrow}$ ($\epsilon_{p_{y}\downarrow}$) from $\boldsymbol{Q}_{x}=(\pi, 0)$ ($\boldsymbol{Q}_{y}=(0,\pi)$) to $\boldsymbol{\Gamma}=(0,0)$ [Fig.\ref{fig2} (a) and (b)]. Additionally, the energy spacing $\epsilon_{s_{\uparrow}}(\boldsymbol{\Gamma})-\epsilon_{p_{x(y)\downarrow}}(\boldsymbol{\Gamma})$ can be tuned by the effective Zeeman splitting $m_z$. The remaining effect of spin-flipped hopping contributes to the SOC in the $x-y$ plane, resulting in the four lowest energy points in the single-particle spectrum: $\boldsymbol{k}_{0}^{1}=(k_{c},k_{c})$, $\boldsymbol{k}_{0}^{2}=(-k_{c},k_{c})$, $\boldsymbol{k}_{0}^{3}=(-k_{c},-k_{c})$ and $\boldsymbol{k}_{0}^{4}=(k_{c},-k_{c})$, which are connected by $C_{4}$ symmetry [Fig.\ref{fig2} (c)]. 

\subsection{Interacting part}
We next consider a contact potential $V(\boldsymbol{r}-\boldsymbol{r}^{\prime})=g_{\sigma\sigma^{\prime}}\delta(\boldsymbol{r}-\boldsymbol{r}^{\prime})$ ($\sigma,\sigma^{\prime}=\uparrow,\downarrow$), with the corresponding interaction Hamiltonian given by 
\begin{equation} 
\hat{H}_{\mathrm{int}}=\frac{1}{2}\sum_{\sigma,\sigma^{\prime}}\int\mathrm{d}\boldsymbol{r}\mathrm{d}\boldsymbol{r}^{\prime}\Psi_{\sigma}^{\dagger}(\boldsymbol{r})\Psi_{\sigma^{\prime}}^{\dagger}(\boldsymbol{r}^{\prime})V(\boldsymbol{r}-\boldsymbol{r}^{\prime})\Psi_{\sigma^{\prime}}(\boldsymbol{r}^{\prime})\Psi_{\sigma}(\boldsymbol{r}),\label{10}
\end{equation} 
where $\Psi_{\sigma}^{\dagger}(\boldsymbol{r})=\sum_{\boldsymbol{i},l}\phi_{l,\sigma}^{\ast}(\boldsymbol{r}-\boldsymbol{r}_{\boldsymbol{i}})b_{\boldsymbol{i},l,\sigma}^{\dagger}$ ($l=s,p_{x},p_{y};\sigma=\uparrow,\downarrow$). Here, only the single center integral terms in the above equation are retained
\begin{equation}  
\hat{H}_{\mathrm{int}}=\sum_{\sigma,\sigma^{\prime}}\sum_{l_{1},l_{2},l_{3},l_{4}}\sum_{\boldsymbol{i}}\frac{U_{l_{1}l_{2}l_{3}l_{4}}^{\sigma\sigma^{\prime}}}{2}b_{\boldsymbol{i},l_{1},\sigma}^{\dagger}b_{\boldsymbol{i},l_{2},\sigma^{\prime}}^{\dagger}b_{\boldsymbol{i},l_{3},\sigma^{\prime}}b_{\boldsymbol{i},l_{4},\sigma},\label{11}
\end{equation}
with $U_{l_{1}l_{2}l_{3}l_{4}}^{\sigma\sigma^{\prime}}=g_{\sigma\sigma^{\prime}}\int\mathrm{d}\boldsymbol{r}\phi_{l_{1},\sigma}^{\ast}(\boldsymbol{r}-\boldsymbol{r}_{\boldsymbol{i}})\phi_{l_{2},\sigma^{\prime}}^{\ast}(\boldsymbol{r}-\boldsymbol{r}_{\boldsymbol{i}})\phi_{l_{3},\sigma^{\prime}}(\boldsymbol{r}-\boldsymbol{r}_{\boldsymbol{i}})\phi_{l_{4},\sigma}(\boldsymbol{r}-\boldsymbol{r}_{\boldsymbol{i}})$.  Since we only focus on the orbital physics of the ($s_{\uparrow}$, $p_{x\downarrow}$, $p_{y\downarrow}$) subspace, Eq.(\ref{11}) can be reduced to 
\begin{align} 
\hat{H}_{\mathrm{int}} = & \frac{U_s}2\sum_in_{is\uparrow}(n_{is\uparrow}-1)+\frac{U_p}2\sum_{i\nu}n_{ip_{\nu\downarrow}}(n_{ip_{\nu\downarrow}}-1) 
+U_{sp}\sum_in_{is\uparrow}n_{ip\downarrow}+2\widetilde{U}_{p}\sum_in_{ip_x\downarrow}n_{ip_y\downarrow} \nonumber\\
&+\frac{\widetilde{U}_{p}}2\sum_i\left(b_{ip_{x\downarrow}}^\dagger b_{ip_x\downarrow}^\dagger b_{ip_y\downarrow}b_{ip_y\downarrow}+\text{ H.c }\right).\label{12}
\end{align}
Here,
\begin{align} 
&U_{s}=g\int\mathrm{d}\boldsymbol{r}\phi_{s,\uparrow}^{4}(\boldsymbol{r}-\boldsymbol{r}_{\boldsymbol{i}}),  \nonumber\\
&U_{sp}=g\int\mathrm{d}\boldsymbol{r}\phi_{p_{\nu},\downarrow}^{2}(\boldsymbol{r}-\boldsymbol{r}_{\boldsymbol{i}})\phi_{s,\uparrow}^{2}(\boldsymbol{r}-\boldsymbol{r}_{\boldsymbol{i}}), \nonumber\\
&U_{p}=g\int\mathrm{d}\boldsymbol{r}\phi_{p_{\nu},\downarrow}^{4}(\boldsymbol{r}-\boldsymbol{r}_{\boldsymbol{i}}),  \nonumber\\ &\widetilde{U}_{p}=g\int\mathrm{d}\boldsymbol{r}\phi_{p_{\nu},\downarrow}^{2}(\boldsymbol{r}-\boldsymbol{r}_{\boldsymbol{i}})\phi_{p_{\widetilde{\nu}},\downarrow}^{2}(\boldsymbol{r}-\boldsymbol{r}_{\boldsymbol{i}}), \hspace{1em} \widetilde{\nu} \neq \nu . 
\end{align}
Note that we take $g_{\uparrow\uparrow}\approx g_{\uparrow\downarrow}\approx g_{\downarrow\downarrow}=g$. In the harmonic approximation, $\widetilde{U}_{p}/U_{p}=1/3$, and Eq.~\eqref{12} can be rewritten as
\begin{equation} 
\hat{H}_{\mathrm{int}}=\sum_{\boldsymbol{i}}\bigg\{\frac{U_{s}}{2}n_{\boldsymbol{i},s,\uparrow}(n_{\boldsymbol{i},s,\uparrow}-1)+U_{sp}n_{\boldsymbol{i},s,\uparrow}n_{\boldsymbol{i},p,\downarrow}+\frac{U_{p}}{2}\Big[n_{\boldsymbol{i},p,\downarrow}(n_{\boldsymbol{i},p,\downarrow}-\frac{2}{3})-\frac{1}{3}L_{\boldsymbol{i},z}^{2}\Big]\bigg\},\label{13}
\end{equation}
where $n_{\boldsymbol{i},p,\downarrow}=n_{\boldsymbol{i},p_{x},\downarrow}+n_{\boldsymbol{i},p_{y},\downarrow}$ and $L_{\boldsymbol{i},z}=-\mathrm{i}b_{\boldsymbol{i},p_{x},\downarrow}^{\dagger}b_{\boldsymbol{i},p_{y},\downarrow}+\mathrm{H.c.}$.
Remarkably, the emergence of angular momentum order $\langle L_{i,z} \rangle\neq 0$ will reduce the $p$-orbital interaction energy.

\subsection{Model parameters}
Here we provide the appropriate model parameters
\begin{equation}
V_{0}=5.0E_{r},\quad M_{0}=1.0E_{r},\quad g=0.01E_{r}.
\end{equation}
For tight-binding Hamiltonian: $t_{s}=0.0658E_{r}$, $t_{p}^{\parallel}=0.4228E_{r}$, $t_{p}^{\perp}=0.0658E_{r}$, $t_{so}=0.1059E_{r}$, $\mu_{s}=5.1909E_{r}$, $\mu_{p}=5.9049E_{r}$, $U_{s}=0.1044E_{r}$, $U_{sp}=0.0428E_{r}$, $U_{p}=0.0618E_{r}$, $\tilde{U}_{p}=0.0175E_{r}$. Moreover, $m_{z}$ can be adjusted by two-photon detuning $\delta$, with $m_{z}=0.00E_{r}\sim0.18 E_{r}$. Density of condensation bosons: $N_{0}/N=2\sim3$.
%---------------------------------------------------------------------------------------------------------------------------------

\section{Variational minimization}

The variational ground-state ansatz $|g\rangle \sim e^{\sqrt{N_{0}}b^{\dagger}}|\mathrm{vac}\rangle$, where  $b^{\dagger}=\sum_{n=1}^{4}\sum_{l_{\sigma}}\gamma_{\boldsymbol{k}_{c}^{n}}\beta_{\boldsymbol{k}_{c}^{n},l_{\sigma}}b_{\boldsymbol{k}_{c}^{n},l_{\sigma}}^{\dagger}$ for the many-body system is introduced in the main text. Within this ansatz, the energy density functional of the system is parametrized as 
\begin{align}
\mathcal{E}(\boldsymbol{k}_{c}^{n},\gamma_{\boldsymbol{k}_{c}^{n}},\beta_{\boldsymbol{k}_{c}^{n}})  &\equiv 
\langle g|\hat{\mathcal{H}}_{0}+\hat{\mathcal{H}}_{\mathrm{int}}|g\rangle/N \nonumber \\
&= n_0\sum_{n=1}^{4}\Big\{ |\gamma_{\boldsymbol{k}_{c}^{n}}|^{2}\Big[|\beta_{\boldsymbol{k}_{c}^{n},s_{\uparrow}}|^{2}\epsilon_{s\uparrow}(\boldsymbol{k}_{c}^{n})+|\beta_{\boldsymbol{k}_{c}^{n},p_{x\downarrow}}|^{2}\epsilon_{p_{x\downarrow}}(\boldsymbol{k}_{c}^{n})+|\beta_{\boldsymbol{k}_{c}^{n},p_{y\downarrow}}|^{2}\epsilon_{p_{y\downarrow}}(\boldsymbol{k}_{c}^{n})\Big]\nonumber \\
&\hspace{4em} -2\mathrm{i}t_{so}\sin \boldsymbol{k}_{c,x}^{n}\cdot|\gamma_{\boldsymbol{k}_{c}^{n}}|^{2}(\beta_{\boldsymbol{k}_{c}^{n},s_{\uparrow}}^{\ast}\beta_{\boldsymbol{k}_{c}^{n},p_{x\downarrow}}-\beta_{\boldsymbol{k}_{c}^{n},p_{x\downarrow}}^{\ast}\beta_{\boldsymbol{k}_{c}^{n},s_{\uparrow}})\nonumber \\
&\hspace{4em}-2\mathrm{i}t_{so}\sin \boldsymbol{k}_{c,y}^{n}\cdot|\gamma_{\boldsymbol{k}_{c}^{n}}|^{2}(\beta_{\boldsymbol{k}_{c}^{n},s_{\uparrow}}^{\ast}\beta_{\boldsymbol{k}_{c}^{n},p_{y\downarrow}}-\beta_{\boldsymbol{k}_{c}^{n},p_{y\downarrow}}^{\ast}\beta_{\boldsymbol{k}_{c}^{n},s_{\uparrow}})\Big\}\nonumber \\
&\hspace{1em}+n_0^2\sum_{\boldsymbol{k}_{1}+\boldsymbol{k}_{2}=\boldsymbol{k}_{3}+\boldsymbol{k}_{4}}\Big[  \frac{U_{s}}{2}\gamma_{\boldsymbol{k}_{3}}^{\ast}\beta_{\boldsymbol{k}_{3},s_{\uparrow}}^{\ast}\gamma_{\boldsymbol{k}_{4}}^{\ast}\beta_{\boldsymbol{k}_{4},s_{\uparrow}}^{\ast}\gamma_{\boldsymbol{k}_{2}}\beta_{\boldsymbol{k}_{2},s_{\uparrow}}\gamma_{\boldsymbol{k}_{1}}\beta_{\boldsymbol{k}_{1},s_{\uparrow}}\nonumber \\
&\hspace{4em} +  U_{sp}\sum_{\nu=x,y}\gamma_{\boldsymbol{k}_{3}}^{\ast}\beta_{\boldsymbol{k}_{3},p_{\nu\downarrow}}^{\ast}\gamma_{\boldsymbol{k}_{4}}^{\ast}\beta_{\boldsymbol{k}_{4},s_{\uparrow}}^{\ast}\gamma_{\boldsymbol{k}_{2}}\beta_{\boldsymbol{k}_{2},s_{\uparrow}}\gamma_{\boldsymbol{k}_{1}}\beta_{\boldsymbol{k}_{1},p_{\nu\downarrow}}\nonumber \\
&\hspace{4em}+  \frac{U_{p}}{2}\sum_{\nu=x,y}\gamma_{\boldsymbol{k}_{3}}^{\ast}\beta_{\boldsymbol{k}_{3},p_{\nu\downarrow}}^{\ast}\gamma_{\boldsymbol{k}_{4}}^{\ast}\beta_{\boldsymbol{k}_{4},p_{\nu\downarrow}}^{\ast}\gamma_{\boldsymbol{k}_{2}}\beta_{\boldsymbol{k}_{2},p_{\nu\downarrow}}\gamma_{\boldsymbol{k}_{1}}\beta_{\boldsymbol{k}_{1},p_{\nu\downarrow}}\nonumber \\
&\hspace{4em}+  \frac{\widetilde{U}_{p}}{2}(\gamma_{\boldsymbol{k}_{3}}^{\ast}\beta_{\boldsymbol{k}_{3},p_{x\downarrow}}^{\ast}\gamma_{\boldsymbol{k}_{4}}^{\ast}\beta_{\boldsymbol{k}_{4},p_{x\downarrow}}^{\ast}\gamma_{\boldsymbol{k}_{2}}\beta_{\boldsymbol{k}_{2},p_{y\downarrow}}\gamma_{\boldsymbol{k}_{1}}\beta_{\boldsymbol{k}_{1},p_{y\downarrow}}+\mathrm{H.c.})\nonumber \\
&\hspace{4em}+  2\widetilde{U}_{p}\gamma_{\boldsymbol{k}_{3}}^{\ast}\beta_{\boldsymbol{k}_{3},p_{y\downarrow}}^{\ast}\gamma_{\boldsymbol{k}_{4}}^{\ast}\beta_{\boldsymbol{k}_{4},p_{x\downarrow}}^{\ast}\gamma_{\boldsymbol{k}_{2}}\beta_{\boldsymbol{k}_{2},p_{x\downarrow}}\gamma_{\boldsymbol{k}_{1}}\beta_{\boldsymbol{k}_{1},p_{y\downarrow}}\Big],\label{13.1}
\end{align}
 where $\boldsymbol{k}_{1}\sim \boldsymbol{k}_{4} \in \{\boldsymbol{k}_{c}^{n}, n=1\sim4\}$. $\boldsymbol{k}_{c}^{n}$, $\gamma_{\boldsymbol{k}_{c}^{n}}$, and $\beta_{\boldsymbol{k}_{c}^{n}}$ are variational parameters. 

In this section, we present the numerical implementation of the simulated annealing (SA) algorithm ~\cite{vanLaarhoven1987}— a thermodynamically inspired global optimization method — to locate the global minimum of $\mathcal{E}(\boldsymbol{k}_{c}^{n},\gamma_{\boldsymbol{k}_{c}^{n}},\beta_{\boldsymbol{k}_{c}^{n}})$. The SA algorithm employs temperature-controlled stochastic exploration, progressively converging to the global optimum by mimicking thermal annealing processes.  
The SA protocol consists of:

\begin{itemize}
    \item \textbf{Initialization}: Select a set of $C_4$-symmetry related variational momenta $\{\boldsymbol{k}_{c}^{n}, n=1\sim4\}$, randomly generate initial parameters  $\gamma_{\boldsymbol{k}_{c}^{n}}$ and $\beta_{\boldsymbol{k}_{c}^{n}}$, with initial temperature \( T_{\text{max}} \sim 10^3 \).
    
    \item \textbf{Metropolis sampling}:
    For each temperature \( T \), a perturbation solution ($\gamma_{\boldsymbol{k}_{c}^{n}}'$,  $\beta_{\boldsymbol{k}_{c}^{n}}'$) is stochastically generated near ($\gamma_{\boldsymbol{k}_{c}^{n}}$,  $\beta_{\boldsymbol{k}_{c}^{n}}$) , with perturbation magnitudes temperature-dependent. Specifically, randomly generate another set of solutions ( $\gamma_{\boldsymbol{k}_{c}^{n}}^a$,  $\beta_{\boldsymbol{k}_{c}^{n}}^a$), thereby generating perturbation solution \( \big(\gamma_{\boldsymbol{k}_{c}^{n}}'=(1-\delta)\cdot \gamma_{\boldsymbol{k}_{c}^{n}}+\delta \cdot \gamma_{\boldsymbol{k}_{c}^{n}}^a,   \beta_{\boldsymbol{k}_{c}^{n}}'=(1-\delta)\cdot\beta_{\boldsymbol{k}_{c}^{n}}+\delta\cdot \beta_{\boldsymbol{k}_{c}^{n}}^a \big)\) near ($\gamma_{\boldsymbol{k}_{c}^{n}}$,  $\beta_{\boldsymbol{k}_{c}^{n}}$) . Here, $\delta\approx 0.15$ ( $T \ge 10^{-3}$ ) or $\delta\approx 0.015$ ( $T < 10^{-3}$ ) . 
    The acceptance probability for the perturbative solution is denoted as
    \begin{equation}
        P_{\text{accept}} = \begin{cases}
            1, & \Delta E \leq 0 \\
            \exp(-\Delta E / T), & \Delta E > 0
        \end{cases}
    \end{equation}
    where $\Delta E$ = $\mathcal{E}$($\boldsymbol{k}_{c}^{n},\gamma_{\boldsymbol{k}_{c}^{n}}',\beta_{\boldsymbol{k}_{c}^{n}}'$) - $\mathcal{E}$($\boldsymbol{k}_{c}^{n},\gamma_{\boldsymbol{k}_{c}^{n}},\beta_{\boldsymbol{k}_{c}^{n}}$).
    
    \item \textbf{Cooling schedule}: 
   Reduce temperature via \( T_{k+1} = \alpha T_k \) (\( \alpha \sim 0.95 \)), terminating when \( T \leq T_{\text{min}} (\sim 10^{-6}) \) and solution convergence is achieved.  Besides, a restart mechanism, i.e., temperature resets to \( T_{\text{max}} \) when \( T < T_{\text{re}} \) (reset threshold \(\sim 10^{-3}\)) up to \( N_{\text{re}} (\sim 10 ) \) times is used to mitigate local minima.

   \item \textbf{Ground state identification}:
   Change $\{\boldsymbol{k}_{c}^{n}, n=1\sim4\}$ and repeat the previous steps to identify the variational parameter values corresponding to the global minimum of the energy functional $\mathcal{E}$($\boldsymbol{k}_{c}^{n},\gamma_{\boldsymbol{k}_{c}^{n}},\beta_{\boldsymbol{k}_{c}^{n}}$) as the BEC condensation momenta \( \{\widetilde{\boldsymbol{k}}_{c}^{n}\}\), as well as condensation parameters \(\{\widetilde{\gamma}_{\boldsymbol{k}_{c}^{n}},  \widetilde{\beta}_{\boldsymbol{k}_{c}^{n}} \} \).

\end{itemize}

%---------------------------------------------------------------------------------------------------------------------------------
\section{Bogoliubov mean field theory}
As presented in the main text, we discover two types of exotic condensed ground states of bosons: the uniform angular momentum superfluid phase and the two-dimensional spin-orbital supersolid phase, along with their respective non-trivial quasiparticle excitations. In this section, we will utilize the Bogoliubov-de Gennes (BdG) mean field theory to obtain the corresponding BdG Hamiltonian based on these two ground state phases.
\subsection{Uniform angular momentum superfluid phase}
%Let's first examine the Bogoliubov excitations of the UAMS. The time-reversal symmetry is broken due to the uniform distribution of angular momentum order in real space, which implies that the excitation spectrum of Bogoliubov quasi-particles based on superfluid ground state will open a gap at the \textbf{$\boldsymbol{\Gamma}$} point and obtain a non-zero Chern number. We will demonstrate this in the following.

First, we derive the Bogoliubov excitation spectrum based on uniform angular momentum superfluid phase (UAMSF). We perform a Fourier transformation on the total Hamiltonian $\hat{H}=\hat{H}_{0}+\hat{H}_{\mathrm{int}}$, using $b_{\mathrm{\boldsymbol{i}},l,\sigma}=\frac{1}{\sqrt{N}}\sum_{\boldsymbol{k}}b_{\boldsymbol{k},l,\sigma}e^{\mathrm{i}\boldsymbol{k}\cdot\mathrm{\boldsymbol{r}_{i}}}$. Consequently, the total Hamiltonian in momentum space is given by 
\begin{align}
\hat{\mathcal{H}}=&\sum_{\boldsymbol{k}}\Big[  \epsilon_{s\uparrow}(\boldsymbol{k})b_{\boldsymbol{k},s,\uparrow}^{\dagger}b_{\boldsymbol{k},s,\uparrow}+\epsilon_{p_{x}\downarrow}(\boldsymbol{k})b_{\boldsymbol{k},p_{x},\downarrow}^{\dagger}b_{\boldsymbol{k},p_{x},\downarrow}+\epsilon_{p_{y}\downarrow}(\boldsymbol{k})b_{\boldsymbol{k},p_{y},\downarrow}^{\dagger}b_{\boldsymbol{k},p_{y},\downarrow}\nonumber \\
&-  (2\mathrm{i}t_{so}\sin \boldsymbol{k}_{x}b_{\boldsymbol{k},s,\uparrow}^{\dagger}b_{\boldsymbol{k},p_{x},\downarrow}+2\mathrm{i}t_{so}\sin \boldsymbol{k}_{y}b_{\boldsymbol{k},s,\uparrow}^{\dagger}b_{\boldsymbol{k},p_{y},\downarrow}+\mathrm{H.c.})\Big]\nonumber \\
&+\frac{1}{N}\sum_{\boldsymbol{k}_{1}+\boldsymbol{k}_{2}=\boldsymbol{k}_{3}+\boldsymbol{k}_{4}}\Big[  \frac{U_{s}}{2}b_{\boldsymbol{k}_{3},s,\uparrow}^{\dagger}b_{\boldsymbol{k}_{4},s,\uparrow}^{\dagger}b_{\boldsymbol{k}_{2},s,\uparrow}b_{\boldsymbol{k}_{1},s,\uparrow}+U_{sp}\sum_{\nu=x,y}b_{\boldsymbol{k}_{3},p_{\nu},\downarrow}^{\dagger}b_{\boldsymbol{k}_{4},s,\uparrow}^{\dagger}b_{\boldsymbol{k}_{2},s,\uparrow}b_{\boldsymbol{k}_{1},p_{\nu},\downarrow}\nonumber \\
&+  \frac{U_{p}}{2}\sum_{\nu=x,y}b_{\boldsymbol{k}_{3},p_{\nu},\downarrow}^{\dagger}b_{\boldsymbol{k}_{4},p_{\nu},\downarrow}^{\dagger}b_{\boldsymbol{k}_{2},p_{\nu},\downarrow}b_{\boldsymbol{k}_{1},p_{\nu},\downarrow}+2\widetilde{U}_{p}b_{\boldsymbol{k}_{3},p_{y},\downarrow}^{\dagger}b_{\boldsymbol{k}_{4},p_{x},\downarrow}^{\dagger}b_{\boldsymbol{k}_{2},p_{x},\downarrow}b_{\boldsymbol{k}_{1},p_{y},\downarrow}\nonumber \\
&+  \frac{\widetilde{U}_{p}}{2}(b_{\boldsymbol{k}_{3},p_{x},\downarrow}^{\dagger}b_{\boldsymbol{k}_{4},p_{x},\downarrow}^{\dagger}b_{\boldsymbol{k}_{2},p_{y},\downarrow}b_{\boldsymbol{k}_{1},p_{y},\downarrow}+b_{\boldsymbol{k}_{3},p_{y},\downarrow}^{\dagger}b_{\boldsymbol{k}_{4},p_{y},\downarrow}^{\dagger}b_{\boldsymbol{k}_{2},p_{x},\downarrow}b_{\boldsymbol{k}_{1},p_{x},\downarrow})\Big].\label{14}
\end{align} 
%where $\epsilon_{s\uparrow}(\boldsymbol{k})=\mu_{s}+m_{z}-2t_{s}(\cos \boldsymbol{k}_{x}+\cos \boldsymbol{k}_{y})$, $\epsilon_{p_{x}\downarrow}(\boldsymbol{k})=\mu_{p}-m_{z}-2t_{p}^{\parallel}\cos \boldsymbol{k}_{x}-2t_{p}^{\perp}\cos \boldsymbol{k}_{y}$, and $\epsilon_{p_{y}\downarrow}(\boldsymbol{k})=\mu_{p}-m_{z}-2t_{p}^{\perp}\cos \boldsymbol{k}_{x}-2t_{p}^{\parallel}\cos \boldsymbol{k}_{y}$ represent the energy bands for $s_{\uparrow}$, $p_{x\downarrow}$, and $p_{y\downarrow}$ states, respectively.

We have known that the ground-state wave function for the UAMSF takes the form
\begin{equation}
|\mathrm{G}\rangle = e^{\sqrt{N_{0}}(\beta_{\boldsymbol{\Gamma},p_{x\downarrow}}b_{\boldsymbol{\Gamma},p_{x},\downarrow}^{\dagger}+\beta_{\boldsymbol{\Gamma},p_{y\downarrow}}b_{\boldsymbol{\Gamma},p_{y},\downarrow}^{\dagger})}|\mathrm{vac}\rangle, 
\hspace{2em} \beta_{\boldsymbol{\mathrm{\Gamma}},p_{x\downarrow}}=\mathrm{-i}\beta_{\boldsymbol{\mathrm{\Gamma}},p_{y\downarrow}}.
\end{equation}
It is easy to see that 
\begin{align}
\langle\mathrm{G}|b_{\boldsymbol{k},s,\uparrow}|\mathrm{G}\rangle =  0, \hspace{1em}
\langle\mathrm{G}|b_{\boldsymbol{k},p_{x},\downarrow}|\mathrm{G}\rangle= \beta_{\boldsymbol{\Gamma},p_{x\downarrow}}\sqrt{N_{0}}\delta_{\boldsymbol{k},\boldsymbol{\Gamma}},
\hspace{1em}
\langle\mathrm{G}|b_{\boldsymbol{k},p_{y},\downarrow}|\mathrm{G}\rangle= \beta_{\boldsymbol{\Gamma},p_{y\downarrow}}\sqrt{N_{0}}\delta_{\boldsymbol{k},\boldsymbol{\Gamma}}.
\end{align}
Under the mean field approximation~\cite{zhai2021ultracold}, we can obtain the BdG Hamiltonian $\hat{\mathcal{H}}_{\mathrm{BdG}}=\frac{1}{2}\sum_{\boldsymbol{k}}\Psi(\boldsymbol{k})^{\dagger}\mathcal{H}_{\mathrm{BdG}}(\boldsymbol{k})\Psi(\boldsymbol{k})$, where the Nambu basis is chosen as $\Psi(\boldsymbol{k})=$$\left(\begin{array}{cccccc}
b_{\boldsymbol{k},s,\uparrow} & b_{\boldsymbol{k},p_{x},\downarrow} & b_{\boldsymbol{k},p_{y},\downarrow} & b_{-\boldsymbol{k},s,\uparrow}^{\dagger} & b_{-\boldsymbol{k},p_{x},\downarrow}^{\dagger} & b_{-\boldsymbol{k},p_{y},\downarrow}^{\dagger}\end{array}\right)^T$, and the BdG Hamiltonian matrix reads
\begin{equation}
\mathcal{H}_{\mathrm{BdG}}(\boldsymbol{k})=\left(\begin{array}{cc}
\mathrm{h}_{0}(\boldsymbol{k}) & \mathrm{h}_{\mathrm{int}}(\boldsymbol{k})\\
\mathrm{h}_{\mathrm{int}}^{\dagger}(\boldsymbol{k}) & \mathrm{h}_{0}^{\ast}(-\boldsymbol{k})
\end{array}\right).
\label{17}
\end{equation}
Note that the diagonal element $\mathrm{h}_{0}(\boldsymbol{k}) =\mathcal{H}_{0}(\boldsymbol{k})+\mathcal{H}_{0,\mathrm{int}}^{\prime}-\mu$, where the first term takes the form of 
\begin{equation}
\mathcal{H}_{0}(\boldsymbol{k})=\left(\begin{array}{ccc}
\epsilon_{s\uparrow}(\boldsymbol{k}) & -\mathrm{i}2t_{so}\sin\boldsymbol{k}_x & -\mathrm{i}2t_{so}\sin\boldsymbol{k}_y\\
\mathrm{i}2t_{so}\sin\boldsymbol{k}_x & \epsilon_{p_{x}\downarrow}(\boldsymbol{k}) & 0\\
\mathrm{i}2t_{so}\sin\boldsymbol{k}_y & 0 & \epsilon_{p_{y}\downarrow}(\boldsymbol{k})
\end{array}\right),
\end{equation}
and the second term reads $\mathcal{H}_{0,\mathrm{int}}^{\prime}=\textbf{Diag}\left(\begin{array}{ccc}h_{11}, & h_{22}, & h_{33}\end{array}\right)$ with $h_{11}= 2U_{sp}n_0|\beta_{\boldsymbol{\Gamma},p_{x\downarrow}}|^{2}$, $h_{22}= h_{33}= 2(U_{p}+\widetilde{U}_{p})n_0|\beta_{\boldsymbol{\Gamma},p_{x\downarrow}}|^{2}$, and $n_0=\frac{N_{0}}{N}$.
Moreover, the third term $\mu$ denotes the chemical potential. As for the off-diagonal element in Eq.(\ref{17}), 
\begin{equation}
\mathrm{h}_{\mathrm{int}}(\boldsymbol{k})=n_0\beta_{\boldsymbol{\Gamma},p_{x\downarrow}}^{2}\left(\begin{array}{ccc}
0 & 0 & 0\\
0 &  U_{p} - \widetilde{U}_{p} & \text{i}2\widetilde{U}_{p}\\
0 & \text{i}2\widetilde{U}_{p} &  \widetilde{U}_{p} - U_{p}  
\end{array}\right).
\end{equation}
Performing the Bogoliubov transformation, we have $T_{\boldsymbol{k}}^{\dagger}\mathcal{H}_{\mathrm{BdG}}(\boldsymbol{k})T_{\boldsymbol{k}}=E_{\boldsymbol{k}}$, where the para-unitary matrix $T_{\boldsymbol{k}}$ satisfies $T_{\boldsymbol{k}}^{\dagger}\tau_{z}T_{\boldsymbol{k}}=\tau_{z}$ (\textbf{$\tau_{z}=\sigma_{z}\otimes\mathrm{\boldsymbol{I}}_{3\times3}$}), and the diagonal terms of $E_{\boldsymbol{k}}$ represent the excitation spectrum. Furthermore, the $j^{\mathrm{th}}$ column of $T_{\boldsymbol{k}}$ denotes the eigenstate $|T_{\boldsymbol{k}}^{j}\rangle$ of the $j^{\mathrm{th}}$ band. To investigate the topological properties of the excitation spectrum, we can calculate the corresponding topological invariant:
\begin{equation}
\C_{j}=\frac{1}{2\pi}\int d\boldsymbol{k}B_{j}(\boldsymbol{k}),
\end{equation}
where the Berry curvature reads
\begin{equation}
B_{j}(\boldsymbol{k})=\partial_{k_{x}}\mathcal{A}_{y}^{j}(\boldsymbol{k})-\partial_{k_{y}}\mathcal{A}_{x}^{j}(\boldsymbol{k}),
\end{equation}
and the Berry connection is given by 
\begin{equation}
\mathcal{A}_{\nu}^{j}(\boldsymbol{k})=-\mathrm{i}\langle T_{\boldsymbol{k}}^{j}|\tau_{z}\partial_{k_{\nu}}|T_{\boldsymbol{k}}^{j}\rangle.
\end{equation}
The Chern number of each band is shown in the Fig.3(b) in the main text.

\subsection{Two-dimensional spin-orbital supersolid phase}
\begin{figure}[tb]
\centering
\includegraphics[scale=0.22]{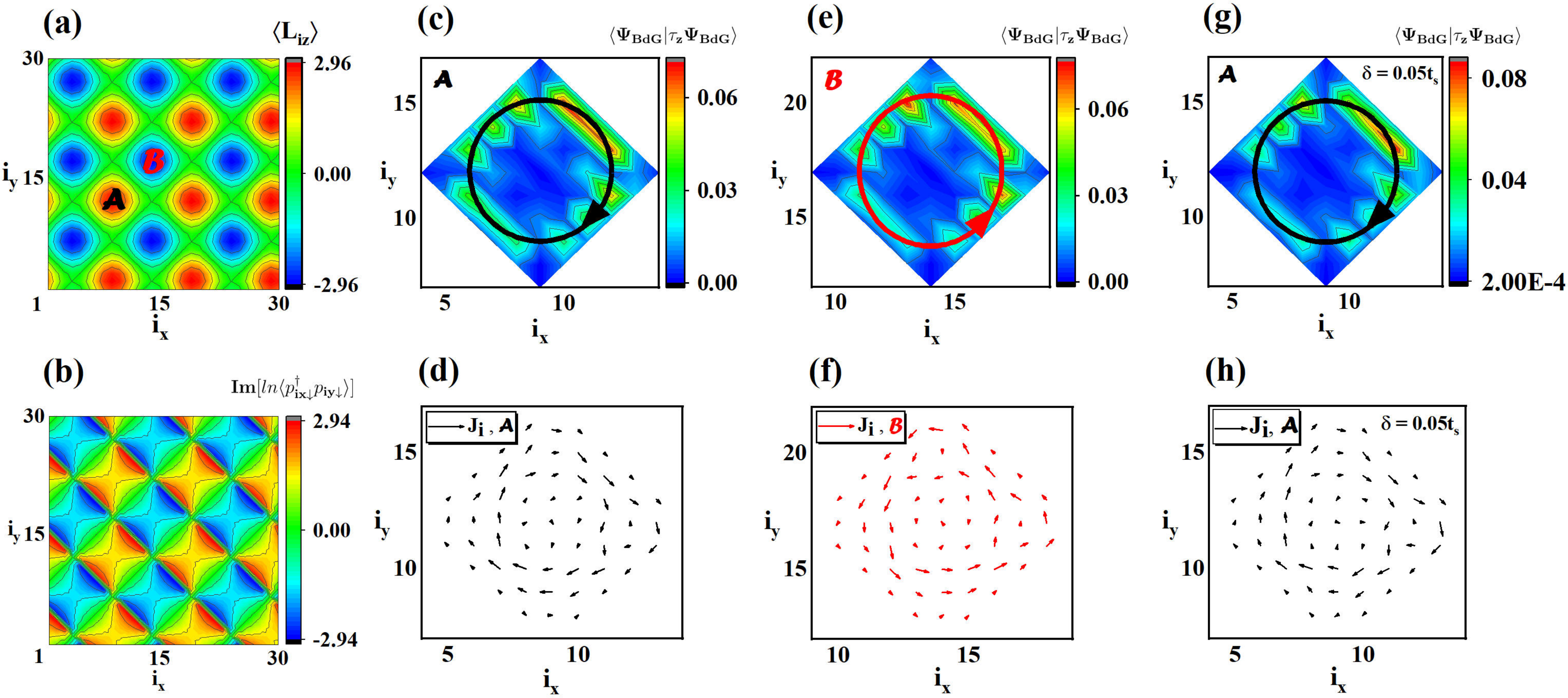}
\caption{ (a) Spatial pattern of orbital angular momentum order $\langle L_{iz}\rangle$ for 2D SOSS. Here, we take the positive (A) and negative (B) clusters as examples to show the topological quasiparticle excitations on the cluster boundary. (b) Spatial distribution of the relative phase difference $\theta_i=\text{Im}\Big[\mathrm{ln}\langle b_{\boldsymbol{i},p_{x},\downarrow}^{\dagger}b_{\boldsymbol{i},p_{y},\downarrow}\rangle\Big]$ between the $p_x$ and $p_y$ orbitals. There exist excited states that are distributed along the cluster boundary [(c) and (e)]. (d) and (f) show the expectation values of the current operators for the corresponding excited states, indicating that such boundary quasiparticle excitations of clusters A and B propagate clockwise and counterclockwise, respectively. Furthermore, we introduce the on-site random disorders with magnitude 0.05$t_{s}$ to the positive cluster A. As shown in (g) and (h), the edge excitation remains robust, manifesting that it is a topological boundary excitation. However, the topological boundary excitations at A and B have opposite chiralities, which results in the zero chiral current globally. In conclusion, the quasiparticle excitations for the 2D SOSS embody the properties of “topological
in each cluster, trivial in the overall bulk”.}
\label{fig3}
\end{figure}

For the two-dimensional spin-orbital supersolid phase (SOSS), Bose-Einstein condensation(BEC) occurs equally at four momentum points $\tilde{\boldsymbol{k}}_c^n$ ($n=1\sim 4$), and the wave function for this ground state reads 
\begin{equation} 
|\mathrm{G}\rangle = e^{\sqrt{N_{0}}\sum_{n}\gamma_{\tilde{\boldsymbol{k}}_{c}^{n}}(\beta_{\tilde{\boldsymbol{k}}_{c}^{n},s_{\uparrow}}b_{\tilde{\boldsymbol{k}}_{c}^{n},s,\uparrow}^{\dagger}+\beta_{\tilde{\boldsymbol{k}}_{c}^{n},p_{x\downarrow}}b_{\tilde{\boldsymbol{k}}_{c}^{n},p_{x},\downarrow}^{\dagger}+\beta_{\tilde{\boldsymbol{k}}_{c}^{n},p_{y\downarrow}}b_{\tilde{\boldsymbol{k}}_{c}^{n},p_{y},\downarrow}^{\dagger})}|\mathrm{vac}\rangle, \label{20}
\end{equation}
with parameters fulfilling $|\gamma_{\widetilde{\boldsymbol{k}}_{c}^{n}}|^{2}=\frac{1}{4}$ and $\beta_{\widetilde{\boldsymbol{k}}_{c}^{n},p_{x\downarrow}}=-(-1)^n\beta_{\widetilde{\boldsymbol{k}}_{c}^{n},p_{y\downarrow}}$.
As discussed in the main text, this ground state features a staggered pattern of positive and negative angular momentum clusters [Fig.\ref{fig3} (a)], resulting in zero net angular momentum in the real space. Here, we focus on a single angular momentum cluster to study its quasiparticle topological excitation. Since the lattice translational symmetry is broken within the angular momentum cluster, we can analyze the topological excitation in real space. 

Similarly, the BEC order parameters for the ground state given in Eq.\eqref{20} are 
\begin{align}
\langle\mathrm{G}|b_{\boldsymbol{i},s,\uparrow}|\mathrm{G}\rangle= & \sqrt{n_0}\sum_{n}\gamma_{\tilde{\boldsymbol{k}}_{c}^{n}}\beta_{\tilde{\boldsymbol{k}}_{c}^{n},s_{\uparrow}}e^{\mathrm{i}\tilde{\boldsymbol{k}}_{c}^{n}\cdot\mathrm{\boldsymbol{r}_{i}}},
\nonumber \\
\langle\mathrm{G}|b_{\boldsymbol{i},p_{x},\downarrow}|\mathrm{G}\rangle= & \sqrt{n_0}\sum_{n}\gamma_{\tilde{\boldsymbol{k}}_{c}^{n}}\beta_{\tilde{\boldsymbol{k}}_{c}^{n},p_{x\downarrow}}e^{\mathrm{i}\tilde{\boldsymbol{k}}_{c}^{n}\cdot\mathrm{\boldsymbol{r}_{i}}},
\nonumber \\
\langle\mathrm{G}|b_{\boldsymbol{i},p_{y},\downarrow}|\mathrm{G}\rangle= & \sqrt{n_0}\sum_{n}\gamma_{\tilde{\boldsymbol{k}}_{c}^{n}}\beta_{\tilde{\boldsymbol{k}}_{c}^{n},p_{y\downarrow}}e^{\mathrm{i}\tilde{\boldsymbol{k}}_{c}^{n}\cdot\mathrm{\boldsymbol{r}_{i}}}.
\end{align}
With the Bogoliubov mean-field approximation, the BdG Hamiltonian in real space can be written as 
\begin{align}
\hat{H}_{\mathrm{BdG}}= \hat{H}_0
+n_0\sum_{\boldsymbol{i}}\sum_{n_{1},n_{2}}\Big\{ & [2U_{s}\beta_{s_{\uparrow}}^{n_{1}\ast}\beta_{s_{\uparrow}}^{n_{2}}+U_{sp}(\beta_{p_{x\downarrow}}^{n_{1}\ast}\beta_{p_{x\downarrow}}^{n_{2}}+\beta_{p_{y\downarrow}}^{n_{1}\ast}\beta_{p_{y\downarrow}}^{n_{2}})]\cdot e^{i\cdot(\tilde{\boldsymbol{k}}_{c}^{n_{2}}-\tilde{\boldsymbol{k}}_{c}^{n_{1}})\cdot\boldsymbol{r}_{i}}b_{\boldsymbol{i},s,\uparrow}^{\dagger}b_{\boldsymbol{i},s,\uparrow}\nonumber \\
+ & [2U_{p}\beta_{p_{x\downarrow}}^{n_{1}\ast}\beta_{p_{x\downarrow}}^{n_{2}}+2\tilde{U}_{p}\beta_{p_{y\downarrow}}^{n_{1}\ast}\beta_{p_{y\downarrow}}^{n_{2}}+U_{sp}\beta_{s_{\uparrow}}^{n_{1}\ast}\beta_{s_{\uparrow}}^{n_{2}}]\cdot e^{i\cdot(\tilde{\boldsymbol{k}}_{c}^{n_{2}}-\tilde{\boldsymbol{k}}_{c}^{n_{1}})\cdot\boldsymbol{r}_{i}}b_{\boldsymbol{i},p_{x},\downarrow}^{\dagger}b_{\boldsymbol{i},p_{x},\downarrow}\nonumber \\
+ & [2U_{p}\beta_{p_{y\downarrow}}^{n_{1}\ast}\beta_{p_{y\downarrow}}^{n_{2}}+2\tilde{U}_{p}\beta_{p_{x\downarrow}}^{n_{1}\ast}\beta_{p_{x\downarrow}}^{n_{2}}+U_{sp}\beta_{s_{\uparrow}}^{n_{1}\ast}\beta_{s_{\uparrow}}^{n_{2}}]\cdot e^{i\cdot(\tilde{\boldsymbol{k}}_{c}^{n_{2}}-\tilde{\boldsymbol{k}}_{c}^{n_{1}})\cdot\boldsymbol{r}_{i}}b_{\boldsymbol{i},p_{y},\downarrow}^{\dagger}b_{\boldsymbol{i},p_{y},\downarrow}\nonumber \\
+ & [(2\tilde{U}_{p}\beta_{p_{x\downarrow}}^{n_{1}\ast}\beta_{p_{y\downarrow}}^{n_{2}}+2\tilde{U}_{p}\beta_{p_{y\downarrow}}^{n_{1}\ast}\beta_{p_{x\downarrow}}^{n_{2}})\cdot e^{i\cdot(\tilde{\boldsymbol{k}}_{c}^{n_{2}}-\tilde{\boldsymbol{k}}_{c}^{n_{1}})\cdot\boldsymbol{r}_{i}}b_{\boldsymbol{i},p_{x},\downarrow}^{\dagger}b_{\boldsymbol{i},p_{y},\downarrow}+\mathrm{H.c.}]\nonumber \\
+ & [U_{sp}\beta_{s_{\uparrow}}^{n_{1}\ast}\beta_{p_{x\downarrow}}^{n_{2}}\cdot e^{i\cdot(\tilde{\boldsymbol{k}}_{c}^{n_{2}}-\tilde{\boldsymbol{k}}_{c}^{n_{1}})\cdot\boldsymbol{r}_{i}}b_{\boldsymbol{i},p_{x},\downarrow}^{\dagger}b_{\boldsymbol{i},s,\uparrow}+U_{sp}\beta_{p_{x\downarrow}}^{n_{1}\ast}\beta_{s_{\uparrow}}^{n_{2}}\cdot e^{i\cdot(\tilde{\boldsymbol{k}}_{c}^{n_{2}}-\tilde{\boldsymbol{k}}_{c}^{n_{1}})\cdot\boldsymbol{r}_{i}}b_{\boldsymbol{i},s,\uparrow}^{\dagger}b_{\boldsymbol{i},p_{x},\downarrow}]\nonumber \\
+ & [U_{sp}\beta_{s_{\uparrow}}^{n_{1}\ast}\beta_{p_{y\downarrow}}^{n_{2}}\cdot e^{i\cdot(\tilde{\boldsymbol{k}}_{c}^{n_{2}}-\tilde{\boldsymbol{k}}_{c}^{n_{1}})\cdot\boldsymbol{r}_{i}}b_{\boldsymbol{i},p_{y},\downarrow}^{\dagger}b_{\boldsymbol{i},s,\uparrow}+\mathrm{H.c.}]\nonumber \\
+ & [\frac{1}{2}U_{s}\beta_{s_{\uparrow}}^{n_{1}\ast}\beta_{s_{\uparrow}}^{n_{2}\ast}\cdot e^{-i\cdot(\tilde{\boldsymbol{k}}_{c}^{n_{2}}+\tilde{\boldsymbol{k}}_{c}^{n_{1}})\cdot\boldsymbol{r}_{i}}b_{\boldsymbol{i},s,\uparrow}b_{\boldsymbol{i},s,\uparrow}+\mathrm{H.c.}]\nonumber \\
+ & [\frac{1}{2}(U_{p}\beta_{p_{x\downarrow}}^{n_{1}\ast}\beta_{p_{x\downarrow}}^{n_{2}\ast}+\tilde{U}_{p}\beta_{p_{y\downarrow}}^{n_{1}\ast}\beta_{p_{y\downarrow}}^{n_{2}\ast})\cdot e^{-i\cdot(\tilde{\boldsymbol{k}}_{c}^{n_{2}}+\tilde{\boldsymbol{k}}_{c}^{n_{1}})\cdot\boldsymbol{r}_{i}}b_{\boldsymbol{i},p_{x},\downarrow}b_{\boldsymbol{i},p_{x},\downarrow}+\mathrm{H.c.}]\nonumber \\
+ & [\frac{1}{2}(U_{p}\beta_{p_{y\downarrow}}^{n_{1}\ast}\beta_{p_{y\downarrow}}^{n_{2}\ast}+\tilde{U}_{p}\beta_{p_{x\downarrow}}^{n_{1}\ast}\beta_{p_{x\downarrow}}^{n_{2}\ast})\cdot e^{-i\cdot(\tilde{\boldsymbol{k}}_{c}^{n_{2}}+\tilde{\boldsymbol{k}}_{c}^{n_{1}})\cdot\boldsymbol{r}_{i}}b_{\boldsymbol{i},p_{y},\downarrow}b_{\boldsymbol{i},p_{y},\downarrow}+\mathrm{H.c.}]\nonumber \\
+ & [2\tilde{U}_{p}\beta_{p_{x\downarrow}}^{n_{1}\ast}\beta_{p_{y\downarrow}}^{n_{2}\ast}\cdot e^{-i\cdot(\tilde{\boldsymbol{k}}_{c}^{n_{2}}+\tilde{\boldsymbol{k}}_{c}^{n_{1}})\cdot\boldsymbol{r}_{i}}b_{\boldsymbol{i},p_{x},\downarrow}b_{\boldsymbol{i},p_{y},\downarrow}+\mathrm{H.c.}]\nonumber \\
+ & [U_{sp}\beta_{p_{x\downarrow}}^{n_{1}\ast}\beta_{s_{\uparrow}}^{n_{2}\ast}\cdot e^{-i\cdot(\tilde{\boldsymbol{k}}_{c}^{n_{2}}+\tilde{\boldsymbol{k}}_{c}^{n_{1}})\cdot\boldsymbol{r}_{i}}b_{\boldsymbol{i},p_{x},\downarrow}b_{\boldsymbol{i},s,\uparrow}+U_{sp}\beta_{s_{\uparrow}}^{n_{1}}\beta_{p_{x\downarrow}}^{n_{2}}\cdot e^{i\cdot(\tilde{\boldsymbol{k}}_{c}^{n_{2}}+\tilde{\boldsymbol{k}}_{c}^{n_{1}})\cdot\boldsymbol{r}_{i}}b_{\boldsymbol{i},s,\uparrow}^{\dagger}b_{\boldsymbol{i},p_{x},\downarrow}^{\dagger}]\nonumber \\
+ & [U_{sp}\beta_{p_{y\downarrow}}^{n_{1}\ast}\beta_{s_{\uparrow}}^{n_{2}\ast}\cdot e^{-i\cdot(\tilde{\boldsymbol{k}}_{c}^{n_{2}}+\tilde{\boldsymbol{k}}_{c}^{n_{1}})\cdot\boldsymbol{r}_{i}}b_{\boldsymbol{i},p_{y},\downarrow}b_{\boldsymbol{i},s,\uparrow}+\mathrm{H.c.}]\Big\},
\end{align}
where $\beta_{l_{\sigma}}^{n}\equiv\gamma_{\tilde{\boldsymbol{k}}_{c}^{n}}\beta_{\tilde{\boldsymbol{k}}_{c}^{n},l_{\sigma}}$. Here, we focus on the angular momentum clusters A and B in Fig.\ref{fig3} (a) and solve the corresponding real space excitation spectrum to search for the excited states that primarily distribute on the cluster boundary and can withstand impurities or disorders (the essential properties of topological edge states). Calculating the expectation values of the current operators $\hat{J}_{\boldsymbol{i}}$ for the selected excited state, we find that there exists a chiral topological edge excitation [Fig.\ref{fig3} (c) and (e)] at each angular momentum cluster. Notably, the positive (A) and negative (B) clusters exhibit edge excitation with opposite chiralities [Fig.\ref{fig3} (d) and (f)]. 

\begin{figure}[htb]
\centering
\includegraphics[scale=0.85]{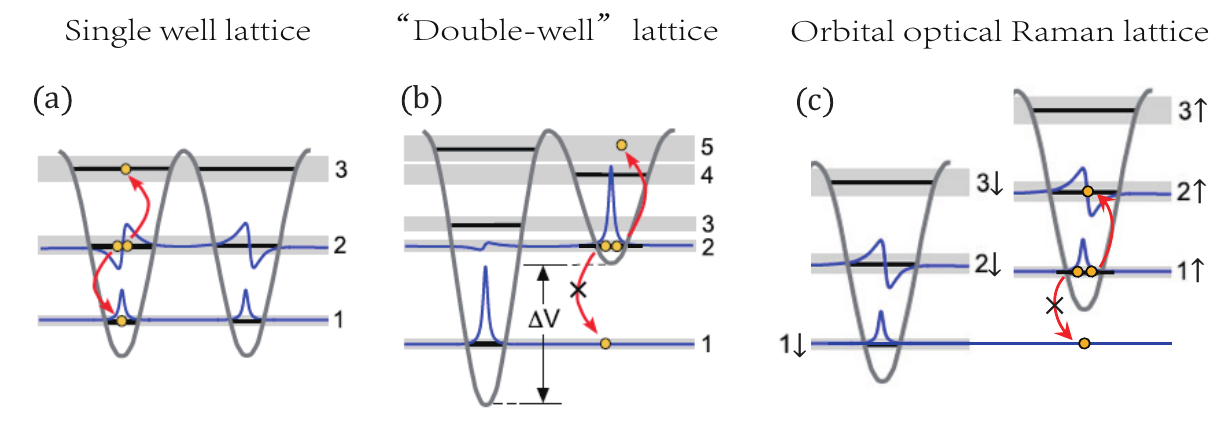}
\caption{ The similarity between the “double-well” lattice and the orbital optical Raman lattice in prolonging the lifetime of $p$-orbital bosons. 
Here, Fig.(a) and (b) are taken from Ref.\cite{kock2016orbital}.
(a). The $p$-orbital bosons are usually unstable and will soon decay through an elastic scattering process $(p,p)\Rightarrow(s,d)$ within a single well. (b). In the “double-well” lattice, the $s$-orbital of the shallow well is close to the $p$-orbital of the deep well. 
As demonstrated in Ref.\cite{kock2016orbital}, such a configuration can be tuned to clearly slow down the decay of the boson population in the mixed $s$-orbital in the shallow well and $p$-orbital in the deep well, but appreciable decay can still proceed according to the elastic scattering process shown in (a). 
%and the population of $s$-orbital bosons will delay the elastic scattering process shown in (a)~\cite{kock2016orbital}.
(c). In the orbital optical Raman lattice, Raman coupling can effectively bring the $s_{\uparrow}$ and $p_{\downarrow}$ orbitals closer together and facilitate the nearest-neighbor $p_{\downarrow}$-$s_{\uparrow}$ hopping. This causes bosons to populate the $s_{\uparrow}$-orbital, thereby further substantially suppressing the decay process $(p_{\downarrow},p_{\downarrow})\Rightarrow(s_{\downarrow},d_{\downarrow})$ compared with the already successful step in (b).
 }
\label{fig4}
\end{figure}

\section{Lifetime of the exotic many-body ground states}
The uniform angular momentum superfluid phase and the two-dimensional spin-orbital supersolid phase are metastable states because they occupy the $p$-band. $P$-orbital bosons are generally unstable and decay under interaction, even when the latter are weak. In the spinless cases, an elastic decay channel that conserves total energy involves two bosons initially in the $p$ band scattering into a final state where one boson occupies the $s$ band and the other the $d$ band [Fig.\ref{fig4} (a)]. The currently used “double-well" lattice scheme effectively extends the lifetime of $p$-orbital bosons. As shown in Fig.\ref{fig4} (b), a deep well and a shallow well are arranged so that the $p$-orbital in the deep well is nearly resonant with the $s$ orbital in the shallow well. Bosons can tunnel from the deep-well $p$-orbital to the nearest-neighbor shallow-well $s$ orbital, and because the two $s$-orbitals have negligible overlap, the shallow-well $s$-orbital bosons do not decay like the deep-well $p$-orbital bosons~\cite{kock2016orbital}. This yields a relatively stable high-orbital many-body state. To date, many successful experiments on quantum gases have employed the double-well lattice scheme---e.g.~\cite{wirth2011evidence,PhysRevLett.121.265301,jin2021evidence,wang2021evidence}, with $p$-orbital lifetimes ranging from tens to hundreds of microseconds. Our orbital optical Raman lattice plays an analogous role: the sublattice degree of freedom in the double-well lattice is replaced by a spin degree of freedom. Although two $p_{\downarrow}$ bosons can scatter into an $s_{\downarrow}$ and a $d_{\downarrow}$ boson, the Raman-induced nearest-neighbor $p_{\downarrow}$–$s_{\uparrow}$ transition populates $s_{\uparrow}$ orbitals. As illustrated in Fig.\ref{fig4}(c), these $s_{\uparrow}$ bosons do not decay because spin is conserved, so the high-orbital states remain stable. Consequently, we expect the lifetimes of the $p$-orbital many-body phases in our scheme to lie in the same range. 

As for heating, several experiments on optical Raman lattices—such as~\cite{Wu2016Science,sunPRL2018,sunPRL2018_1,wang2021realization}—demonstrate that Raman-induced heating can be well controlled. Moreover, Fig.1(a) in the main text shows that our orbital optical Raman lattice is simpler than these previous set-ups, so heating will not impede experimental implementation or affect the main physical results.

\bibliographystyle{unsrt}
\bibliography{Supplementary_Material}